\documentclass[aps,prb,reprint,groupedaddress]{revtex4-1}

\usepackage{amsmath}
\usepackage{amssymb}
\usepackage{amsfonts}
\usepackage{bbold}
\usepackage{graphicx}
\usepackage{hyperref}

\hypersetup{
  colorlinks   = true, 
  urlcolor     = blue, 
  linkcolor    = blue, 
  citecolor   = blue 
}

\newcommand{\ket}[1]{\left| #1 \right\rangle}

\newcommand{\ketbra}[2]{\left| #1 \right\rangle\!\!\left\langle #2 \right|}

\begin{document}

\title{Optimal dispersive readout of a spin qubit with a microwave resonator}

\author{B. D'Anjou}
\author{Guido Burkard}

\affiliation{Department of Physics, University of Konstanz, D-78457 Konstanz, Germany}



\begin{abstract}
Strong coupling of semiconductor spin qubits to superconducting microwave resonators was recently demonstrated~[X. Mi et al., \href{https://www.doi.org/10.1038/nature25769}{Nature 555, 599 (2018)}; A. J. Landig et al., \href{https://www.doi.org/10.1038/s41586-018-0365-y}{Nature 560, 179 (2018)}; N. Samkharadze et
al., \href{https://www.doi.org/10.1126/science.aar4054}{Science 359, 1123 (2018)}; T. Cubaynes et al., \href{https://www.doi.org/10.1038/s41534-019-0169-4}{npj Quantum Inf. 5, 47 (2019)}]. These breakthroughs pave the way for quantum information processing that combines the long coherence times of solid-state spin qubits with the long-distance connectivity, fast control, and fast high-fidelity quantum-non-demolition readout of existing superconducting qubit implementations. Here, we theoretically analyze and optimize the dispersive readout of a single spin in a semiconductor double quantum dot (DQD) coupled to a microwave resonator via its electric dipole moment. The strong spin-photon coupling arises from the motion of the electron spin in a local magnetic field gradient. We calculate the signal-to-noise ratio (SNR) of the readout accounting for both Purcell spin relaxation and spin relaxation arising from intrinsic electric noise within the semiconductor. We express the maximum achievable SNR in terms of the cooperativity associated with these two dissipation processes. We find that while the cooperativity increases with the strength of the dipole coupling between the DQD and the resonator, it does not depend on the strength of the magnetic field gradient. We then optimize the SNR as a function of experimentally tunable DQD parameters. We identify wide regions of parameter space where the unwanted backaction of the resonator photons on the qubit is small. Moreover, we find that the coupling of the resonator to other DQD transitions can enhance the SNR by at least a factor of two, a ``straddling'' effect~[J. Koch et al., \href{https://doi.org/10.1103/PhysRevA.76.042319}{Phys. Rev. A 76, 042319 (2007)}] that occurs only at nonzero energy detuning of the DQD double-well potential. We estimate that with current technology, single-shot readout fidelities in the range $82-95\%$ can be achieved within a few $\mu\textrm{s}$ of readout time without requiring the use of Purcell filters.
\end{abstract}


\maketitle


\section{Introduction}

Spins in the solid state have long been hailed as a promising platform for quantum information processing~\cite{loss1998,kane1998}. Indeed, their isolation from their electric environment and, in the case of isotopically purified silicon, from their magnetic environment, can lead to significantly enhanced coherence times compared to other implementations~\cite{tyryshkin2012,saeedi2013,bargill2013,muhonen2014,watson2017-2,bradley2019}. Such long coherence times enable high-fidelity control which, combined with the ability to perform single-shot qubit readout~\cite{elzerman2004,barthel2009,jiang2009,morello2010,robledo2011,pla2013,nakajima2017,pakkiam2018,west2019,crippa2019,urdampilleta2019,harveycollard2018,zheng2019,nakajima2019}, makes spins in the solid state a natural choice for scalable quantum technologies. The appeal of solid-state spins has recently been further increased by the successful experimental demonstration of strong coupling between spins and superconducting microwave resonators (also referred to as cavities)~\cite{mi2018,landig2018,samkharadze2018,cubaynes2019,kubo2010,hou2019,dold2019}. Strong coupling between spins and microwave photons could allow spin qubits to benefit from, among other things, the long-distance connectivity~\cite{majer2007,sillanpaa2007,vanwoerkom2018,borjans2019-2}, fast and high-fidelity control~\cite{steffen2006,lucero2008,chow2009,barends2014,rol2019}, and high-fidelity quantum-non-demolition readout~\cite{lupascu2006,mallet2009,liu2014,hover2014,jeffrey2014,walter2017} which have so far been successfully achieved in superconducting qubit implementations~\cite{blais2004}.

The resonator-assisted dispersive readout of a single-electron solid-state spin qubit, in particular, has already been demonstrated, although not in the single-shot regime~\cite{mi2018}. For important applications such as quantum error correction and feedback control of quantum states, however, it is desirable to be able to perform quantum-non-demolition readout of the spin state in a single-shot and with high fidelity. Due to the inherent difficulty in achieving strong spin-photon coupling, however, performing a fast and high-fidelity single-shot readout is likely to prove more challenging than for superconducting qubits. While Hamiltonian engineering methods have been proposed to circumvent weak-coupling limitations, they often require simultaneous qubit control~\cite{beaudoin2017} which adds a layer of complexity to the readout. Similarly, the use of auxiliary resonator modes has also been proposed to relax the constraints of strong coupling, but such schemes rely on the engineering of spectrally close pairs of modes in multidimensional resonators~\cite{troiani2019}~or of special symmetries in multiresonator systems~\cite{zhang2019}. Recent work has proposed to circumvent weak electric dipole moments in multielectron quantum dots by instead coupling the resonator field to the quantum capacitance of the qubit energy dispersion~\cite{ruskov2019}, but this effect is often suppressed by large energy gaps and may require parametric driving of the qubit to achieve strong enough dispersive coupling. As spin-qubit devices enter the strong spin-photon coupling regime, therefore, it is of great interest to optimize the performance of the standard resonator-assisted dispersive readout which has been so widely and successfully used in the context of superconducting qubits.

In this work, we theoretically optimize the performance of the dispersive readout of a spin qubit assisted by a single mode of a microwave resonator. We focus on the case of a single electron spin in a double quantum dot (DQD), where the orbital and spin degrees of freedom are hybridized using a transverse magnetic field gradient~\cite{cottet2010,hu2012,beaudoin2016,benito2017,benito2019-2,burkard2019}. This so-called flopping-mode spin qubit~\cite{benito2019-2} has already entered the strong spin-photon coupling regime through a combination of a large magnetic field gradient and of a large DQD electric dipole coupling with the resonator~\cite{mi2018,samkharadze2018}. Moreover, the ability to address the spin with electric fields instead of magnetic fields~\cite{haikka2017} makes this setup attractive for integration with semiconductor technologies. We derive an expression for the maximum signal-to-noise ratio (SNR) achievable for dispersive spin readout in these devices. We account for the intrinsic relaxation of the spin due to coupling of the semiconductor environment to the electric dipole of the electron, such as relaxation via emission of a phonon. In particular, we show that the maximum achievable SNR is directly proportional to the cooperativity associated with the Purcell spin relaxation and the intrinsic spin relaxation. Interestingly, we show that the cooperativity does not depend on the strength of the magnetic field gradient for these dissipation processes. This means that while increasing the field gradient reduces the readout time, it does not improve the maximum achievable SNR. We then describe how to choose the tunable parameters of the DQD in order to achieve an optimal SNR. Our systematic analysis of transition-inducing terms in the Hamiltonian enables us to identify regions of parameter space where the deleterious backaction of the resonator photons on the qubit state is small. Furthermore, we find that there can be flexibility in the choice of parameters for a given SNR, freeing the parameter space for the optimization of other qubit performance metrics. We also find that at nonzero energy detuning of the DQD double-well potential, the SNR can be enhanced by at least a factor of two due to the existence of a so-called straddling regime~\cite{koch2007,boissonneault2012,inomata2012,zhu2013} arising from the coupling of the resonator to transitions that simultaneously change the molecular wave function and the spin. Our analysis shows that the single-shot readout regime is well within reach of current technology. The achievable single-shot readout fidelities range from $82$ to $95\%$ with the help of quantum limited amplifiers but without requiring the use of Purcell filters. Our work provides the theoretical framework to achieve fast, high-fidelity, quantum-non-demolition readout of single solid-state spins in the near future.

This paper is structured as follows. In Section~\ref{sec:system}, a model of the DQD and its coupling to the resonator is introduced. Sec.~\ref{sec:dispersiveHamiltonian} discusses the dispersive approximation as well as the dispersive Hamiltonian for the DQD. In Sec.~\ref{sec:dispersiveReadout}, the SNR is defined and the performance of the dispersive readout is theoretically optimized. Moreover, single-shot readout fidelity estimates are given for the experimental parameters of Ref.~\onlinecite{mi2018}. The results are summarized in Sec.~\ref{sec:conclusions}.

\section{System and model \label{sec:system}}

\subsection{Hamiltonian \label{sec:Hamiltonian}}

\subsubsection{Double-quantum-dot Hamiltonian}

\begin{figure}
\includegraphics[width = \columnwidth]{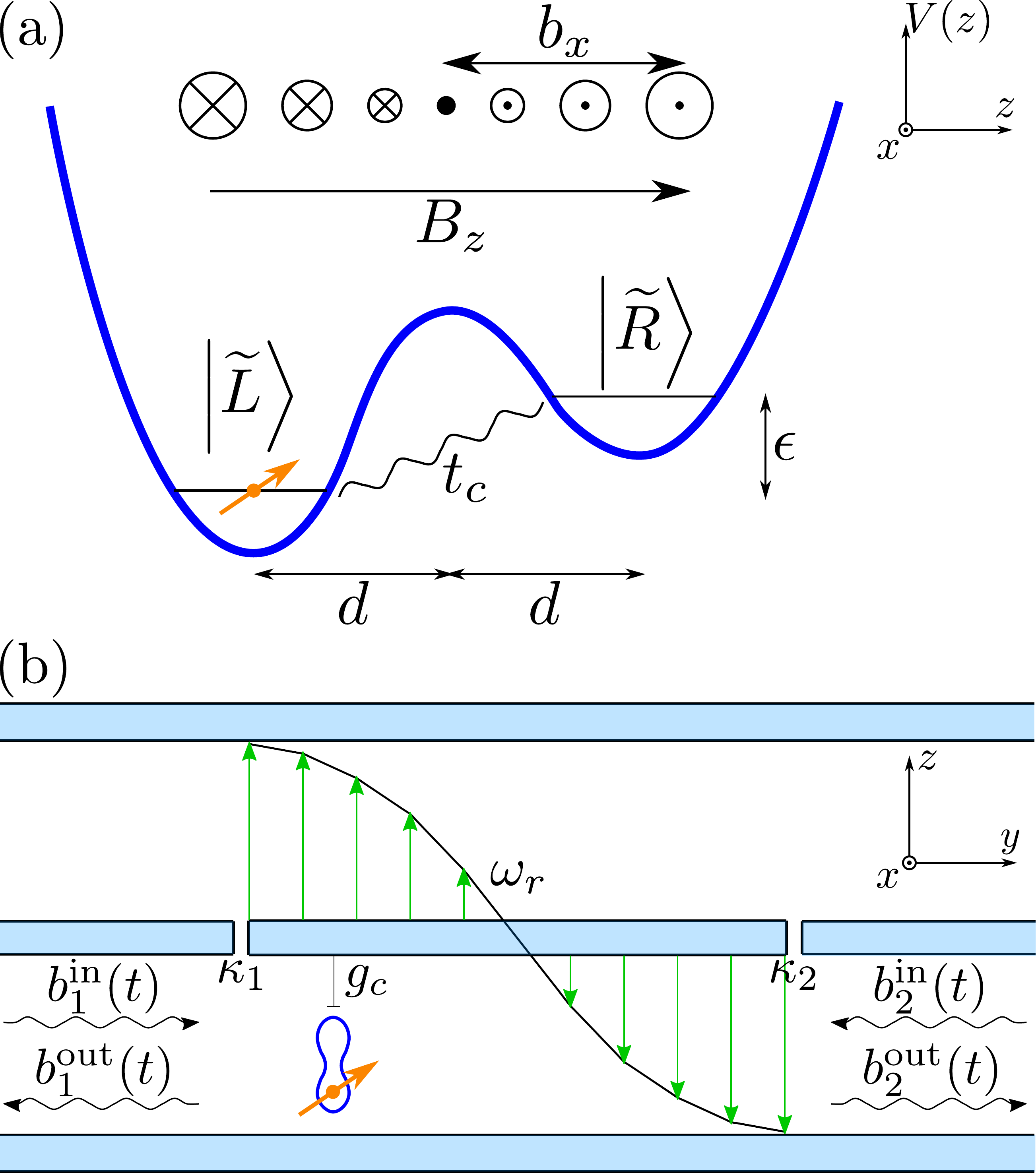}
\caption{(a) Schematic representation of the double-well potential $V(z)$ forming the DQD. The detuning and tunnel coupling between the right dot orbital $\ket{\widetilde{R}}$ and left dot orbital $\ket{\widetilde{L}}$ are $\epsilon$ and $t_c$, respectively. The DQD is subject to a longitudinal magnetic field $B_z \hat{z}$ and an external transverse magnetic field gradient $\partial_z B_x(z) \hat{x}$ such that the field $B_x$ varies by $b_x = \partial_z B_x(z) \times d$ over the half interdot distance $d$. (b) Setup for the dispersive readout of a DQD embedded in a two-port microwave resonator with resonance frequency $\omega_r$. The DQD and the resonator electric field (green arrows) interact via the electric dipole coupling $g_c$. The resonator can be driven in the $i\textrm{th}$ port by an input field $b_i^{\textrm{in}}(t)$. The output fields $b_i^{\textrm{out}}(t)$ then carry information on the state of the DQD. The leakage rates of ports $1$ and $2$ to their respective feedlines are $\kappa_1$ and $\kappa_2$.
\label{fig:fig1}}
\end{figure}
We consider a DQD defined by a double-well potential $V(z)$ whose wells are separated by a distance $2d$, as depicted in Fig.~\ref{fig:fig1}(a). The two lowest-energy orbitals of the right and left wells are labeled $\ket{\widetilde{R}}$ and $\ket{\widetilde{L}}$, respectively, with corresponding energies $\epsilon_R$ and $\epsilon_L$. The energy detuning between the right and left orbitals is $\epsilon = \epsilon_R - \epsilon_L$, and the tunnel coupling between them is $t_c>0$. Moreover, a uniform longitudinal magnetic field is applied along the axis of the quantum dot (the $z$ axis). This induces a Zeeman energy splitting $\hbar \gamma_e B_z$ of the electronic spin states $\ket{\widetilde{\uparrow}}$ and $\ket{\widetilde{\downarrow}}$, where $\gamma_e$ is the electron gyromagnetic ratio. In addition, a transverse position-dependent magnetic field $B_x(z)$ is applied along the $x$ axis using, e.g., a proximal micromagnet. As the electron moves across the DQD, it therefore experiences a magnetic field variation of order $b_x = \partial_z{B_x(z)}\times d$. This hybridizes the spin and charge degrees of freedom, enabling electrical control and readout of the spin. In the following, we set $\hbar = 1$ and $\gamma_e = 1$. The resulting DQD Hamiltonian is
\begin{align}
\begin{split}
 &H_d = H_m + H_Z, \\
 &H_m = \frac{\epsilon}{2} \widetilde{\tau}_z + t_c \widetilde{\tau}_x, \\
 &H_Z = \frac{B_z}{2}\widetilde{\sigma}_z + \frac{b_x}{2}\widetilde{\tau}_z \widetilde{\sigma}_x.
\end{split}
 \label{eq:hamiltonianDQD}
\end{align}
In Eq.~\eqref{eq:hamiltonianDQD}, $H_m$ is the molecular Hamiltonian of the DQD and $H_Z$ is the Zeeman Hamiltonian. Moreover, the $\widetilde{\tau}_i$ are the Pauli matrices in the $\left\{\ket{\widetilde{R}},\ket{\widetilde{L}}\right\}$ basis and the $\widetilde{\sigma}_i$ are the Pauli matrices in the $\left\{\ket{\widetilde{\uparrow}},\ket{\widetilde{\downarrow}}\right\}$ basis. 
It is also convenient to introduce the eigenstates $\ket{\widetilde{\pm}}$ of the molecular Hamiltonian $H_m$. They satisfy $H_m \ket{\widetilde{\pm}} = \pm \frac{\Omega}{2} \ket{\widetilde{\pm}}$, where $\Omega = \sqrt{(2t_c)^2+\epsilon^2} = 2t_c \sec\theta$ is the molecular energy gap and where $\theta = \arctan\left(\epsilon/2t_c\right)$ is the molecular mixing angle. Note that the description of electronic motion in terms of the two lowest-energy orbitals is only valid in the limit where $\Omega$ is much smaller than the single-dot orbital splitting, whether it originates from confinement or from valley splitting.

\subsubsection{Double-quantum-dot-resonator interaction}
The electric field of the resonator couples directly to the electric dipole moment of the electron, as shown schematically in Fig.~\ref{fig:fig1}(b). Due to the interaction of the spin and orbit degrees of freedom in Eq.~\eqref{eq:hamiltonianDQD}, the resonator photons can drive spin transitions. The Hamiltonian of the combined resonator and DQD system is
\begin{align}
\begin{split}
 &H = H_d + H_r + V , \\
 &H_r = \omega_r a^\dagger a, \\
 &V = g_c \widetilde{\tau}_z \left(a + a^\dagger \right).
 \label{eq:hamiltonian}
\end{split}
\end{align}
In Eq.~\eqref{eq:hamiltonian}, $H_r$ is the free Hamiltonian for a single mode of the resonator, $V$ is the dipole interaction Hamiltonian between the electron and the resonator, and $a$ annihilates a photon in the resonator. The resonance frequency of the resonator is $\omega_r > 0$ and the strength of the dipole coupling is $g_c$.

\subsubsection{Probe-resonator interaction \label{sec:probeResonatorInteraction}}

We assume that the resonator can be probed through two input ports, which we label port $1$ and port $2$. This is depicted in Fig.~\ref{fig:fig1}(b). Photons can leak in and out of the resonator~\footnote{Any intrinsic resonator losses that result in photons being emitted somewhere other than the input and output ports can simply be modeled by including an additional port with leakage rate $\kappa_0$. Therefore, the formalism presented below remains completely general in the presence of such losses. In this manuscript, we assume that the intrinsic losses are small compared to the leakage through the input ports.} through the $i\textrm{th}$ port at rate $\kappa_i$, resulting in a total leakage rate $\kappa = \sum_i \kappa_i$. Accordingly, the resonator can be populated with photons by irradiating the input ports at frequency $\omega_\textrm{in}\approx \omega_r$. Under this near-resonance condition, we may describe the interaction of the input radiation with the resonator in the rotating-wave approximation:
\begin{align}
 V_{\textrm{in}}(t) = i \sum_{i} \sqrt{\kappa_i} \left[b_i^{\textrm{in}}(t)^\dagger a - b_i^{\textrm{in}}(t) a^\dagger \right]. \label{eq:inputInteraction}
\end{align}
The quantum input fields $b_i^{\textrm{in}}(t)$ in Eq.~\eqref{eq:inputInteraction} are the ones derived in the input-output theory of Gardiner and Collett~\cite{gardiner1985}. They consist of a classical drive $\beta_i^{\textrm{in}}(t)$ with added noise. More precisely, we have
\begin{align}
 b_i^{\textrm{in}}(t) = \beta_i^{\textrm{in}}(t) + \delta b_i^{\textrm{in}}(t).
\end{align}
Here, we assume that the noise is Gaussian and white~\footnote{More precisely, it is assumed that the correlation time of the noise is much smaller than the typical timescale for the system evolution in the frame rotating at the probe frequency $\omega_\textrm{in}$. The correlation time should also be much smaller than the inverse detector bandwidth.}. In the absence of squeezing of the inputs, the moments of $\delta b_i^{\textrm{in}}(t)$ are
\begin{align}
\begin{split}
 &\left\langle \delta b_i^{\textrm{in}}(t)^\dagger \delta b_i^{\textrm{in}}(t') \right\rangle = \bar{N}\delta(t-t'),\\
 &\left\langle \delta b_i^{\textrm{in}}(t) \delta b_i^{\textrm{in}}(t')^\dagger \right\rangle = (\bar{N}+1)\delta(t-t'),\\
 &\left\langle \delta b_i^{\textrm{in}}(t) \delta b_i^{\textrm{in}}(t') \right\rangle = 0.
 \label{eq:inputStatistics}
\end{split}
\end{align}
In Eq.~\eqref{eq:inputStatistics}, $\bar{N}$ is the average number of thermal noise photons~\footnote{Even though thermal noise is not white, it is typically approximately white in the neighborhood of the probe frequency $\omega_\textrm{in}$.} at frequency $\omega_\textrm{in}$, which we assume to be the same for both ports. For $\bar{N}=0$, the noise in the input field arises purely from vacuum fluctuations.

The output fields $b_i^{\textrm{out}}(t)$ are given by the input-output relations:
\begin{align}
 b_i^{\textrm{out}}(t) = b_i^{\textrm{in}}(t) + \sqrt{\kappa_i} a(t). \label{eq:inputOutput}
\end{align}
The noise in the output field is in general not white because it inherits the temporal correlations in the dynamics of the resonator and DQD. In the dispersive limit discussed in Sec.~\ref{sec:dispersiveLimit}, the main effect of the DQD is to modify the frequency of the resonator. As a result, the resonator Hamiltonian remains approximately quadratic and the covariance of the output noise is only weakly modified. If follows that the output noise is white with the same moments as in Eq.~\eqref{eq:inputStatistics} when the system is in a steady state~\footnote{If the value of $\bar{N}$ is not the same in both ports, there is a net flow of noise photons from high-noise ports to low-noise ports through the resonator. In that case, the output noise acquires a finite correlation time $\sim \kappa^{-1}$ and is a function of the transmission and reflection coefficients of the resonator. The observed output noise may then be modeled as white provided that the detector bandwidth is smaller than $\kappa$.}. We assume that $b_i^{\textrm{out}}(t)$ is sent through a phase-preserving amplifier and then measured with the help of a homodyne detector~\footnote{The readout can also be performed with the help of a heterodyne detector, at the cost of the additional vacuum noise which is unavoidably introduced by attempting to simultaneously measure two noncommuting quadratures of a quantum field.} whose local oscillator has phase $\varphi$. The detector records a photocurrent $I^\varphi_i(t) = \beta_i^{\textrm{out},\varphi}(t) + \delta I_i^\varphi(t)$, where $\beta_i^{\textrm{out},\varphi}(t) = \frac{1}{2}\left\langle b_i^{\textrm{out}}(t)e^{-i \varphi}+ b_i^{\textrm{out}}(t)^\dagger e^{i \varphi}\right\rangle$ is the $\varphi$ quadrature of the output field. The autocorrelation function of the photocurrent noise in the steady state is then
\begin{align}
 \left\langle \delta I^\varphi_i(t) \delta I^\varphi_i(t')\right\rangle = \frac{2N_{\textrm{hom}}+1}{4} \delta(t-t').\label{eq:homodyneNoise}
\end{align}
Here, $N_{\textrm{hom}} = \bar{N} + N_{\textrm{amp}}$ is the total noise in the homodyne signal accounting for the $N_{\textrm{amp}}$ effective noise photons added in the amplification chain. It follows from Eq.~\eqref{eq:homodyneNoise} that a given quadrature of $b_i^{\textrm{out}}(t)$ integrated over a time interval $t$ is determined with precision
\begin{align}
 \sigma_{\textrm{hom}}(t) = \frac{1}{\sqrt{R t}}. \label{eq:stdHomodyne}
\end{align}
Here, $R = 4/(2N_{\textrm{hom}}+1)$ is the rate of change of the inverse noise variance. In the following, we assume that the input noise is limited by vacuum fluctuations, $\bar{N} \ll 1$.

\subsection{Double-quantum-dot eigenbasis and spin qubit \label{sec:DQDEigenbasis}}

\begin{figure}
\includegraphics[width = 0.7\columnwidth]{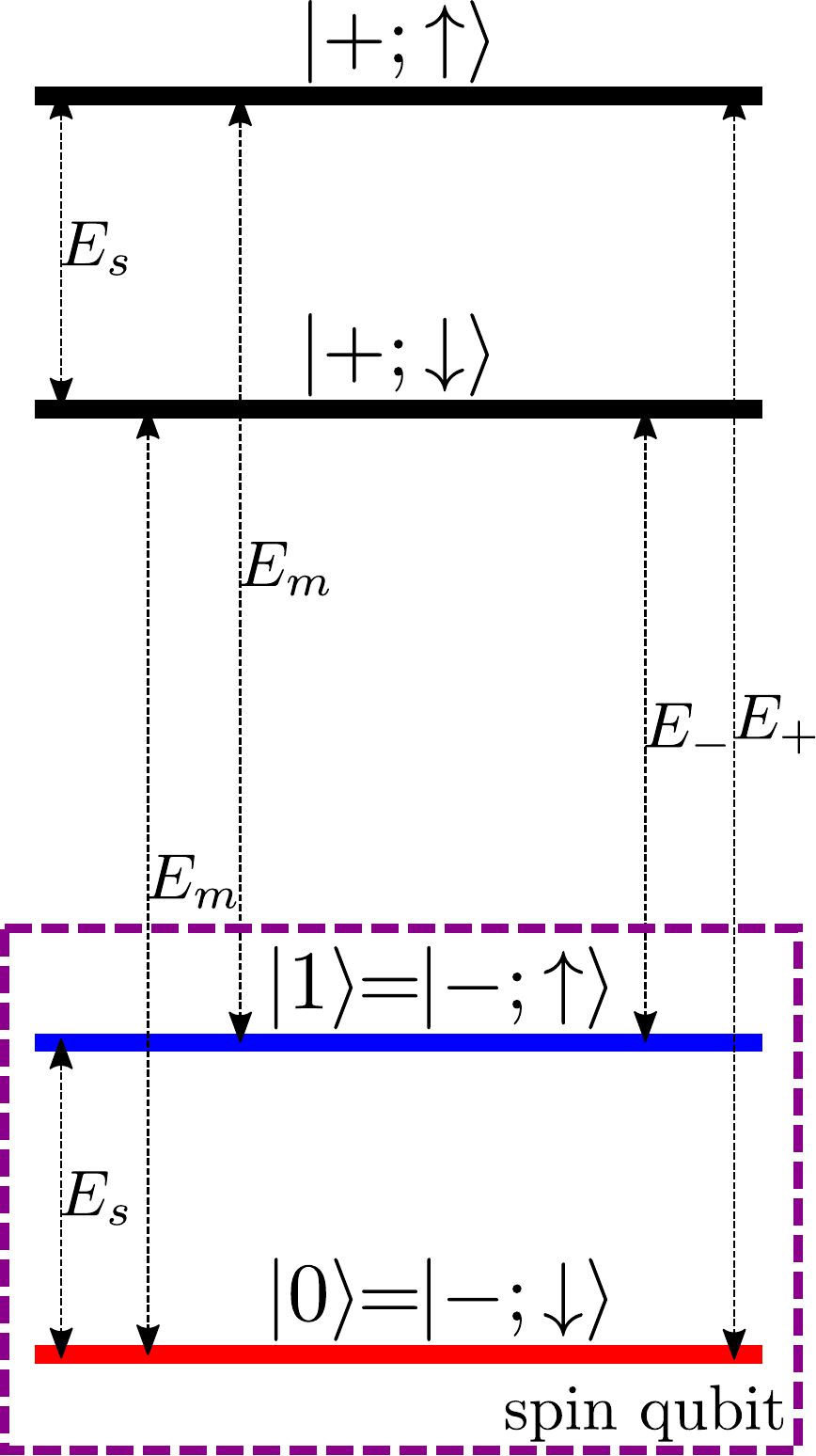}
\caption{Energy-level diagram of the DQD in the basis dressed by the magnetic field gradient. The dressed eigenstates are labeled $\left\{\ket{+;\uparrow}, \ket{+;\downarrow}, \ket{-;\uparrow}, \ket{-;\downarrow}\right\}$. All possible transition frequencies $E_j$ between the eigenstates are indicated. The spin-qubit states $\ket{1}$ and $\ket{0}$ are chosen to be $\ket{-;\uparrow}$ and $\ket{-;\downarrow}$, respectively (dashed magenta box). More precisely, the spin qubit is formed by the states $\ket{1}$ and $\ket{0}$ dressed by the resonator (see Sec.~\ref{sec:effectiveSpinQubitHamiltonian}). \label{fig:fig2}}
\end{figure}

The DQD Hamiltonian, Eq.~\eqref{eq:hamiltonianDQD}, can be diagonalized exactly as detailed in Appendix~\ref{app:exactDiagonalizationDQD}. Expressed in its eigenbasis, the Hamiltonian $H_d$ takes the form
\begin{align}
 H_d = \frac{E_m}{2} \tau_z + \frac{E_s}{2}\sigma_z, \label{eq:hamiltonianDQDDiag}
\end{align}
where the $\tau_i$ and the $\sigma_i$ are now Pauli matrices in the eigenbasis $\ket{\tau_z;\sigma_z}$ of $H_d$ dressed by the field gradient. Here $\tau_z = \pm$ labels the dressed ``molecular-like'' states and $\sigma_z = \uparrow \left(\downarrow\right)$ labels the dressed ``spin-like'' states~\footnote{Note that here, $\tau_z$ and $\sigma_z$ are defined such that the molecular-like eigenstates and spin-like eigenstates are always those which have a mostly molecular character and mostly spin character, respectively. An alternative definition is to choose $\tau_z$ and $\sigma_z$ such that the two lowest-energy eigenstates always correspond to the same value of $\tau_z$, such as in Ref.~\onlinecite{benito2019}. In that case, the Hamiltonian takes the form $H_d = \frac{E_\tau}{2}\tau_z + \frac{E_\sigma}{2}\sigma_z$, where $E_\tau = E_m$ and $E_\sigma = E_s$ when $2 t_c > B_z$, and $E_\tau = E_s$ and $E_\sigma = E_m$ when $2t_c < B_z$.}. Exact expressions for the molecular-like and spin-like Larmour frequencies $E_m$ and $E_s$ are derived in Appendix~\ref{app:exactDiagonalizationDQD}. The energy-level diagram of the DQD is illustrated in Fig.~\ref{fig:fig2}, where we have also introduced the transition frequencies $E_\pm = E_m \pm E_s$.
In the following, we consider the spin qubit formed from the two dressed `spin-like' eigenstates spanning the molecular ground state. Specifically, we choose the computational basis $\left\{\ket{1},\ket{0}\right\} = \left\{\ket{-;\uparrow},\ket{-;\downarrow}\right\}$. Despite their spin-like character, the electric dipole matrix element between these two states is finite and transitions between them can be induced electrically. In particular, the DQD-resonator interaction of Eq.~\eqref{eq:hamiltonian} is written in the new basis as
\begin{align}
\begin{split}
 &V = \mathcal{V} (a + a^\dagger), \\
 &\mathcal{V} = -g_m \tau_x + g_s \tau_z \sigma_x + g_+ \left(\tau_+ \sigma_+ + \tau_- \sigma_-\right) \\
 & \;\;\;\;\;\;\;\;\; + g_- \left(\tau_+ \sigma_- + \tau_- \sigma_+\right) + g_{mp} \tau_z + g_{sp} \sigma_z.
\end{split} \label{eq:interactionDiag}
\end{align}
Here, $\left\{g_m,g_s,g_+,g_-\right\}$ are the coupling strengths of the resonator to the DQD transitions of frequencies $\left\{E_m,E_s,E_+,E_-\right\}$ illustrated in Fig.~\ref{fig:fig2}. In addition, $g_{mp}$ and $g_{sp}$ are couplings arising from the finite dc electric polarizabilities of the molecular electric dipole and of the spin, respectively. Exact expressions for the $g_i$ are given in Appendix~\ref{app:exactDiagonalizationDQD}. In Eq.~\eqref{eq:interactionDiag}, the term $g_s \tau_z \sigma_x (a + a^\dagger)$ exchanges energy between the resonator and the spin qubit. It can thus be exploited for resonator-assisted qubit control and readout. As will be discussed in Sec.~\ref{sec:optimizationDQDParameters}, the couplings $g_\pm$ {of the resonator to the transitions of frequencies $E_\pm$} can also be harnessed to improve readout performance.

In the remainder of this article, we focus on the limit of weak field gradient. In particular, we assume that the direction of the spin quantization axis is not substantially modified by the presence of the field gradient, $|b_x \sin \theta | \ll |B_z|$. Moreover, we assume that the admixture of spin and orbit is weak, $|b_x \cos \theta| \ll \min\left(|\Omega - B_z|,|\Omega+B_z|\right)$. Under these conditions, the dressed molecular and spin Larmour frequencies are
\begin{align}
\begin{split}
 &E_m \approx \Omega + \frac{b_x}{2} \cos\theta \sin \bar{\phi}, \\
 &E_s \approx B_z - \frac{B_z}{2}\frac{b_x}{2 t_c}\left( 1 - \frac{\epsilon^2}{B_z^2} \right) \sin \bar{\phi}, \\
 &\sin\bar{\phi} \approx \frac{2t_c b_x}{\Omega^2 - B_z^2},
\end{split} \label{eq:approximateFrequencies}
\end{align}
where $\bar{\phi} \ll \pi/2$ is the effective spin-orbit mixing angle arising from the field gradient. Approximate expressions may also be obtained for the couplings $g_i$. In particular, the molecular-photon coupling $g_m$ and spin-photon coupling $g_s$ become
\begin{align}
\begin{split}
 g_m \approx g_c \cos \theta \cos \bar{\phi}, \;\;\;\; g_s \approx g_c \cos\theta \sin \bar{\phi}.
\end{split} \label{eq:spinAndChargeCouplings}
\end{align}

\section{Dispersive Hamiltonian \label{sec:dispersiveHamiltonian}}

\subsection{Dispersive limit \label{sec:dispersiveLimit}}

Dispersive readout of the spin is performed by probing the resonator near its resonance frequency, $\omega_\textrm{in} \approx \omega_r$, and observing the spin-dependent phase of the output field. For many quantum information processing tasks, it is highly desirable that the readout perturbs the system as little as possible. To minimize such unwanted backaction on the system, we work in the so-called dispersive limit. In that limit, all DQD-resonator interaction terms in Eq.~\eqref{eq:interactionDiag} are off-resonant. A given interaction term is off-resonant if its magnitude is smaller than the detuning of the resonator with the transition it induces. Let $\left<n\right> \approx 4\kappa_i|\beta_i^{\textrm{in}}|^2/\kappa^2$ be the average number of photons entering the resonator from port $i$. Noting that the resonator field is of order $a \sim \sqrt{\left<n\right>}$, it is easily verified that a term $\propto g_j$ in Eq.~\eqref{eq:interactionDiag} is off-resonant if $\left<n\right>$ is smaller than the so-called critical photon number,
\begin{align}
 n_{c,j} \approx \frac{1}{4}\textrm{max}\left(|\eta_j|,|\eta'_j|\right)^{-2}. \label{eq:criticalPhotonNumbers}
\end{align}
Here, $|\eta_j| \ll 1$ and $|\eta'_j| \ll 1$ are the small dimensionless parameters that control the dispersive limit. They have the form (see Appendix~\ref{app:fullDispersive} for details)
\begin{align}
 \eta_j = \frac{2 E_j g_j}{\omega_r^2 - E_j^2}, \;\;\;\; \eta'_j = \frac{\omega_r}{E_j} \eta_j.\label{eq:dispersiveParameters}
\end{align}
The condition $\left<n\right> < n_{c,j}$ ensures that the probe photons excite a given transition between eigenstates of $H$ at a rate that is smaller than the relaxation rate for that transition~\cite{boissonneault2009}. Thus, the condition $\left<n\right> < n_{c,j}$ ensures that probe photons close to resonance with a given transition $j$ excite that transition with negligible probability~\footnote{Note that pure dephasing of a transition $j$ in the bare double-quantum-dot eigenbasis also enables probe photons to induce transitions in the basis dressed by the resonator~\cite{boissonneault2008,boissonneault2009,slichter2012}. While the associated transition rate is also suppressed in the ratio $\left<n\right> /n_{c,j}$, it can in principle be made larger than the relaxation rate of the transition if the fluctuations that induce the dephasing have a large spectral weight at the frequencies $\pm |\omega_r - E_j|$ (typically several MHz in the present work). For low frequency charge and nuclear noise typical of spin-qubit environments, these contributions are likely to be small. A detailed analysis of these processes is beyond the scope of this work and we thus ignore the effect of pure dephasing throughout.}. A detailed analysis of these transition rates and their effect on readout is beyond the scope of this work. In the following, we will reserve the symbol $n_c$ for the critical photon number of the spin transition, $n_c \equiv n_{c,s}$.

In the dispersive limit, the Hamiltonian, Eq.~\eqref{eq:hamiltonian}, can be diagonalized to first order in $g_c$ using a Schrieffer-Wolff transformation. The resulting dispersive Hamiltonian for the DQD-resonator interaction is derived in Appendix~\ref{app:fullDispersive} and has the form
\begin{align}
 H_{\textrm{dis}} = H_0 + V_{\textrm{dis}} + V_{\textrm{tr}}. \label{eq:dispersiveHamiltonianFull}
\end{align}
Here, $H_0 = H_d + H_r$ is the free Hamiltonian. The interaction is separated into a dispersive part $V_{\textrm{dis}}$ that commutes with $H_0$ and a transition-inducing part $V_{\textrm{tr}}$ that does not commute with $H_0$. The dispersive interaction has the form:
\begin{align}
\begin{split}
 V_{\textrm{dis}} = - \frac{1}{2} \chi_0 \tau_z \sigma_z -\left( \chi_m \tau_z + \chi_s \sigma_z \right)\left(a^\dagger a + \frac{1}{2}\right). \label{eq:dispersiveInteractionFull}
\end{split}
\end{align}
Here, $\chi_m \tau_z$ and $\chi_s \sigma_z$ are the dispersive energy shifts of the resonator frequency due to coupling with the molecular electric dipole and the spin, respectively. In addition, $-\chi_0 \tau_z \sigma_z/2$ is an Ising-like dispersive interaction between the molecular electric dipole and the spin. Expressions for $\chi_m$, $\chi_s$, and $\chi_0$ are given in Appendix~\ref{app:fullDispersive}. As discussed in Sec.~\ref{sec:effectiveSpinQubitHamiltonian}, the spin dispersive shift, $\chi_s \sigma_z$, may be exploited for dispersive readout of the spin. Contrary to the dispersive interaction, the off-diagonal term $V_{\textrm{tr}}$ induces transitions between the eigenstates of $H_0$. Specifically, $V_{\textrm{tr}}$ can generate all the DQD transitions of Fig.~\ref{fig:fig2} via the exchange of either $0$ or $2$ photons with the resonator. Thus, a transition term inducing a transition $j$ can be neglected if its magnitude is smaller than $\omega_r$, $|E_j|$, and $|2\omega_r \pm E_j|$ (a concrete example is given in Appendix~\ref{app:transitionTerms}). These off-resonance conditions can be seen as a higher-order dispersive approximation. They ensure that the operators appearing in Eq.~\eqref{eq:dispersiveInteractionFull} are expressed in a basis that is close to the true eigenbasis of the full system Hamiltonian $H$. If the transition term becomes resonant, the resulting change of basis enables probe photons to generate new transitions between the system eigenstates. The above off-resonance conditions are typically satisfied in the dispersive limit (though not always, see Fig.~\ref{fig:fig5}). We will therefore ignore the transition term in the following analysis until stated otherwise.

\subsection{Effective spin-qubit Hamiltonian \label{sec:effectiveSpinQubitHamiltonian}}
In the absence of photon-induced DQD transitions, the dispersive Hamiltonian, Eq.~\eqref{eq:dispersiveInteractionFull}, may safely be projected into the logical subspace of the spin qubit to obtain an effective dispersive Hamiltonian for the spin qubit, in the form (up to an irrelevant constant):
\begin{align}
 H^{\textrm{eff}}_{\textrm{dis}} = (\omega_r' - \chi_s \sigma_z) a^\dagger a + \frac{1}{2}(E_s'-\chi_s) \sigma_z. \label{eq:projectedDispersiveHamiltonian}
\end{align}
Here, $\omega_r' = \omega_r + \chi_m$ and $E_s' = E_s + \chi_0$ are renormalized resonator and spin-qubit frequencies, respectively. In addition, $\chi_s \sigma_z$ is the spin-state-dependent dispersive shift of the resonator frequency which enables dispersive readout. The full expression for the dispersive shift is
\begin{align}
 \chi_s = \frac{2 E_s g_s^2}{\omega_r^2-E_s^2}+\frac{E_+ g_+^2}{\omega_r^2 - E_+^2} - \frac{E_- g_-^2}{\omega_r^2 - E_-^2}. \label{eq:dispersiveShiftFull}
\end{align}
When the resonator is close to resonance with the spin transition but far detuned from $E_+$ and $E_-$, the dispersive shift takes the more familiar form $\chi_s \approx g_s^2/\Delta$, where $\Delta = \omega_r - E_s$ is the spin-resonator detuning. We will assume that this is the case for most of the analysis of Sec.~\ref{sec:dispersiveReadout}. In Sec.~\ref{sec:optimizationDQDParameters}, however, we will see that the various contributions in Eq.~\eqref{eq:dispersiveShiftFull} can interfere constructively and thereby significantly improve readout performance. This mirrors the so-called straddling regime of superconducting qubits~\cite{koch2007,boissonneault2012,inomata2012,zhu2013}. Finally, note that the renormalization of the resonator and spin frequencies are unimportant for the optimization of the dispersive readout. As discussed in Sec.~\ref{sec:readoutContrast}, the readout response only depends on the detuning between the probe frequency $\omega_\textrm{in}$ and the renormalized resonator frequency $\omega_r'$. Thus, the renormalization of the $\omega_r$ can always be compensated by adjusting $\omega_\textrm{in}$. Moreover, inspection of the expression for $\chi_0$ given in Appendix~\ref{app:fullDispersive} shows that $\chi_0 \lesssim \chi_s \ll \Delta$ near the DQD-resonator resonances. Therefore, the renormalization of the spin frequency may safely be neglected.

All operators appearing in Eq.~\eqref{eq:projectedDispersiveHamiltonian}, and in particular $\sigma_z$, are dressed by the DQD-resonator interaction to first order in $g_c$. Thus, the spin qubit we consider is in fact formed by the states $\left\{\ket{-;\uparrow},\ket{-;\downarrow}\right\}$ dressed by resonator photons. In the regime where both the resonator and the qubit are near-resonant with the probe, the effective driving Hamiltonian in the dressed basis takes the form [see the discussion following Eq.~\eqref{eq:transformationA} in Appendix~\ref{app:fullDispersive}]
\begin{align}
\begin{split}
 V^{\textrm{eff}}_{\textrm{in}}(t) = &i \sum_i \sqrt{\kappa_i} \left[b_i^{\textrm{in}}(t)^\dagger a - b_i^{\textrm{in}}(t) a^\dagger \right] \\
 &+ i\frac{g_s}{\Delta} \sum_i \sqrt{\kappa_i} \left[b_i^{\textrm{in}}(t)^\dagger \sigma_- - b_i^{\textrm{in}}(t) \sigma_+ \right]. \label{eq:dispersiveInputInteraction}
\end{split} 
\end{align}
The second term enables the direct exchange of energy between the spin qubit and the resonator environment. In particular, the spin qubit may relax via the Purcell emission of a photon in the resonator ports (see Sec.~\ref{sec:qubitRelaxation}). Correspondingly, the input-output relation of Eq.~\eqref{eq:inputOutput} becomes
\begin{align}
 b_i^{\textrm{out}}(t) = b_i^{\textrm{in}}(t) + \sqrt{\kappa_i}a + \sqrt{\kappa_i}\frac{g_s}{\Delta}\sigma_-, \label{eq:inputOutput2}
\end{align}
where the last term describes output radiation emitted by coherent spin oscillations. When performing dispersive readout, the detector is typically locked in to the frequency $\omega_\textrm{in}\approx \omega_r$. Thus, the qubit emission is filtered out provided the detector bandwidth is smaller than the spin-resonator detuning $|\Delta|$. Even if this were not the case, the qubit necessarily loses all coherence as soon as the two qubit states can be distinguished due to the fundamental quantum backaction introduced by readout. We will therefore ignore the last term in what follows. The expectation values of the output fields are then given by
\begin{align}
 \beta_i^{\textrm{out}}(t) = \beta_i^{\textrm{in}}(t) + \sqrt{\kappa_i}\left<a\right>. \label{eq:inputOutput3}
\end{align}

\section{Dispersive readout of the spin qubit \label{sec:dispersiveReadout}}

\subsection{Equation of motion \label{sec:dispersiveEOM}}
We start our analysis of the dispersive readout by discussing the dynamics of the resonator. Throughout the remainder of the text, we work in the frame rotating with the probe frequency $\omega_\textrm{in}$. In this frame, the dispersive Hamiltonian of Eq.~\eqref{eq:projectedDispersiveHamiltonian} takes the form
\begin{align}
 H^{\textrm{eff}}_{\textrm{dis}} = (\delta_c - \chi_s \sigma_z) a^\dagger a + \frac{1}{2}(\delta_s-\chi_s) \sigma_z,
\end{align}
where $\delta_c = \omega_r'-\omega_\textrm{in}$ and $\delta_s = E_s'-\omega_\textrm{in}$ are the detunings of the probe from the resonator and the spin qubit, respectively. The interaction of Eq.~\eqref{eq:dispersiveInputInteraction} remains unchanged. The resulting (It{\= o}) Langevin equation of motion~\cite{gardiner1985} for the resonator field is
\begin{align}
\begin{split}
 da = -i(\delta_c - \chi_s \sigma_z) a\,dt - \frac{\kappa}{2} a\,dt - &\sum_i \sqrt{\kappa_i} b_i^{\textrm{in}}(t)dt \\
 & - \frac{\kappa g_s}{2\Delta} \sigma_- dt . \label{eq:dynamicsResonator}
\end{split}
\end{align}
The first term describes the dispersive motion of the resonator field, the second term describes resonator damping, the third term describes driving of the resonator through its ports, and the last term describes driving of the resonator by coherent oscillations of the spin qubit. This latter term contributes small oscillations of amplitude $\sim (\kappa g_s/ \Delta^2) \left<\sigma_-\right>$ to the resonator field in the dispersive limit (optimal readout occurs in the regime $|\Delta| > \kappa$, see Fig.~\ref{fig:fig4}). Moreover, these oscillations disappear as soon as the readout dephases the qubit. We therefore neglect them in what follows. The equation of motion for the expectation value of the resonator field becomes
\begin{align}
 \dot{\left<a\right>} = -i\left<(\delta_c - \chi_s \sigma_z) a\right> &- \frac{\kappa}{2} \left<a\right> - \sum_i \sqrt{\kappa_i} \beta_i^{\textrm{in}}(t). \label{eq:dynamicsResonator2}
\end{align}

\subsection{Readout contrast \label{sec:readoutContrast}}
In order to analyze the readout performance, it is not necessary to solve Eq.~\eqref{eq:dynamicsResonator2}. Instead, we first consider the purely ``quantum-non-demolition'' scenario in which $\sigma_z$ is a constant of motion, $\sigma_z(t)\approx \sigma_z(0)$. Although this assumption clearly cannot be exact due to, e.g., qubit relaxation, it leads to a simple and useful definition of the readout contrast.

\begin{figure}
 \includegraphics[width = \columnwidth]{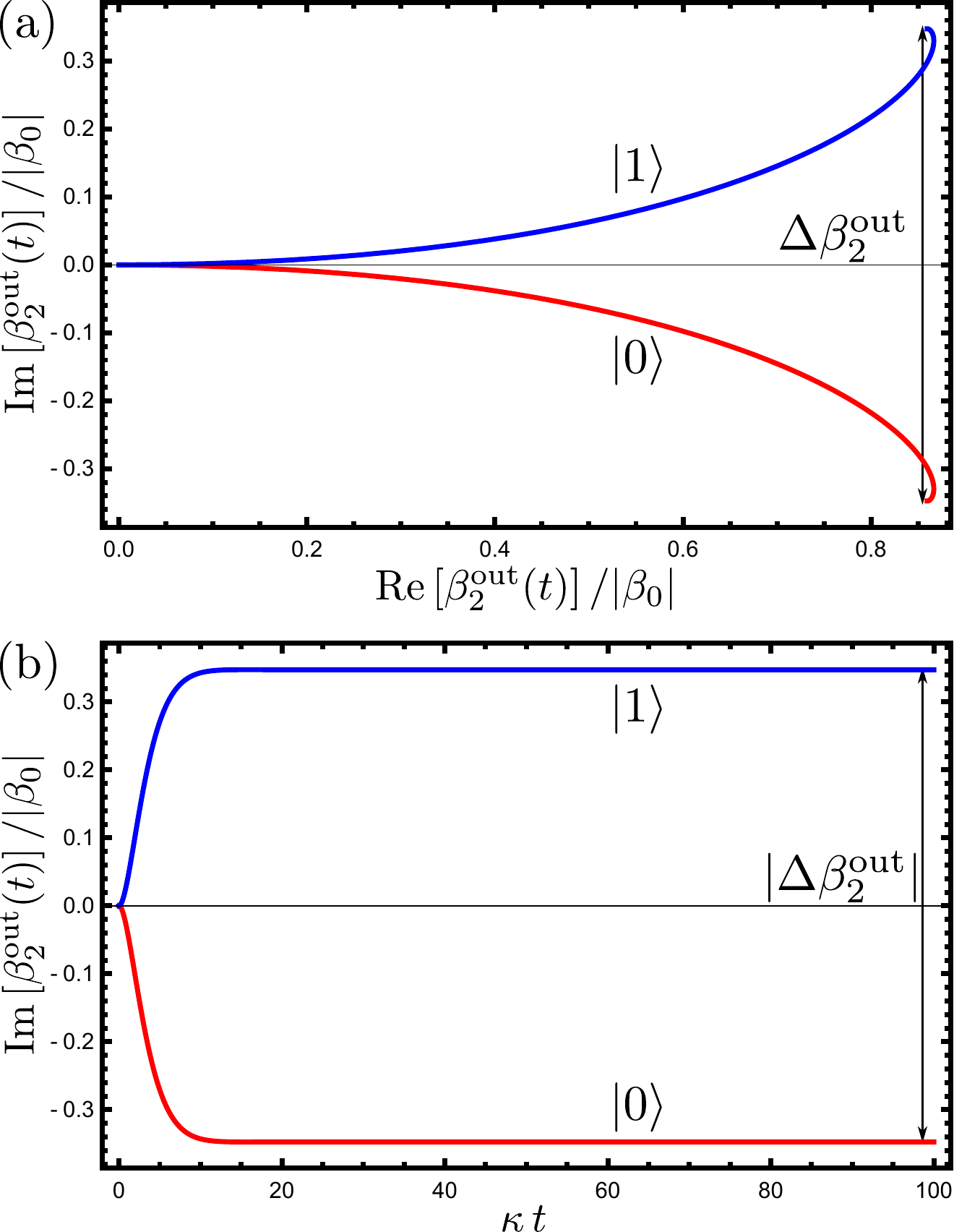}
\caption{Typical dispersive readout response of the resonator for a continuous resonant drive of amplitude $\beta_0 = -|\beta_0|$. Plot (a) shows the trajectory of the output field in phase space while plot (b) shows the time evolution of the quadrature relevant for qubit readout. In both plots, the resonator is initally empty. The solid lines show the response for the qubit states $\ket{1}$ (blue) and $\ket{0}$ (red) in the absence of qubit transitions. They are obtained by solving Eqs.~\eqref{eq:inputOutput3} and \eqref{eq:nonDemolitionEOM} with $\chi_s = 0.2\kappa$ for both qubit states. The signals for the two qubit states are separated by $\Delta\beta_2^{\textrm{out}}$ in the steady state. \label{fig:fig3}}
\end{figure}

Under the above assumption, we may substitute $\sigma_z = \pm 1$ into Eq.~\eqref{eq:dynamicsResonator2} and obtain
\begin{align}
 \dot{\left<a\right>} = -i(\delta_c \mp \chi_s)\left<a\right> - \frac{\kappa}{2}\left<a\right> - \sum_i \sqrt{\kappa_i}\beta_i^{\textrm{in}}(t). \label{eq:nonDemolitionEOM}
\end{align}
Solving Eq.~\eqref{eq:nonDemolitionEOM} and substituting the solution into Eq.~\eqref{eq:inputOutput3} then yields the output field for each qubit state. We assume that the resonator is initially empty. It is then probed continuously through port $1$ only, $\beta_1^{\textrm{in}}(t)=\beta_0$ and $\beta_2^{\textrm{in}}(t)=0$. Finally, the output field is measured in port $i$. Typical trajectories for the transmitted output field obtained using Eqs.~\eqref{eq:inputOutput3} and \eqref{eq:nonDemolitionEOM} are depicted in Fig.~\ref{fig:fig3}.
The relevant quantity for readout performance is the steady-state contrast $\Delta \beta_i^{\textrm{out}} = \lim_{t \to \infty} \left[\beta_i^{\textrm{out}}(t)|_{\sigma_z = +1} - \beta_i^{\textrm{out}}(t)|_{\sigma_z = -1}\right]$ between the output fields corresponding to the two qubit states. Solving Eqs.~\eqref{eq:inputOutput3} and \eqref{eq:nonDemolitionEOM} gives the squared magnitude of the contrast:
\begin{align}
 |\Delta \beta_i^{\textrm{out}}|^2 = \kappa_i \left<n\right> D.\label{eq:squareContrast}
\end{align}
Here, $\left<n\right> = 4\kappa_1 |\beta_0|^2/\kappa^2$ is the number of steady-state photons in the resonator directly on resonance. The quantity $D$ may be thus interpreted as the fraction of input photons that contribute to the readout contrast. We find
\begin{align}
 D = \frac{\kappa^2 \chi_s^2}{\left[(\kappa/2)^2+(\delta_c-\chi_s)^2\right]\left[(\kappa/2)^2+(\delta_c+\chi_s)^2\right]}. \label{eq:squareContrastFunction}
\end{align}
We choose the input frequency $\omega_\textrm{in}$ to maximize the contrast of Eq.~\eqref{eq:squareContrastFunction}. The optimal resonator-probe detuning is
\begin{align}
 \delta_c =
 \left\{
 \begin{array}{cll}
 0 & \textrm{if} & |\chi_s| < \kappa/2\\
 \pm \sqrt{\chi_s^2-(\kappa/2)^2} & \textrm{if} & |\chi_s| > \kappa/2
 \end{array}
 \right. .
\end{align}
At the optimum, Eq.~\eqref{eq:squareContrastFunction} becomes a function $D(x)$ of $x=\chi_s/\kappa$ only:
\begin{align}
 D(x) 
 = 
 \left\{
 \begin{array}{cll}
 \frac{16 x^2}{\left(1 + 4 x^2\right)^2} & \textrm{if} & x^2 < 1/4\\
 1 & \textrm{if} & x^2 > 1/4
 \end{array}
 \right. . \label{eq:photonFraction}
\end{align}

\subsection{Qubit relaxation \label{sec:qubitRelaxation}}

The assumption that the qubit state remains the same at all times is of course not physical. In practice, the qubit state necessarily relaxes on a timescale given by the inverse qubit relaxation rate $\gamma^{-1}$. We account for two distinct relaxation processes. The first is the Purcell relaxation via emission of a photon in the resonator environment. Under our assumption $\bar{N} \ll 1$, this process occurs at the rate
\begin{align}
 \gamma_{\textrm{pu}} = \kappa \left(\frac{g_s}{\Delta}\right)^2. \label{eq:purcellRelaxationRate}
\end{align}
The second process we consider is the relaxation due to electric fluctuations coupling to the electric dipole of the electron~\cite{borjans2019}, most notably relaxation with the emission of a phonon. Such relaxation processes have the general form
\begin{align}
 \gamma_{\textrm{el}} = \gamma_m(E_s) \left( \frac{g_s}{g_c} \right)^2. \label{eq:intrinsicRelaxationRate}
\end{align}
Here $\gamma_m(E_s)$ is a molecular-electric-dipole relaxation rate which depends on the DQD parameters through the spin-qubit frequency $E_s$ only. Moreover, the factor $(g_s/g_c)^2$ accounts for the hybridization of the molecular electric dipole and the spin (see Appendix~\ref{app:exactDiagonalizationDQD}). For dispersive readout, the spin frequency remains in the neighborhood of the resonator frequency, $E_s \approx \omega_r$. Thus, we set $\gamma_m \approx \gamma_m(\omega_r)$ in what follows. Since these two relaxation processes are due to coupling with independent reservoirs, they can be added to leading order in $g_s$. The total relaxation rate is then
\begin{align}
 \gamma = \kappa \left(\frac{g_s}{\Delta}\right)^2 + \gamma_m \left( \frac{g_s}{g_c} \right)^2. \label{eq:relaxationRate}
\end{align}
There exist corrections to Eq.~\eqref{eq:relaxationRate} due to spin transitions induced by probe photons. As discussed in Sec.~\ref{sec:dispersiveLimit}, however, these corrections are suppressed by an integer power of $\left<n\right>/n_c < 1$~\cite{boissonneault2009} and can therefore be neglected in the dispersive limit considered here. Moreover, we remark that there are additional spin relaxation channels when $|E_s| > |E_m|$, in which case the spin may indirectly relax to its ground state via an intermediate state~\cite{srinivasa2013}. Such processes are not included in Eq.~\eqref{eq:relaxationRate}. The following analysis is therefore restricted to the case $|E_s| < |E_m|$. Finally, note that in order to read out the spin, it is necessary that the relaxation rate be smaller than the resonator leakage rate, $\gamma \ll \kappa$.

\subsection{Signal-to-noise ratio \label{sec:signalToNoise}}

Due to noise in the homodyne signal, Eq.~\eqref{eq:stdHomodyne}, it is not possible to perfectly discriminate the output signals for the two qubit states in a finite time. Here, the readout time $t$ is limited by the inverse qubit relaxation time $\gamma^{-1}$. Thus, a useful measure of distinguishability of the two qubit states is the (power) SNR $\mathcal{S}_i$, defined as the ratio of the squared half-contrast observed in resonator port $i$ to the variance of the homodyne signal integrated over a time $t = \gamma^{-1}$~\cite{gambetta2007,gambetta2008,danjou2014,danjou2017-2}:

\begin{align}
 \mathcal{S}_i \equiv \frac{|\Delta \beta_i^{\textrm{out}}|^2}{4 \sigma_{\textrm{hom}}^2\left(t=\gamma^{-1}\right)} = \frac{r_i}{\gamma}. \label{eq:definitionSNR}
\end{align}

Here, we have defined the measurement rate
\begin{align}
 r_i \equiv R \frac{|\Delta \beta_i^{\textrm{out}}|^2}{4}. \label{eq:definitionMeasurementRate}
\end{align}
The measurement rate $r_i$ can be interpreted as the rate at which an observer at the $i\textrm{th}$ port of the resonator acquires information about the qubit state.

To estimate the SNR and measurement rate, we first recall that the critical photon number of the spin transition is $n_c = \Delta^2/4g_s^2$. Moreover, we recall that $g_s^2 \approx \chi_s\Delta$ under the assumption that only the spin transition is close to resonance with the resonator. Using these expressions and Eqs.~\eqref{eq:squareContrast} and \eqref{eq:relaxationRate}, Eqs.~\eqref{eq:definitionSNR} and \eqref{eq:definitionMeasurementRate} are rewritten as
\begin{align}
\begin{split}
 \mathcal{S}_i &= \frac{R}{4} \frac{\kappa_i}{\kappa} \frac{\left<n\right>}{n_c} \times \frac{1}{4x^2} D(x)G(y), \\
 \frac{r_i}{\kappa} &= \frac{R}{4} \frac{\kappa_i}{\kappa} \frac{\left<n\right>}{n_c} \times \frac{y}{4x} D(x). \label{eq:expressionSNR}
\end{split}
\end{align}
In Eq.~\eqref{eq:expressionSNR}, we defined the dimensionless parameters $x = \chi_s/\kappa$ and $y = \Delta/\kappa$ and
\begin{align}
 G(y) &= \left(\frac{1}{C} + \frac{1}{y^2}\right)^{-1}. \label{eq:functionsSNR}
\end{align}
Here $C$ is the cooperativity
\begin{align}
 C = \frac{\chi_s^2}{\gamma_{\textrm{pu}}\gamma_{\textrm{el}}} = \frac{g_s^2}{\kappa \gamma_{\textrm{el}}} = \frac{g_c^2}{\kappa \gamma_m}. \label{eq:definitionCooperativity}
\end{align}
We note that the cooperativity does not depend on the strength of the transverse magnetic field gradient $b_x$ hybridizing the electric dipole with the spin~\cite{benito2019}. This is because the dispersive shift $\chi_s$, the Purcell relaxation rate $\gamma_{\textrm{pu}}$, and the intrinsic relaxation rate $\gamma_{\textrm{el}}$ are all proportional to $b_x^2$. As a result, the maximum value of the SNR derived in Sec.~\ref{sec:optimizationDispersive}, Eq.~\eqref{eq:maximumAchievableSNR}, is also independent of $b_x$. This occurs because while reducing $b_x$ at fixed $\left<n\right>/n_c$ reduces the readout contrast, Eq.~\eqref{eq:squareContrast}, it proportionally reduces the spin-relaxation rate, Eq.~\eqref{eq:relaxationRate}. As a consequence, the loss in contrast can be exactly compensated by integrating the homodyne photocurrent for a longer time. Note that this does not mean that a large field gradient is unnecessary for readout. While reducing $b_x$ does not change the maximum achievable SNR, Eq.~\eqref{eq:maximumAchievableSNR}, it significantly reduces the measurement rate, Eq.~\eqref{eq:quantumLimitedRate}, and thus the measurement speed. As soon as the SNR reaches its maximum value, it therefore becomes undesirable to further reduce $b_x$. Following the discussion of Sec.~\ref{sec:optimizationDispersive}, it is easily verified that the SNR reaches its maximum value as soon as $g_s \lesssim \kappa C^{1/4}$. The magnetic field gradient $b_x$ should therefore be small enough so that $g_s \lesssim \kappa C^{1/4}$, but not lower.

The relevance of the SNR and measurement rate, Eq.~\eqref{eq:expressionSNR}, arises from the fact that they fully and monotonically determine the single-shot readout fidelity and the optimal readout time in the regime where the two states cannot be accurately discriminated in a time $\kappa^{-1}$. This is the case when $\gamma \ll r_i < \kappa$, i.e., $1 \ll \mathcal{S}_i < \kappa/\gamma$. In that regime, the single-shot readout fidelity $F_i$ (defined as the average probability of successful readout) and the optimal readout time $t_i^{\textrm{opt}}$ are approximately given by~\cite{gambetta2007,danjou2014,danjou2017-2}
\begin{align}
\begin{split}
 &F_i \approx 1 - \frac{1}{2\mathcal{S}_i} \ln \mathcal{S}_i, \;\;\;\; t^{\textrm{opt}}_i \approx \frac{2}{r_i} \ln \mathcal{S}_i. \label{eq:expressionFidelity}
\end{split}
\end{align}
Thus, optimizing the SNR automatically optimizes the single-shot readout fidelity. Even though the values of the SNR discussed below are of order $\mathcal{S}_i \gtrsim 1$, we have verified that Eq.~\eqref{eq:expressionFidelity} gives estimates similar to those obtained with a more detailed analysis~\cite{gambetta2007,danjou2014,danjou2017-2}. Note that when $r_i > \kappa$ and $\mathcal{S}_i > \kappa/\gamma$ (see the region enclosed by the dashed black line in Fig.~\ref{fig:fig4}), the transient behavior depicted in Fig.~\ref{fig:fig3} becomes important when estimating the fidelity. In that regime, Eq.~\eqref{eq:expressionFidelity} must be modified. The effect of such transient behavior can be taken into account within the theory of matched filtering~\cite{kay1998,ryan2015,harveycollard2018}. Also note that the infidelity $1-F_i$ is proportional to the probability $\sim \gamma t_i^{\textrm{opt}}$ of the qubit relaxing within time $t_i^\textrm{opt}$. It follows that for a readout time $t=t_i^\textrm{opt}$, it is necessary to have a high SNR, $\mathcal{S}_i \gg 1$, in order to have a quantum-non-demolition readout, $\sigma_z(t_i^\textrm{opt})\approx \sigma_z(0)$.

\subsection{Optimization of the dispersive parameters \label{sec:optimizationDispersive}}
We now turn our attention to the optimization of the dispersive parameters, $\chi_s$ and $\Delta$. In the present analysis, we assume that only the spin transition is close to resonance with the resonator so that the dispersive shift has the usual form $\chi_s \approx g_s^2/\Delta$. This greatly simplifies the optimization and gives the correct order of magnitude for the SNR and measurement rate. The effect of the corrections to the dispersive shift appearing in Eq.~\eqref{eq:dispersiveShiftFull} are discussed separately in Sec.~\ref{sec:optimizationDQDParameters}.

\begin{figure*}
\includegraphics[width = \textwidth]{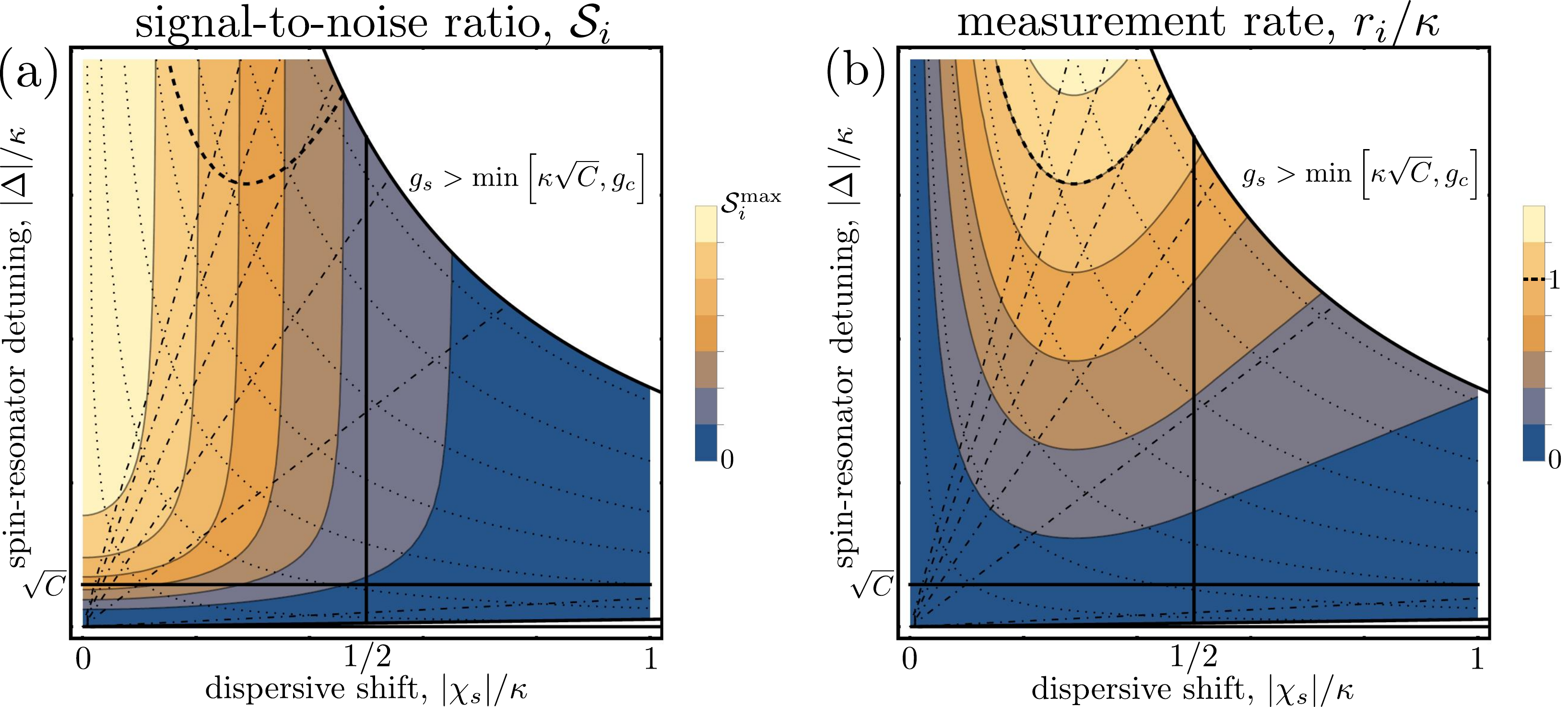}
\caption{Optimization landscape of (a) the SNR $\mathcal{S}_i$ and (b) the dimensionless measurement rate $r_i/\kappa$ plotted using Eq.~\eqref{eq:expressionSNR}. The SNR saturates when $|\chi_s|/\kappa < 1/2$ and $|\Delta|/\kappa > \sqrt{C}$, where it takes its maximum value $\mathcal{S}_i^{\textrm{max}} = R (\kappa_i/\kappa) (\left<n\right>/n_c) \; C$. Note that the SNR remains finite as $|\chi_s|/\kappa \rightarrow 0$ because the relaxation rate $\gamma \rightarrow 0$ compensates the loss in readout contrast in that regime. In reality, other spin-relaxation processes not considered here or other experimental limitations on the probe power cause the SNR to vanish at $|\chi_s|/\kappa = 0$. The dot-dashed black lines are contours of constant probe power $\propto n_c = \Delta/4\chi_s$ and the dotted black lines are the contours of constant $g_s = \sqrt{\chi_s \Delta}$. Due to the constraints $g_s \ll g_c$ and $\gamma \ll \kappa$, the value of $g_s$ is upper bounded either by $g_c$ or by $\kappa \sqrt{C}$. This is indicated by the white area in the upper right corner of each plot. The dashed black line is the contour $r_i = \kappa \Rightarrow \mathcal{S}_i=\kappa/\gamma$. The white region and the dashed black line are plotted for $C = 34.3$, $g_c/\kappa = 22.2$, and $\mathcal{S}_i^\textrm{max} = 3.43$.
\label{fig:fig4}}
\end{figure*}

We assume that the leakage rates are fixed and that the probe power is increased proportionally to the critical photon number of the spin transition $n_c$, i.e., the ratio $\left<n\right>/n_c \ll 1$ is kept constant. This ensures that as many photons as possible are put into the resonator for a given level of (small) disturbance to the qubit state. Maximizing the SNR given in Eq.~\eqref{eq:expressionSNR} then amounts to maximizing the quantity $D(x)G(y)/4x^2$. The optimization landscape is depicted in Fig.~\ref{fig:fig4} for both the SNR and the measurement rate. We also indicate contours of constant $g_s = \sqrt{\chi_s \Delta}$ and $n_c = \Delta/4\chi_s$. As will become clear below, these parameters sometimes provide a more convenient parametrization of the SNR and measurement rate.

As seen in Eqs.~\eqref{eq:expressionSNR} as well as in Fig.~\ref{fig:fig4}(a), the optimal SNR occurs for
\begin{align}
|\chi_s| \ll \frac{\kappa}{2}, \;\;\; |\Delta| \gg \kappa \sqrt{C}, \label{eq:saturationRegion}
\end{align}
where it saturates to its maximum value
\begin{align}
 \mathcal{S}_i^{\textrm{max}} = R \frac{\kappa_i}{\kappa} \frac{\left<n\right>}{n_c} \; C. \label{eq:maximumAchievableSNR}
\end{align}
It follows from Eq.~\eqref{eq:saturationRegion} that the critical photon number (and thus the probe power) must reach a high-enough value in order to achieve the optimum, namely $n_c \gg \sqrt{C}/2$. This also means that the optimum occurs deep in the dispersive regime, $n_c \gg 1$, when $C \gg 1$. Even though $n_c \gg \sqrt{C}/2$ is sufficient to saturate the SNR, further increasing $n_c$ can increase the measurement rate $r_i$, as can be seen in Fig.~\ref{fig:fig4}(b). In particular, for a fixed $|\chi_s|/\kappa \ll 1/2$, it is possible to achieve $\mathcal{S} \approx \mathcal{S}_i^{\textrm{max}}$ with the measurement rate scaling linearly with $n_c$ (for fixed $\left<n\right>/n_c$):
\begin{align}
 \frac{r_i}{\kappa} = 4 R \frac{\kappa_i}{\kappa} \frac{\left<n\right>}{n_c} \times \left(\frac{\chi_s}{\kappa}\right)^2 n_c. \label{eq:quantumLimitedRate}
\end{align}
We remark that the scaling of $r_i$ with $(\chi_s/\kappa)^2 \left<n\right>$ is expected from the fundamental limit set by quantum backaction~\cite{blais2004}. It is clear from Fig.~\ref{fig:fig4} that there is a trade-off between SNR and measurement rate. If a SNR of $\mathcal{S}_i = 9\mathcal{S}_i^{\textrm{max}}/16$ is deemed sufficient, for instance, then the dispersive shift need not be smaller than $|\chi_s| = \kappa/2\sqrt{3}$. The measurement rate is then $r_i \approx 3R \,\kappa_i\left<n\right>/16$, much larger than Eq.~\eqref{eq:quantumLimitedRate}.

Since the readout contrast $|\Delta\beta_i^\textrm{out}|$ decreases as $|\chi_s|/\kappa \rightarrow 0$, it may seem counterintuitive that the SNR saturates to a finite value in that limit. This occurs because the spin-relaxation processes that we consider, Eqs.~\eqref{eq:purcellRelaxationRate} and \eqref{eq:intrinsicRelaxationRate}, are also suppressed as $|\chi_s|/\kappa \rightarrow 0$. As a result, the reduction in contrast at fixed $\left<n\right>/n_c$ is compensated by a longer integration time $\propto \gamma^{-1}$ [see also the discussion following Eq.~\eqref{eq:definitionCooperativity}]. In reality, however, the SNR eventually starts to decrease as $|\chi_s|/\kappa \rightarrow 0$ either because additional spin relaxation processes not considered here become dominant or because other experimental constraints on the probe power become relevant. As made clear by Figs.~\ref{fig:fig4}(a) and \ref{fig:fig4}(b), however, the regime where these modifications become important is in general not desirable since it decreases the measurement rate without appreciably increasing the SNR.

There are other constraints that put limits on the values of $\chi_s$ and $\Delta$. In particular, the spin-resonator coupling $g_s = \sqrt{\chi_s \Delta}$ cannot be arbitrarily high for two distinct reasons. First, $g_s$ is limited by the bare dipole coupling $g_c$. Second, the readout must necessarily operate in a regime where $\gamma/\kappa \ll 1$. Using Eq.~\eqref{eq:relaxationRate}, we find that this latter constraint limits the coupling to $g_s \ll \kappa \sqrt{C}$. These constraints are indicated by the white area in Fig.~\ref{fig:fig4}. The optimal SNR of Eq.~\eqref{eq:maximumAchievableSNR} can nevertheless be achieved for any fixed value of $g_s$ provided that the probe power is high enough, $n_c \gg (g_s/\kappa)^2$. In the same limit, the measurement rate saturates:
\begin{align}
\frac{r_i}{\kappa} < R\frac{\kappa_i}{\kappa}\frac{\left<n\right>}{n_c} \left(\frac{g_s}{\kappa}\right)^2.
\end{align}
In practice, limits on the probe power might make it impossible to achieve the maximum SNR or measurement rate. For instance, $\left<n\right>$ may become limited by the critical photon number of the other DQD transitions. In such cases, the formalism of Sec.~\ref{sec:signalToNoise} may still be used to optimize readout under the appropriate constraints. However, we note that the present analysis must be modified when the spin-resonator detuning becomes comparable to the resonator frequency, in which case the spin and the resonator can no longer be assumed to be near resonance and $\gamma_m(E_s)$ can no longer be assumed to be frequency independent. This only occurs in the ultrastrong coupling regime, $\sqrt{n_c}g_s \lesssim \omega_r$. The analysis must also be modified to account for all terms in the dispersive shift of Eq.~\eqref{eq:dispersiveShiftFull} when the resonator is simultaneously close to resonance with $E_s$ and $E_+$ or $E_-$. The effect of these additional terms is discussed in Sec.~\ref{sec:optimizationDQDParameters}.

\subsection{Optimization of the double-quantum-dot parameters \label{sec:optimizationDQDParameters}}

As discussed in Sec.~\ref{sec:optimizationDispersive}, the dispersive parameters $\chi_s$ and $\Delta$ (and thus $g_s$) can be chosen to optimize the SNR and measurement rate under experimental constraints. There remains to find the set of tunable DQD parameters that correspond to the chosen values of $\chi_s$ and $\Delta$.

\begin{figure*}
\includegraphics[width = \textwidth]{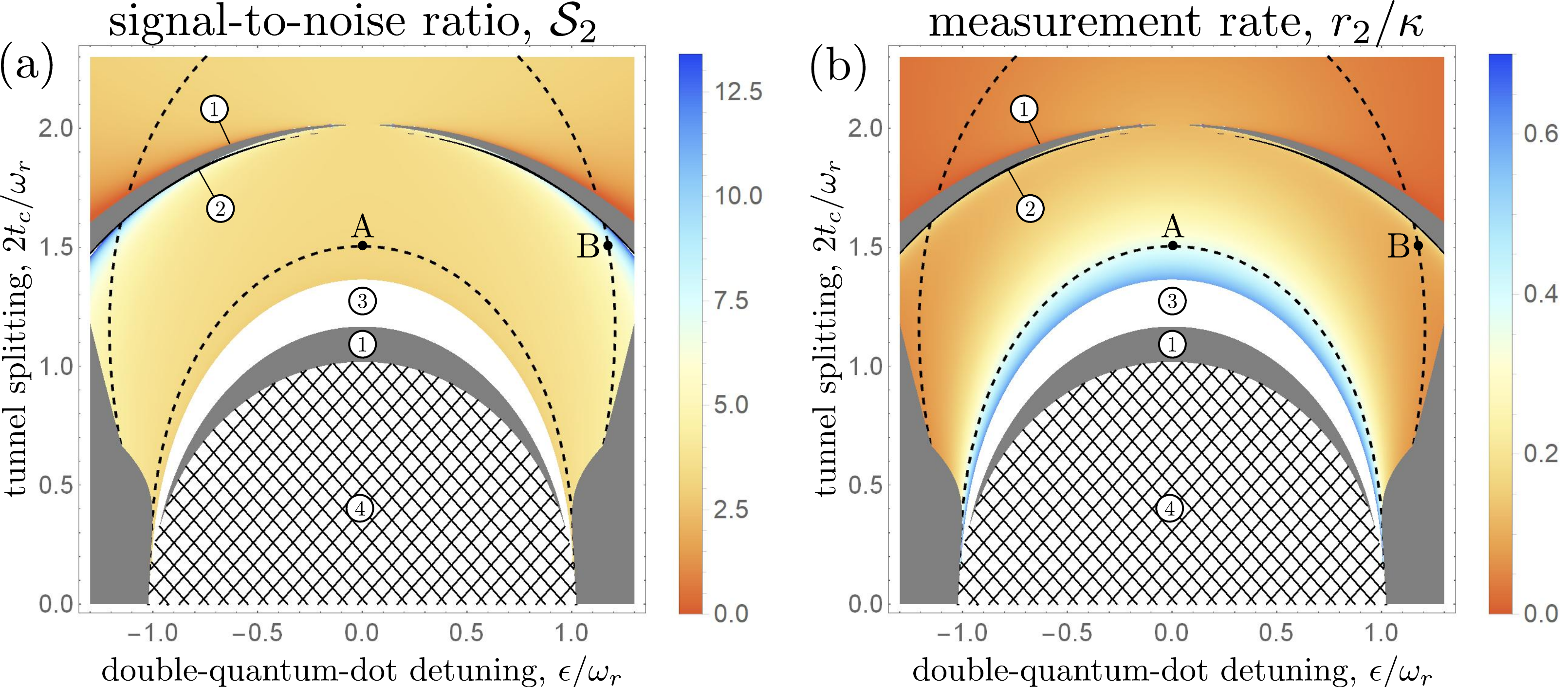}
\caption{Optimization landscape for (a) the SNR $\mathcal{S}_2$ and (b) the measurement rate $r_2$. Points A and B indicate the two numerical examples discussed in the text. The dashed black lines are contours of constant $g_s$ passing by $A$ and $B$ and are given by Eq.~\eqref{eq:molecularEnergyAngleRelationship}. The system parameters are similar to those measured in Ref.~\onlinecite{mi2018}, namely, $\omega_r = 2\pi\times(5.8\,\textrm{GHz})$, $\kappa = 2\kappa_1 = 2\kappa_2 = 2\pi\times(1.8\,\textrm{MHz})$, $b_x = 2\pi\times(420\,\textrm{MHz})$, and $g_c = 2\pi\times(40\,\textrm{MHz})$. We extract the molecular relaxation rate $\gamma_m = (6.1\,\textrm{ns})^{-1}$ by comparing the measured relaxation time in Ref.~\onlinecite{mi2018} with Eq.~\eqref{eq:relaxationRate}. This yields a cooperativity of $C \approx 34$. The noise is assumed to be quantum limited, $\bar{N}=0$ and $N_{\textrm{amp}} = 1/2$, and the number of photons in the resonator is fixed to $\left<n\right> = 0.1 n_c$. We fix $B_z = \omega_r + 10\kappa \sqrt{C} \approx 2\pi\times (5.9\,\textrm{GHz})$ to ensure that the optimal region of Fig.~\ref{fig:fig4}, $|\Delta| \gg \kappa \sqrt{C}$, is accessible. The contours are then plotted using Eqs.~\eqref{eq:definitionSNR} and \eqref{eq:definitionMeasurementRate} with $\Delta = \omega_r - E_s$ and $\chi_s$ given by Eq.~\eqref{eq:dispersiveShiftFull}. Here, the frequencies $E_j$ and the couplings $g_j$ are calculated as a function of $\epsilon$ and $t_c$ from the expressions given in Appendix~\ref{app:exactDiagonalizationDQD}. The grayscale areas with numbered circles indicate regions of parameter space where the various assumptions made in the text break down. In region 1 (dark gray), the dispersive approximation breaks down, $n_{c,j} < 10$ or $\left<n\right>/n_{c,j} > 0.1$ for all couplings $g_j$ in Eq.~\eqref{eq:interactionDiag} (except for the spin transition for which $\left<n\right> = 0.1 n_c$ everywhere). In region 2 (black), the transition terms $V_{\textrm{tr}}$ discussed in Sec.~\ref{sec:dispersiveLimit} become resonant (see Appendix~\ref{app:transitionTerms} for details). In region 3 (white), the relaxation rate $\gamma$ is larger than $\kappa/5$. In region 4 (tiled), the spin Larmour frequency is larger than the molecular Larmour frequency, $|E_s| > |E_m|$. \label{fig:fig5} }
\end{figure*}

The optimal magnetic field is determined by requiring that $\omega_r - E_s = \Delta$. In the limit of a weak field-gradient and weak spin-orbit admixture considered here, the spin-resonator detuning $\Delta$ is typically chosen to be much larger than the correction to $E_s$ appearing in Eq.~\eqref{eq:approximateFrequencies}. In that case, the optimal magnetic field is approximately
\begin{align}
 B_z \approx \omega_r - \Delta.
\end{align}
Having thus fixed $B_z$, the optimal values of the DQD energy detuning $\epsilon$ and of the tunnel coupling $t_c$ for the chosen values of $\chi_s$ and $\Delta$ are determined by requiring that $g_s$ is a constant $g_s(\epsilon,t_c) = \sqrt{\chi_s \Delta}$. Using Eqs.~\eqref{eq:approximateFrequencies} and \eqref{eq:spinAndChargeCouplings}, this leads to the following relationship between $\epsilon$ and $t_c$:
\begin{align}
\begin{split}
 \frac{2t_c}{B_z} = \frac{1}{2}\mu \cos^3\theta \pm \cos\theta\sqrt{1+\frac{1}{4}\mu^2 \cos^4\theta}. \label{eq:molecularEnergyAngleRelationship}
\end{split}
\end{align}
Here, $\tan \theta = \epsilon/2t_c$ and
\begin{align}
 \mu = \frac{g_c}{g_s}\frac{b_x}{B_z}.
\end{align}
Equation~\eqref{eq:molecularEnergyAngleRelationship} defines contours in the $\left(\epsilon,2t_c\right)$ plane as a function of the parameter $\mu$. Two such contours are indicated by the dashed lines in Fig.~\ref{fig:fig5}. Within the approximation $\chi_s \approx g_s^2/\Delta$, every point on such a contour yields the same SNR and measurement rate, with the SNR (measurement rate) increasing (decreasing) with increasing $\mu$. Thus, the qubit readout can be operated with a similar performance over a wide range of DQD energy detunings $\epsilon$ provided that the tunnel coupling $t_c$ is adjusted to remain on the chosen contour. In particular, such freedom can be used to operate the readout at ``sweet spots'' of the qubit energy dispersion, where the coherence time of the qubit is expected to be longer~\cite{reed2016,martins2016}.

Until now, we have assumed that only the spin transition contributes to the spin dispersive shift, $\chi_s \approx g_s^2/\Delta$. We now relax this assumption to include all terms in Eq.~\eqref{eq:dispersiveShiftFull}. The optimization landscape for the SNR and measurement rate beyond the assumption $\chi_s \approx g_s^2/\Delta$ is shown in Fig.~\ref{fig:fig5} as a function of $\epsilon$ and $2t_c$. It is plotted at fixed $\left<n\right>/n_c$ using the definitions of the SNR and measurement rate, Eqs.~\eqref{eq:definitionSNR} and \eqref{eq:definitionMeasurementRate}, with the dispersive shift $\chi_s$ given by Eq.~\eqref{eq:dispersiveShiftFull}. Here the values of the spin-qubit frequency $E_s$ and the spin-resonator coupling $g_s$ are calculated from the exact expressions given in Appendix~\ref{app:exactDiagonalizationDQD}. The numerical values of the parameters are the ones given in Sec.~\ref{sec:fidelityEstimates}. Figure~\ref{fig:fig5} shows that when the resonator frequency $\omega_r$ becomes close to the transition at frequency $E_{-}$ (e.g., point B), the contours get distorted compared to what is predicted by Eq.~\eqref{eq:molecularEnergyAngleRelationship}. This is due to the corrections to the dispersive shift appearing in Eq.~\eqref{eq:dispersiveShiftFull}. In the absence of these corrections, any reduction in the spin-photon coupling $g_s$ leads to a simultaneous reduction of the readout contrast and of the relaxation rate. For fixed $\left<n\right> /n_c$, these two effects compensate each other exactly and the SNR saturates to the value $\mathcal{S}_i^{\textrm{max}}$. However, the corrections to the dispersive shift (and thus to the readout contrast) in Eq.~\eqref{eq:dispersiveShiftFull} occur without a corresponding change in the qubit relaxation rate. This means that the SNR and measurement rate are enhanced on one side of the $E_-$ transition and suppressed on the other, depending on the relative arrangement of the resonator and DQD transition frequencies. The enhancement regime was termed the ``straddling regime'' in the theory of superconducting qubits~\cite{koch2007,boissonneault2012,zhu2013}. Figure~\ref{fig:fig5} shows that the straddling regime for the present spin qubit only occurs at large DQD energy detunings $\epsilon$, when the DQD-resonator couplings $g_\pm$ become finite.

The regions where our model breaks down are also indicated in Fig.~\ref{fig:fig5}. In particular, the regions where the dispersive assumptions of Sec.~\ref{sec:dispersiveLimit} are no longer valid are indicated (regions 1 and 2). In region 1, the resonator photon number exceeds the critical photon number for DQD transitions other than the spin transition, while in region 2, the transition term $V_{\textrm{tr}}$ causes unwanted DQD transitions via the absorption of two photons (see Appendix~\ref{app:transitionTerms} for details). It follows that the backaction of resonator photons on the qubit state is small far from regions 1 and 2. Therefore, the readout is approximately quantum-non-demolition, $\sigma_z(t) \approx \sigma_z(0)$, far from regions 1 and 2 provided that the qubit does not relax with high probability during the readout time $t$, $\gamma t \ll 1$. The region of parameter space where the qubit relaxation rate $\gamma$ becomes comparable to $\kappa$ is also plotted (region 3). Far from this region, the probability that the qubit state relaxes during a time $\kappa^{-1}$ becomes small, ensuring that readout is possible. Finally, the region where the spin-like Larmour frequency is larger than the molecular-like Larmour frequency, $|E_s| > |E_m|$, is indicated (region 4). This region is excluded since that regime enables additional spin-relaxation channels that are not considered here (see Sec.~\ref{sec:qubitRelaxation}).

\subsection{Single-shot readout fidelity estimates \label{sec:fidelityEstimates}}

To determine the best possible performance of current technologies, we estimate the achievable SNR and measurement rate for parameters similar to those measured in Ref.~\onlinecite{mi2018}. These are $\omega_r = 2\pi\times(5.8\,\textrm{GHz})$, $\kappa = 2\pi\times(1.8\,\textrm{MHz})$, $b_x = 2\pi\times(420\,\textrm{MHz})$, and $g_c = 2\pi\times(40\,\textrm{MHz})$. The value of the molecular relaxation rate $\gamma_m$ is extracted by fitting the relaxation time measured in Ref.~\onlinecite{mi2018} to Eq.~\eqref{eq:relaxationRate}. We find $\gamma_m = (6.1\,\textrm{ns})^{-1}$. This leads to a cooperativity $C \approx 34$ for this device. We assume that the transmitted field is measured through a symmetric resonator, $\kappa_1 = \kappa_2 = \kappa/2$. Moreover, we assume that the amplification and detection processes are quantum limited, i.e., that they add the minimum number of noise photons $N_{\textrm{amp}}=1/2$ allowed by quantum mechanics for amplifier gains much larger than unity~\cite{clerk2010,mutus2014,roy2015}. In addition, we take the average number of photons in the resonator at resonance to be a tenth of the critical photon number for the spin transition, $\left<n\right>=0.1n_c$. The theoretical maximum achievable SNR in transmission, Eq.~\eqref{eq:maximumAchievableSNR}, is then $\mathcal{S}_2^{\textrm{max}} \approx 3.4$. According to Eq.~\eqref{eq:expressionFidelity}, this corresponds to a single-shot readout fidelity $F_2 \approx 82 \%$.

The above fidelity can be achieved within a readout time that is comparable to $\kappa$. To illustrate this, we work at zero DQD energy detuning $\epsilon = 0$ and tunnel splitting $2t_c = 1.5\omega_r = 2\pi \times (8.7\,\textrm{GHz})$. This corresponds to point A in Fig.~\ref{fig:fig5}. For these parameters, the SNR is $\mathcal{S}_2 = 0.97\mathcal{S}_2^{\textrm{max}}$. Moreover, the measurement rate is $r_2 = 0.37 \kappa$. According to Eq.~\eqref{eq:expressionFidelity}, this corresponds to a fidelity $F_2 \approx 82 \%$ achievable with optimal readout time $t^{\textrm{opt}}_2 \approx 6.4\kappa^{-1}$. The above parameters correspond to a relaxation rate $\gamma = 0.11 \kappa$, a spin-photon coupling $g_s = 2\pi\times(3.5\,\textrm{MHz})$, and a critical photon number $n_c = 174$. To achieve this performance, $\left<n\right> \approx 17$ photons must therefore be introduced into the resonator mode. 

At point B in Fig.~\ref{fig:fig5}, the tunnel splitting has the same value $2t_c = 1.5\omega_r = 2\pi \times (8.7\,\textrm{GHz})$ but the DQD energy detuning is now increased to $\epsilon = 1.175\omega_r = 2\pi \times (6.815\,\textrm{GHz})$ to enter the straddling regime. Without the presence of an additional transition at frequency $E_-$, this would simply reduce the measurement rate to $r_2 = 0.05\kappa$ without appreciably increasing the SNR. Because of the straddling effect, however, the SNR increases to twice its theoretical maximum value, $\mathcal{S}_2 = 1.97\mathcal{S}_2^{\textrm{max}}$, while the measurement rate is twice what it would have been without the straddling effect, $r_2 = 0.1\kappa$. According to Eq.~\eqref{eq:expressionFidelity}, this increases the readout fidelity to $F_2 \approx 86\%$ achievable within a readout time $t^{\textrm{opt}}_2 \approx 38\kappa^{-1}$. The above parameters correspond to a relaxation rate $\gamma = 0.015\kappa$, a spin-photon coupling $g_s = 2\pi\times(1.3\,\textrm{MHz})$, and a critical photon number $n_c = 1690$. Thus, a larger number of photons $\left<n\right> = 169$ must be put into the resonator to achieve optimal performance in the straddling regime. Note that at point B in Fig.~\ref{fig:fig5}, the molecular energy gap is $\Omega \approx 46\,\mu\textrm{eV}$, which can be comparable to the valley splitting observed in silicon qubits. If this is the case, then the present analysis must be modified to account for valley physics. We note, however, that the presence of additional valley states is not necessarily detrimental for dispersive readout. Indeed, the coupling to spin-valley transitions to the resonator could potentially contribute constructively to the dispersive shift $\chi_s$ and thereby enhance the straddling effect discussed here.

These estimates suggest that a high single-shot readout fidelity could be achieved in the near future with the help of quantum limited amplifiers and improvements in resonator impedance to boost the DQD-resonator coupling $g_c$. While the readout is not quantum nondemolition for the values of $\gamma t_i^{\textrm{opt}}$ in the examples above, it will become less destructive as the SNR increases (see the discussion at the end of Sec.~\ref{sec:signalToNoise}). It must also be noted that the above estimates are based on a rather conservative value of the ratio $\left<n\right>/n_c$. Indeed, it has been empirically observed that the backaction of the resonator photons on a superconducting qubit can remain small for up to $\left<n\right>\approx 4 n_c$~\cite{jeffrey2014}. If this is also the case here, then the above values of the SNR and measurement rates could be increased by up to a factor of $40$, while the measurement rate could be increased well above $\kappa$. According to Eq.~\eqref{eq:expressionFidelity}, this would lead to a single-shot readout fidelity of 99\% for the parameters of point B. Note, however, that Eq.~\eqref{eq:expressionFidelity} must be modified to yield quantitative predictions in the regime $r_i > \kappa$ due to detrimental effect of the finite rise time of the readout signal depicted in Fig.~\ref{fig:fig3}(b). Solving Eq.~\eqref{eq:nonDemolitionEOM} give transients of the form $\pm (|\Delta \beta|/2) [1-(1+\kappa t/2)e^{-\kappa t/2}]$ for $|\chi| \ll \kappa/2$. Using these expressions and the method outlined in Ref.~\onlinecite{harveycollard2018}, we have verified that setting $\left<n\right> = 4n_c$ leads to $F_i > 95\%$ for point B. For such high values of $\left<n\right>$, however, a quantitative study of the probe backaction is required to fully validate the fidelity estimate. Such an analysis goes beyond the scope of this work. Finally, we note that the use of Purcell filters~\cite{jeffrey2014,sete2015,walter2017,cleland2019} could significantly reduce qubit relaxation rates and thereby lead to even higher readout fidelities.

\section{Conclusions \label{sec:conclusions}}

In conclusion, we have optimized the dispersive readout of a semiconductor spin qubit in a DQD coupled to a microwave resonator via a transverse magnetic field gradient. Importantly, our analysis accounts for intrinsic relaxation of the spin due to electric noise in the semiconductor environment. We have given an expression for the maximum achievable SNR in terms of the cooperativity associated with the Purcell emission and the intrinsic relaxation. This expression also encapsulates the dependence on the amplifier noise and the probe power. We find that for the relaxation processes considered, the cooperativity increases with the coupling $g_c$ between the electric dipole and the resonator but is independent of the strength of the transverse magnetic field gradient $b_x$. Moreover, we have described how to choose the experimentally tunable parameters of the DQD to optimize the SNR. Our analysis enables us to identify the regions of parameter space where the backaction of the resonator photons on the qubit state is small. To do this, we systematically study all terms in the Hamiltonian that induce transitions between the DQD eigenstates and require that they be off-resonant (see Fig.~\ref{fig:fig5}). In addition, we find that it is possible to operate the readout with a similar performance for a wide range of tunable DQD parameters. Such flexibility is important because it frees up the parameter space for the optimization of other qubit performance metrics. Moreover, we find that transitions that simultaneously change the molecular wave function and the spin can be exploited to enhance the SNR by at least a factor of two. This ``straddling'' effect occurs only at nonzero energy detuning of the DQD double-well potential. Finally, we estimate that single-shot readout fidelities in the range $82-95\%$ should be achievable within a few $\mu\textrm{s}$ of readout time with current technology.

Our work provides the baseline for benchmarking future improvements, including the use of Purcell filters~\cite{jeffrey2014,sete2015,walter2017,cleland2019}, the development of techniques to circumvent Purcell emission~\cite{sete2013,gard2018,wang2018,harrington2019,peronnin2019,dassonneville2019,ruskov2019} or phonon emission~\cite{agarwal2013,rosen2019}, the use of phase-sensitive amplifiers to selectively amplify the relevant readout quadrature~\cite{castellanosbeltran2008,clerk2010,eddins2019}, pulse shaping~\cite{motzoi2018}, and the development of new (meta-)materials for high-impedance resonators~\cite{hagmann2005,altimiras2013,stockklauser2017,bosco2019}. Another important avenue for future research is to incorporate valley physics~\cite{burkard2016} relevant for, e.g., silicon-based qubits into the present analysis. In particular, it is yet unclear whether the straddling effect discussed here could also benefit from coupling the resonator to valley transitions. A more detailed study of readout backaction in the presence of noise sources relevant to semiconductor qubits is also highly desirable to quantify exactly how strongly the resonator can be driven without disturbing the qubit state.

\begin{acknowledgments}
This work was financially supported by the \href{http://dx.doi.org/10.13039/501100000038}{National Sciences and Engineering Research Council of Canada (NSERC)}. We also acknowledge M. Benito and M. Russ for helpful discussions.
\end{acknowledgments}

\appendix

\section{Exact diagonalization of the double-quantum-dot Hamiltonian \label{app:exactDiagonalizationDQD}}

The DQD Hamiltonian, Eq.~\eqref{eq:hamiltonianDQD}, may be diagonalized exactly in three steps. First, we write Eq.~\eqref{eq:hamiltonianDQD} in the eigenbasis of the molecular Hamiltonian. This is done with the transformation
\begin{align}
 U_0 = \textrm{exp}\left[-i \frac{(\pi/2-\theta)}{2} \widetilde{\tau}_y\right]. \label{eq:unitary0}
\end{align}
The transformed Hamiltonian takes the form
\begin{align}
\begin{split}
 U_0^\dagger H_d U_0 = &\frac{\Omega}{2} \widetilde{\tau}_z + \frac{B_z}{2}\widetilde{\sigma}_z \\
 &+ \frac{b_x \sin\theta}{2}\widetilde{\tau}_z \widetilde{\sigma}_x - \frac{b_x \cos\theta}{2}\widetilde{\tau}_x \widetilde{\sigma}_x.
\end{split}
\end{align}
Second, the spin basis is rotated to match the direction of the total magnetic field $\bold{B} = \left(b_x \sin \theta, 0, B_z\right)$. This is achieved through the unitary transformation
\begin{align}
 U_1 = \textrm{exp}\left( -i \frac{\Phi}{2} \widetilde{\tau}_z \widetilde{\sigma}_y \right). \label{eq:transformation1}
\end{align}
Here $\Phi$ is the angle between the magnetic field and the $z$ axis, satisfying $\tan \Phi = b_x \sin \theta / B_z$. In the doubly transformed basis, the DQD Hamiltonian takes the form
\begin{align}
\begin{split}
 U_1^\dagger U_0^\dagger H_d &U_0 U_1 \\
 &= \frac{\Omega}{2}\widetilde{\tau}_z + \frac{B_z \sec \Phi}{2}\widetilde{\sigma}_z - \frac{b_x \cos \theta}{2}\widetilde{\tau}_x \widetilde{\sigma}_x. \label{eq:diagonalization1}
\end{split}
\end{align}
The Hamiltonian of Eq.~\eqref{eq:diagonalization1} preserves the parity quantum number $\widetilde{\tau}_z \widetilde{\sigma}_z$. Thus, it may be diagonalized separately for each parity. The corresponding unitary transformation is
\begin{align}
\begin{split}
 &U_2 = U_{2+} + U_{2-}, \\
 &U_{2+} = \cos\frac{\phi_+}{2}\, P_+ - \sin\frac{\phi_+}{2}\left(\widetilde{\tau}_- \widetilde{\sigma}_- - \widetilde{\tau}_+ \widetilde{\sigma}_+\right), \\
 &U_{2-} = \cos\frac{\phi_-}{2}\,P_- - \sin\frac{\phi_-}{2}\left(\widetilde{\tau}_- \widetilde{\sigma}_+ - \widetilde{\tau}_+ \widetilde{\sigma}_-\right),
\end{split}
\end{align}
where $P_{\pm} = (1 \pm \widetilde{\tau}_z \widetilde{\sigma}_z)/2$ are the projectors on the subspaces of parity $\pm$, respectively. The effective spin-orbit mixing angles are determined by
\begin{align}
 \tan \phi_\pm = \frac{b_x \cos\theta}{\Omega \pm B_z \sec \Phi}.
\end{align}
Defining the total unitary transformation $U = U_0 U_1 U_2$, the DQD Hamiltonian becomes
\begin{align}
 U^\dagger H_d U = \frac{E_m}{2} \widetilde{\tau}_z + \frac{E_s}{2} \widetilde{\sigma}_z, \label{eq:diagonalization2}
\end{align}
where the molecular and spin Larmour frequencies in the dressed basis are
\begin{align}
\begin{split}
 &E_m = \frac{b_x \cos\theta}{2}\left( \csc \phi_+ + \csc \phi_- \right), \\
 &E_s = \frac{b_x \cos\theta}{2}\left( \csc \phi_+ - \csc \phi_- \right). \label{eq:exactFrequencies}
\end{split}
\end{align}
To indicate the final DQD eigenbasis, the explicit unitary transformation as well as the ``$\widetilde{\;\;\;}$'' are dropped. Equation~\eqref{eq:diagonalization2} then becomes Equation~\eqref{eq:hamiltonianDQDDiag}.

The resonator couples to the DQD via the dimensionless position operator $\zeta = \ketbra{\widetilde{R}}{\widetilde{R}}-\ketbra{\widetilde{L}}{\widetilde{L}}$. The unitary transformation of Eq.~\eqref{eq:diagonalization2} may be used to express $\zeta$ in the DQD eigenbasis as
\begin{align}
 \zeta = \sum_{i,j} \zeta^{(ij)} \tau_i \sigma_j. \label{eq:positionDiag}
\end{align}
Here, the Pauli matrices are labeled with indices $\left\{0,x,y,z\right\}$, where $0$ signifies the identity matrix. The nonzero coefficients $\zeta^{(ij)}$ are found to be
\begin{align}
\begin{split}
 \zeta^{(x0)} =& -\cos\theta \cos\Phi \cos\bar{\phi}, \\
 \zeta^{(zx)} =& \cos \theta \cos\Phi \sin\bar{\phi}, \\
 \zeta^{(xx)} =& \sin \theta\sin\bar{\phi}\cos\frac{\Delta\phi}{2} \\
 & \;\;\; + \cos\theta\sin\Phi \sin\bar{\phi} \sin\frac{\Delta\phi}{2},\\
 \zeta^{(yy)} =& -\sin\theta \cos\bar{\phi} \sin\frac{\Delta\phi}{2} \\
 & \;\;\; + \cos\theta \sin\Phi \cos\bar{\phi} \cos\frac{\Delta\phi}{2}, \\
 \zeta^{(z0)} = & \sin\theta \cos\bar{\phi} \cos\frac{\Delta\phi}{2} \\
 & \;\;\; + \cos\theta \sin\Phi \cos\bar{\phi} \sin\frac{\Delta\phi}{2}, \\
 \zeta^{(0z)} = & -\sin\theta \sin\bar{\phi} \sin\frac{\Delta\phi}{2} \\
 & \;\;\; + \cos\theta \sin\Phi \sin\bar{\phi} \cos\frac{\Delta\phi}{2}. \label{eq:positionDiagCoefficients}
\end{split}
\end{align}
In Eq.~\eqref{eq:positionDiagCoefficients}, we have introduced the average spin-orbit mixing angle $\bar{\phi} = (\phi_+ + \phi_-)/2$ and the difference angle $\Delta\phi = \phi_+ - \phi_-$. Expressions for the DQD-resonator couplings appearing in Eq.~\eqref{eq:interactionDiag} are directly obtained from the $\zeta^{(ij)}$:
\begin{align}
\begin{array}{ll}
 g_m = -g_c \zeta^{(x0)}, &g_s = g_c \zeta^{(zx)},\\
 g_+ = g_c [\zeta^{(xx)}-\zeta^{(yy)}], &g_- = g_c [\zeta^{(xx)}+\zeta^{(yy)}],\\
 g_{mp} = g_c \zeta^{(z0)}, &g_{sp} = g_c \zeta^{(0z)}. \label{eq:exactCouplings}
\end{array}
\end{align}
In the limit of small field gradient discussed in Sec.~\ref{sec:DQDEigenbasis}, Eqs.~\eqref{eq:exactFrequencies} and \eqref{eq:exactCouplings} yield Eqs.~\eqref{eq:approximateFrequencies} and \eqref{eq:spinAndChargeCouplings}, respectively. Moreover, note that the matrix element of the dimensionless position operator $\zeta$ between the two spin-qubit states is proportional to $g_s/g_c$. This means that spin transitions arising from the coupling of electric fields to the electric dipole occur at a rate proportional to $\left(g_s/g_c\right)^2$, as was assumed in Eq.~\eqref{eq:intrinsicRelaxationRate}. Finally, we remark that exact expressions for the transformed spin operators may also be obtained.

\section{Dispersive double-quantum-dot-resonator Hamiltonian \label{app:fullDispersive} }

As derived in Appendix~\ref{app:exactDiagonalizationDQD}, the DQD-resonator Hamiltonian expressed in the DQD eigenbasis is
\begin{align}
\begin{split}
 &H = H_0 + V, \\
 &H_0 = \frac{E_m}{2}\tau_z + \frac{E_s}{2}\sigma_z + \omega_r a^\dagger a, \\
 &V = \mathcal{V} (a + a^\dagger), \\
 &\mathcal{V} = -g_m \tau_x + g_s \tau_z \sigma_x + g_+ \left(\tau_+ \sigma_+ + \tau_- \sigma_-\right) \\
 & \;\;\;\;\;\;\;\;\; + g_- \left(\tau_+ \sigma_- + \tau_- \sigma_+\right) + g_{mp} \tau_z + g_{sp} \sigma_z. \label{eq:hamiltonianSummary}
\end{split} 
\end{align}
The dispersive Hamiltonian $H_{\textrm{dis}}$ is obtained by diagonalizing the system Hamiltonian to first order in $g_c$. This is achieved with the help of a Schrieffer-Wolff transformation:
\begin{align}
 H_{\textrm{dis}} = e^S H e^{-S}. \label{eq:transformationSW}
\end{align}
Here, the generator $S$ of the transformation is the solution of
\begin{align}
 [H_0,S] = V. \label{eq:equationSW}
\end{align}
Expanding the transformation of Eq.~\eqref{eq:transformationSW} to second order in $S$ gives
\begin{align}
 H_{\textrm{dis}} = H_0 + \frac{1}{2}[S,V].
\end{align}
The operators on the right-hand side are expressed in a basis dressed by the resonator.

An expression for $S$ is most conveniently obtained with the ansatz
\begin{align}
 S = \Sigma_I I + \Sigma_Q Q, \label{eq:ansatzSW}
\end{align}
where $I = (a+a^\dagger)/2$ and $Q = -i(a-a^\dagger)/2$ are the quadratures of the resonator field. Substituting Eq.~\eqref{eq:ansatzSW} into Eq.~\eqref{eq:equationSW} and solving for $\Sigma_I$ and $\Sigma_Q$ gives
\begin{align}
\begin{split}
 \Sigma_I = \frac{-2 L_d}{\omega_r^2 - L_d^2} \mathcal{V}, \\
 \Sigma_Q = \frac{-2 i \omega_r}{\omega_r^2 - L_d^2} \mathcal{V}.
\end{split} \label{eq:sigmas}
\end{align}
Here, $L_d$ is the Liouville operator corresponding to $H_d$, $L_d O = [H_d, O]$ for any operator $O$. Substituting the explicit form for $\mathcal{V}$ in Eq.~\eqref{eq:hamiltonianSummary} into Eq.~\eqref{eq:sigmas} yields explicit expressions for $\Sigma_I$ and $\Sigma_Q$:
\begin{align}
\begin{split}
 &\Sigma_I = i \eta_m \tau_y - i \eta_s \tau_z \sigma_y - \eta_+ (\tau_+ \sigma_+ - \tau_- \sigma_-) \\
 & \;\;\;\;\;\;\;\; - \eta_-(\tau_+ \sigma_- - \tau_- \sigma_+), \\
 &\Sigma_Q = i \eta'_m \tau_x - i \eta'_s \tau_z \sigma_x - i \eta'_+ (\tau_+ \sigma_+ + \tau_- \sigma_-) \\
 & \;\;\;\;\;\;\;\; - i \eta'_- (\tau_+ \sigma_- + \tau_- \sigma_+) - i\eta'_{mp} \tau_z - i \eta'_{sp}\sigma_z.
\end{split} \label{eq:explicitExpressionS}
\end{align}
In Eq.~\eqref{eq:explicitExpressionS}, the dispersive parameters $\eta_i$ and $\eta'_i$ are given by Eq.~\eqref{eq:dispersiveParameters}. In the case of the couplings $g_{mp}$ and $g_{sp}$, it is understood that $E_{mp} = 0$ and $E_{sp} = 0$. Using these expressions, the dispersive Hamiltonian takes the form
\begin{align}
\begin{split}
 H_{\textrm{dis}} = H_0 &+ \mathcal{V}_0 + \frac{\mathcal{V}_I}{2}\left(a^\dagger a + \frac{1}{2}\right) \\
 &+ \frac{\mathcal{V}_I}{4}\left({a^\dagger}^2 + a^2\right) +\frac{i \mathcal{V}_Q}{4}\left({a^\dagger}^2 - a^2\right),\label{eq:fullDispersiveHamiltonian}
\end{split}
\end{align}
where
\begin{align}
\begin{split}
 &\mathcal{V}_0 = -\frac{i}{4}\left\{\Sigma_Q,\mathcal{V}\right\} = \sum_{ij} v_0^{(ij)} \tau_i \sigma_j, \\
 &\mathcal{V}_I = \left[ \Sigma_I, \mathcal{V} \right] = \sum_{ij} v_I^{(ij)} \tau_i \sigma_j, \\
 &\mathcal{V}_Q = \left[ \Sigma_Q, \mathcal{V} \right] = \sum_{ij} v_Q^{(ij)} \tau_i \sigma_j. 
\end{split}
\end{align}
It is straightforward to calculate the coefficients $v_0^{(ij)}$, $v_I^{(ij)}$, and $v_Q^{(ij)}$ explicitly. Each of these coefficients is a linear combination of the elements of the dispersive tensors $\chi_{j,k} = g_j \eta_k$ and $\chi'_{j,k} = g_j \eta'_k$. Although we do not write all the coefficients explicitly here, we note that the interaction Hamiltonian in Eq.~\eqref{eq:fullDispersiveHamiltonian} has a part $V_{\textrm{dis}}$ that commutes with $H_0$ and a part $V_{\textrm{tr}}$ that induces transitions between the eigenstates of $H_0$:
\begin{align}
 H_{\textrm{dis}} = H_0 + V_{\textrm{dis}} + V_{\textrm{tr}}.
\end{align}
The dispersive part of the interaction has the form (up to an irrelevant additive constant)
\begin{align}
\begin{split}
 V_{\textrm{dis}} = - \frac{1}{2} \chi_0 \tau_z \sigma_z -\left( \chi_m \tau_z + \chi_s \sigma_z \right)\left(a^\dagger a + \frac{1}{2}\right). \label{eq:dispersiveInteractionFullApp}
\end{split}
\end{align}
Here, we have introduced the dispersive shifts
\begin{align}
\begin{split}
 &\chi_m = \chi_{m,m} + \frac{1}{2}\left( \chi_{+,+} + \chi_{-,-} \right), \\
 &\chi_s = \chi_{s,s} + \frac{1}{2}\left( \chi_{+,+} - \chi_{-,-} \right), \\
 &\chi_0 = \chi'_{mp,sp} + \chi'_{sp,mp} + \frac{1}{2}\left(\chi'_{+,+} - \chi'_{-,-}\right).
\end{split}
\end{align}
The expression for $\chi_s$ corresponds to the one given in Eq.~\eqref{eq:dispersiveShiftFull}. In the case where the transition term is off-resonant with all possible transitions between eigenstates of $H_0$, the dispersive Hamiltonian may be projected into the molecular ground state $\ket{-}$ to obtain an effective dispersive spin Hamiltonian (to an irrelevant additive constant):
\begin{align}
 H_{\textrm{dis}}^{\textrm{eff}} = \left(\omega_r' - \chi_s \sigma_z\right) a^\dagger a + \frac{1}{2}\left(E_s' - \chi_s\right)\sigma_z,
\end{align}
where
\begin{align}
 \omega_r' = \omega_r + \chi_m, \;\;\;\; E_s' = E_s + \chi_0.
\end{align}

The transformation of Eq.~\eqref{eq:transformationSW} may be used to transform other system operators in the new basis. Most importantly for dispersive readout, the resonator field transforms to second order in $S$ as
\begin{align}
 e^{S} a e^{-S} = a - \frac{1}{2}\left(\Sigma_I + i \Sigma_Q\right) - \frac{i}{4}\left[\Sigma_I,\Sigma_Q\right]a. \label{eq:transformationA}
\end{align}
Equations~\eqref{eq:dispersiveInputInteraction} and \eqref{eq:inputOutput2} are respectively obtained by transforming Eqs.~\eqref{eq:inputInteraction} and \eqref{eq:inputOutput} to leading order in $g_s$ using Eq.~\eqref{eq:transformationA}, and then projecting the result onto the spin-qubit subspace in the limit where the resonator is near-resonant with the spin transition.

\section{Neglecting the transition terms \label{app:transitionTerms}}

To ensure that the transition term $V_{\textrm{tr}}$ appearing in Eq.~\eqref{eq:dispersiveHamiltonianFull} causes a negligible change in the system eigenstates, its magnitude $||V_{\textrm{tr}}||$ should be much smaller than its detuning $\Delta_{\textrm{tr}}$ from the transition it induces between two eigenstates of $H_0+ V_{\textrm{dis}}$.

For example, one of the terms in $V_{\textrm{tr}}$ has the form
\begin{align}
\begin{split}
 &V_{\textrm{tr}}^{m,2\textrm{PH}} \\
 &= \frac{1}{4}v_I^{(x0)} \tau_x ({a^\dagger}^2 + a^2) + \frac{i}{4} v_Q^{(y0)}\tau_y ({a^\dagger}^2 - a^2).
\end{split}
\end{align}
This term generates molecular transitions with the absorption or emission of two resonator photons. Its magnitude is of order
\begin{align}
 \|V_{\textrm{tr}}^{m,2\textrm{PH}}\| \approx \textrm{max}\left(v_I^{(x0)},v_Q^{(y0)}\right)\frac{\left<n\right>}{4}.
\end{align}
For a given spin state, the detuning from resonance of an absorbing transition between a state with $n$ photons and $n-2$ photons is approximately (accounting for the frequency shifts induced by $V_{\textrm{dis}}$)
\begin{align}
\begin{split}
 &|\Delta_{\textrm{tr}}^{m,2\textrm{PH}}(\sigma_z,n)| \approx \\
 &\left| 2\omega_r - 2\chi_s\sigma_z \mp \left[E_m -\chi_0 \sigma_z - \chi_m \left(2n-1\right)\right] \right|.
\end{split}
\end{align}
Here the sign $\mp$ correspond to the case where the state $\ket{\pm}$ has higher energy than the state $\ket{\mp}$. For $\left<n\right> \gg 1$, we may set $n\approx \left<n\right>$. We then neglect the transitions provided that the amplitude of the induced Rabi oscillations is smaller than $\sim 10^{-1}$:
\begin{align}
 \textrm{max}_{\sigma_z} \left(\frac{\|V_{\textrm{tr}}^{m,2\textrm{PH}}\|}{|\Delta_{\textrm{tr}}^{m,2\textrm{PH}}(\sigma_z,\left<n\right>)|} \right)^2 < 10^{-1}.
\end{align}
We perform a similar procedure for all contributions to $V_{\textrm{tr}}$. The regions of parameter space where the amplitude of the oscillations become larger than $\sim 10^{-1}$ are indicated by the black area in Fig.~\ref{fig:fig5} (region 2). For the particular parameters of Fig.~\ref{fig:fig5}, the only process in $V_{\textrm{tr}}$ that can cause transitions within the dispersive limit is the two-photon molecular transition discussed above.

\bibliography{bibliography}

\begin{thebibliography}{103}%
\makeatletter
\providecommand \@ifxundefined [1]{%
 \@ifx{#1\undefined}
}%
\providecommand \@ifnum [1]{%
 \ifnum #1\expandafter \@firstoftwo
 \else \expandafter \@secondoftwo
 \fi
}%
\providecommand \@ifx [1]{%
 \ifx #1\expandafter \@firstoftwo
 \else \expandafter \@secondoftwo
 \fi
}%
\providecommand \natexlab [1]{#1}%
\providecommand \enquote  [1]{``#1''}%
\providecommand \bibnamefont  [1]{#1}%
\providecommand \bibfnamefont [1]{#1}%
\providecommand \citenamefont [1]{#1}%
\providecommand \href@noop [0]{\@secondoftwo}%
\providecommand \href [0]{\begingroup \@sanitize@url \@href}%
\providecommand \@href[1]{\@@startlink{#1}\@@href}%
\providecommand \@@href[1]{\endgroup#1\@@endlink}%
\providecommand \@sanitize@url [0]{\catcode `\\12\catcode `\$12\catcode
  `\&12\catcode `\#12\catcode `\^12\catcode `\_12\catcode `\%12\relax}%
\providecommand \@@startlink[1]{}%
\providecommand \@@endlink[0]{}%
\providecommand \url  [0]{\begingroup\@sanitize@url \@url }%
\providecommand \@url [1]{\endgroup\@href {#1}{\urlprefix }}%
\providecommand \urlprefix  [0]{URL }%
\providecommand \Eprint [0]{\href }%
\providecommand \doibase [0]{http://dx.doi.org/}%
\providecommand \selectlanguage [0]{\@gobble}%
\providecommand \bibinfo  [0]{\@secondoftwo}%
\providecommand \bibfield  [0]{\@secondoftwo}%
\providecommand \translation [1]{[#1]}%
\providecommand \BibitemOpen [0]{}%
\providecommand \bibitemStop [0]{}%
\providecommand \bibitemNoStop [0]{.\EOS\space}%
\providecommand \EOS [0]{\spacefactor3000\relax}%
\providecommand \BibitemShut  [1]{\csname bibitem#1\endcsname}%
\let\auto@bib@innerbib\@empty
\bibitem [{\citenamefont {Loss}\ and\ \citenamefont
  {DiVincenzo}(1998)}]{loss1998}%
  \BibitemOpen
  \bibfield  {author} {\bibinfo {author} {\bibfnamefont {D.}~\bibnamefont
  {Loss}}\ and\ \bibinfo {author} {\bibfnamefont {D.~P.}\ \bibnamefont
  {DiVincenzo}},\ }\href {\doibase 10.1103/PhysRevA.57.120} {\bibfield
  {journal} {\bibinfo  {journal} {Phys. Rev. A}\ }\textbf {\bibinfo {volume}
  {57}},\ \bibinfo {pages} {120} (\bibinfo {year} {1998})}\BibitemShut
  {NoStop}%
\bibitem [{\citenamefont {Kane}(1998)}]{kane1998}%
  \BibitemOpen
  \bibfield  {author} {\bibinfo {author} {\bibfnamefont {B.~E.}\ \bibnamefont
  {Kane}},\ }\href {\doibase 10.1038/30156} {\bibfield  {journal} {\bibinfo
  {journal} {Nature}\ }\textbf {\bibinfo {volume} {393}},\ \bibinfo {pages}
  {133} (\bibinfo {year} {1998})}\BibitemShut {NoStop}%
\bibitem [{\citenamefont {Tyryshkin}\ \emph {et~al.}(2012)\citenamefont
  {Tyryshkin}, \citenamefont {Tojo}, \citenamefont {Morton}, \citenamefont
  {Riemann}, \citenamefont {Abrosimov}, \citenamefont {Becker}, \citenamefont
  {Pohl}, \citenamefont {Schenkel}, \citenamefont {Thewalt}, \citenamefont
  {Itoh},\ and\ \citenamefont {Lyon}}]{tyryshkin2012}%
  \BibitemOpen
  \bibfield  {author} {\bibinfo {author} {\bibfnamefont {A.~M.}\ \bibnamefont
  {Tyryshkin}}, \bibinfo {author} {\bibfnamefont {S.}~\bibnamefont {Tojo}},
  \bibinfo {author} {\bibfnamefont {J.~J.~L.}\ \bibnamefont {Morton}}, \bibinfo
  {author} {\bibfnamefont {H.}~\bibnamefont {Riemann}}, \bibinfo {author}
  {\bibfnamefont {N.~V.}\ \bibnamefont {Abrosimov}}, \bibinfo {author}
  {\bibfnamefont {P.}~\bibnamefont {Becker}}, \bibinfo {author} {\bibfnamefont
  {H.-J.}\ \bibnamefont {Pohl}}, \bibinfo {author} {\bibfnamefont
  {T.}~\bibnamefont {Schenkel}}, \bibinfo {author} {\bibfnamefont {M.~L.~W.}\
  \bibnamefont {Thewalt}}, \bibinfo {author} {\bibfnamefont {K.~M.}\
  \bibnamefont {Itoh}}, \ and\ \bibinfo {author} {\bibfnamefont {S.~A.}\
  \bibnamefont {Lyon}},\ }\href {\doibase 10.1038/NMAT3182} {\bibfield
  {journal} {\bibinfo  {journal} {Nat. Mater.}\ }\textbf {\bibinfo {volume}
  {11}},\ \bibinfo {pages} {143} (\bibinfo {year} {2012})}\BibitemShut
  {NoStop}%
\bibitem [{\citenamefont {Saeedi}\ \emph {et~al.}(2013)\citenamefont {Saeedi},
  \citenamefont {Simmons}, \citenamefont {Salvail}, \citenamefont {Dluhy},
  \citenamefont {Riemann}, \citenamefont {Abrosimov}, \citenamefont {Becker},
  \citenamefont {Pohl}, \citenamefont {Morton},\ and\ \citenamefont
  {Thewalt}}]{saeedi2013}%
  \BibitemOpen
  \bibfield  {author} {\bibinfo {author} {\bibfnamefont {K.}~\bibnamefont
  {Saeedi}}, \bibinfo {author} {\bibfnamefont {S.}~\bibnamefont {Simmons}},
  \bibinfo {author} {\bibfnamefont {J.~Z.}\ \bibnamefont {Salvail}}, \bibinfo
  {author} {\bibfnamefont {P.}~\bibnamefont {Dluhy}}, \bibinfo {author}
  {\bibfnamefont {H.}~\bibnamefont {Riemann}}, \bibinfo {author} {\bibfnamefont
  {N.~V.}\ \bibnamefont {Abrosimov}}, \bibinfo {author} {\bibfnamefont
  {P.}~\bibnamefont {Becker}}, \bibinfo {author} {\bibfnamefont {H.-J.}\
  \bibnamefont {Pohl}}, \bibinfo {author} {\bibfnamefont {J.~J.~L.}\
  \bibnamefont {Morton}}, \ and\ \bibinfo {author} {\bibfnamefont {M.~L.~W.}\
  \bibnamefont {Thewalt}},\ }\href {\doibase 10.1126/science.1239584}
  {\bibfield  {journal} {\bibinfo  {journal} {Science}\ }\textbf {\bibinfo
  {volume} {342}},\ \bibinfo {pages} {830} (\bibinfo {year}
  {2013})}\BibitemShut {NoStop}%
\bibitem [{\citenamefont {Bar-Gill}\ \emph {et~al.}(2013)\citenamefont
  {Bar-Gill}, \citenamefont {Pham}, \citenamefont {Jarmola}, \citenamefont
  {Budker},\ and\ \citenamefont {Walsworth}}]{bargill2013}%
  \BibitemOpen
  \bibfield  {author} {\bibinfo {author} {\bibfnamefont {N.}~\bibnamefont
  {Bar-Gill}}, \bibinfo {author} {\bibfnamefont {L.~M.}\ \bibnamefont {Pham}},
  \bibinfo {author} {\bibfnamefont {A.}~\bibnamefont {Jarmola}}, \bibinfo
  {author} {\bibfnamefont {D.}~\bibnamefont {Budker}}, \ and\ \bibinfo {author}
  {\bibfnamefont {R.~L.}\ \bibnamefont {Walsworth}},\ }\href {\doibase
  10.1038/ncomms2771} {\bibfield  {journal} {\bibinfo  {journal} {Nat.
  Commun.}\ }\textbf {\bibinfo {volume} {4}},\ \bibinfo {pages} {1743}
  (\bibinfo {year} {2013})}\BibitemShut {NoStop}%
\bibitem [{\citenamefont {Muhonen}\ \emph {et~al.}(2014)\citenamefont
  {Muhonen}, \citenamefont {Dehollain}, \citenamefont {Laucht}, \citenamefont
  {Hudson}, \citenamefont {Kalra}, \citenamefont {Sekiguchi}, \citenamefont
  {Itoh}, \citenamefont {Jamieson}, \citenamefont {McCallum}, \citenamefont
  {Dzurak},\ and\ \citenamefont {Morello}}]{muhonen2014}%
  \BibitemOpen
  \bibfield  {author} {\bibinfo {author} {\bibfnamefont {J.~T.}\ \bibnamefont
  {Muhonen}}, \bibinfo {author} {\bibfnamefont {J.~P.}\ \bibnamefont
  {Dehollain}}, \bibinfo {author} {\bibfnamefont {A.}~\bibnamefont {Laucht}},
  \bibinfo {author} {\bibfnamefont {F.~E.}\ \bibnamefont {Hudson}}, \bibinfo
  {author} {\bibfnamefont {R.}~\bibnamefont {Kalra}}, \bibinfo {author}
  {\bibfnamefont {T.}~\bibnamefont {Sekiguchi}}, \bibinfo {author}
  {\bibfnamefont {K.~M.}\ \bibnamefont {Itoh}}, \bibinfo {author}
  {\bibfnamefont {D.~N.}\ \bibnamefont {Jamieson}}, \bibinfo {author}
  {\bibfnamefont {J.~C.}\ \bibnamefont {McCallum}}, \bibinfo {author}
  {\bibfnamefont {A.~S.}\ \bibnamefont {Dzurak}}, \ and\ \bibinfo {author}
  {\bibfnamefont {A.}~\bibnamefont {Morello}},\ }\href {\doibase
  10.1038/NNANO.2014.211} {\bibfield  {journal} {\bibinfo  {journal} {Nat.
  Nanotech.}\ }\textbf {\bibinfo {volume} {9}},\ \bibinfo {pages} {986}
  (\bibinfo {year} {2014})}\BibitemShut {NoStop}%
\bibitem [{\citenamefont {Watson}\ \emph {et~al.}(2017)\citenamefont {Watson},
  \citenamefont {Weber}, \citenamefont {Hsueh}, \citenamefont {Hollenberg},
  \citenamefont {Rahman},\ and\ \citenamefont {Simmons}}]{watson2017-2}%
  \BibitemOpen
  \bibfield  {author} {\bibinfo {author} {\bibfnamefont {T.~F.}\ \bibnamefont
  {Watson}}, \bibinfo {author} {\bibfnamefont {B.}~\bibnamefont {Weber}},
  \bibinfo {author} {\bibfnamefont {Y.-L.}\ \bibnamefont {Hsueh}}, \bibinfo
  {author} {\bibfnamefont {L.~C.~L.}\ \bibnamefont {Hollenberg}}, \bibinfo
  {author} {\bibfnamefont {R.}~\bibnamefont {Rahman}}, \ and\ \bibinfo {author}
  {\bibfnamefont {M.~Y.}\ \bibnamefont {Simmons}},\ }\href {\doibase
  10.1126/sciadv.1602811} {\bibfield  {journal} {\bibinfo  {journal} {Sci.
  Adv.}\ }\textbf {\bibinfo {volume} {3}},\ \bibinfo {pages} {e1602811}
  (\bibinfo {year} {2017})}\BibitemShut {NoStop}%
\bibitem [{\citenamefont {Bradley}\ \emph {et~al.}(2019)\citenamefont
  {Bradley}, \citenamefont {Randall}, \citenamefont {Abobeih}, \citenamefont
  {Berrevoets}, \citenamefont {Degen}, \citenamefont {Bakker}, \citenamefont
  {Markham}, \citenamefont {Twitchen},\ and\ \citenamefont
  {Taminiau}}]{bradley2019}%
  \BibitemOpen
  \bibfield  {author} {\bibinfo {author} {\bibfnamefont {C.~E.}\ \bibnamefont
  {Bradley}}, \bibinfo {author} {\bibfnamefont {J.}~\bibnamefont {Randall}},
  \bibinfo {author} {\bibfnamefont {M.~H.}\ \bibnamefont {Abobeih}}, \bibinfo
  {author} {\bibfnamefont {R.~C.}\ \bibnamefont {Berrevoets}}, \bibinfo
  {author} {\bibfnamefont {M.~J.}\ \bibnamefont {Degen}}, \bibinfo {author}
  {\bibfnamefont {M.~A.}\ \bibnamefont {Bakker}}, \bibinfo {author}
  {\bibfnamefont {M.}~\bibnamefont {Markham}}, \bibinfo {author} {\bibfnamefont
  {D.~J.}\ \bibnamefont {Twitchen}}, \ and\ \bibinfo {author} {\bibfnamefont
  {T.~H.}\ \bibnamefont {Taminiau}},\ }\href {\doibase
  10.1103/PhysRevX.9.031045} {\bibfield  {journal} {\bibinfo  {journal} {Phys.
  Rev. X}\ }\textbf {\bibinfo {volume} {9}},\ \bibinfo {pages} {031045}
  (\bibinfo {year} {2019})}\BibitemShut {NoStop}%
\bibitem [{\citenamefont {Elzerman}\ \emph {et~al.}(2004)\citenamefont
  {Elzerman}, \citenamefont {Hanson}, \citenamefont {{Van Beveren}},
  \citenamefont {Witkamp}, \citenamefont {Vandersypen},\ and\ \citenamefont
  {Kouwenhoven}}]{elzerman2004}%
  \BibitemOpen
  \bibfield  {author} {\bibinfo {author} {\bibfnamefont {J.~M.}\ \bibnamefont
  {Elzerman}}, \bibinfo {author} {\bibfnamefont {R.}~\bibnamefont {Hanson}},
  \bibinfo {author} {\bibfnamefont {L.~H.~W.}\ \bibnamefont {{Van Beveren}}},
  \bibinfo {author} {\bibfnamefont {B.}~\bibnamefont {Witkamp}}, \bibinfo
  {author} {\bibfnamefont {L.~M.~K.}\ \bibnamefont {Vandersypen}}, \ and\
  \bibinfo {author} {\bibfnamefont {L.~P.}\ \bibnamefont {Kouwenhoven}},\
  }\href {\doibase 10.1038/nature02693} {\bibfield  {journal} {\bibinfo
  {journal} {Nature}\ }\textbf {\bibinfo {volume} {430}},\ \bibinfo {pages}
  {431} (\bibinfo {year} {2004})}\BibitemShut {NoStop}%
\bibitem [{\citenamefont {Barthel}\ \emph {et~al.}(2009)\citenamefont
  {Barthel}, \citenamefont {Reilly}, \citenamefont {Marcus}, \citenamefont
  {Hanson},\ and\ \citenamefont {Gossard}}]{barthel2009}%
  \BibitemOpen
  \bibfield  {author} {\bibinfo {author} {\bibfnamefont {C.}~\bibnamefont
  {Barthel}}, \bibinfo {author} {\bibfnamefont {D.~J.}\ \bibnamefont {Reilly}},
  \bibinfo {author} {\bibfnamefont {C.~M.}\ \bibnamefont {Marcus}}, \bibinfo
  {author} {\bibfnamefont {M.~P.}\ \bibnamefont {Hanson}}, \ and\ \bibinfo
  {author} {\bibfnamefont {A.~C.}\ \bibnamefont {Gossard}},\ }\href {\doibase
  10.1103/PhysRevLett.103.160503} {\bibfield  {journal} {\bibinfo  {journal}
  {Phys. Rev. Lett.}\ }\textbf {\bibinfo {volume} {103}},\ \bibinfo {pages}
  {160503} (\bibinfo {year} {2009})}\BibitemShut {NoStop}%
\bibitem [{\citenamefont {Jiang}\ \emph {et~al.}(2009)\citenamefont {Jiang},
  \citenamefont {Hodges}, \citenamefont {Maze}, \citenamefont {Maurer},
  \citenamefont {Taylor}, \citenamefont {Cory}, \citenamefont {Hemmer},
  \citenamefont {Walsworth}, \citenamefont {Yacoby}, \citenamefont {Zibrov},\
  and\ \citenamefont {Lukin}}]{jiang2009}%
  \BibitemOpen
  \bibfield  {author} {\bibinfo {author} {\bibfnamefont {L.}~\bibnamefont
  {Jiang}}, \bibinfo {author} {\bibfnamefont {J.~S.}\ \bibnamefont {Hodges}},
  \bibinfo {author} {\bibfnamefont {J.~R.}\ \bibnamefont {Maze}}, \bibinfo
  {author} {\bibfnamefont {P.}~\bibnamefont {Maurer}}, \bibinfo {author}
  {\bibfnamefont {J.~M.}\ \bibnamefont {Taylor}}, \bibinfo {author}
  {\bibfnamefont {D.~G.}\ \bibnamefont {Cory}}, \bibinfo {author}
  {\bibfnamefont {P.~R.}\ \bibnamefont {Hemmer}}, \bibinfo {author}
  {\bibfnamefont {R.~L.}\ \bibnamefont {Walsworth}}, \bibinfo {author}
  {\bibfnamefont {A.}~\bibnamefont {Yacoby}}, \bibinfo {author} {\bibfnamefont
  {A.~S.}\ \bibnamefont {Zibrov}}, \ and\ \bibinfo {author} {\bibfnamefont
  {M.~D.}\ \bibnamefont {Lukin}},\ }\href {\doibase 10.1126/science.1176496}
  {\bibfield  {journal} {\bibinfo  {journal} {Science}\ }\textbf {\bibinfo
  {volume} {326}},\ \bibinfo {pages} {267} (\bibinfo {year}
  {2009})}\BibitemShut {NoStop}%
\bibitem [{\citenamefont {Morello}\ \emph {et~al.}(2010)\citenamefont
  {Morello}, \citenamefont {Pla}, \citenamefont {Zwanenburg}, \citenamefont
  {Chan}, \citenamefont {Tan}, \citenamefont {Huebl}, \citenamefont
  {M{\"o}tt{\"o}nen}, \citenamefont {Nugroho}, \citenamefont {Yang},
  \citenamefont {van Donkelaar}, \citenamefont {Alves}, \citenamefont
  {Jamieson}, \citenamefont {Escott}, \citenamefont {Hollenberg}, \citenamefont
  {Clark},\ and\ \citenamefont {Dzurak}}]{morello2010}%
  \BibitemOpen
  \bibfield  {author} {\bibinfo {author} {\bibfnamefont {A.}~\bibnamefont
  {Morello}}, \bibinfo {author} {\bibfnamefont {J.~J.}\ \bibnamefont {Pla}},
  \bibinfo {author} {\bibfnamefont {F.~A.}\ \bibnamefont {Zwanenburg}},
  \bibinfo {author} {\bibfnamefont {K.~W.}\ \bibnamefont {Chan}}, \bibinfo
  {author} {\bibfnamefont {K.~Y.}\ \bibnamefont {Tan}}, \bibinfo {author}
  {\bibfnamefont {H.}~\bibnamefont {Huebl}}, \bibinfo {author} {\bibfnamefont
  {M.}~\bibnamefont {M{\"o}tt{\"o}nen}}, \bibinfo {author} {\bibfnamefont
  {C.~D.}\ \bibnamefont {Nugroho}}, \bibinfo {author} {\bibfnamefont
  {C.}~\bibnamefont {Yang}}, \bibinfo {author} {\bibfnamefont {J.~A.}\
  \bibnamefont {van Donkelaar}}, \bibinfo {author} {\bibfnamefont {A.~D.~C.}\
  \bibnamefont {Alves}}, \bibinfo {author} {\bibfnamefont {D.~N.}\ \bibnamefont
  {Jamieson}}, \bibinfo {author} {\bibfnamefont {C.~C.}\ \bibnamefont
  {Escott}}, \bibinfo {author} {\bibfnamefont {L.~C.~L.}\ \bibnamefont
  {Hollenberg}}, \bibinfo {author} {\bibfnamefont {R.~G.}\ \bibnamefont
  {Clark}}, \ and\ \bibinfo {author} {\bibfnamefont {A.~S.}\ \bibnamefont
  {Dzurak}},\ }\href {\doibase 10.1038/nature09392} {\bibfield  {journal}
  {\bibinfo  {journal} {Nature}\ }\textbf {\bibinfo {volume} {467}},\ \bibinfo
  {pages} {687} (\bibinfo {year} {2010})}\BibitemShut {NoStop}%
\bibitem [{\citenamefont {Robledo}\ \emph {et~al.}(2011)\citenamefont
  {Robledo}, \citenamefont {Childress}, \citenamefont {Bernien}, \citenamefont
  {Hensen}, \citenamefont {Alkemade},\ and\ \citenamefont
  {Hanson}}]{robledo2011}%
  \BibitemOpen
  \bibfield  {author} {\bibinfo {author} {\bibfnamefont {L.}~\bibnamefont
  {Robledo}}, \bibinfo {author} {\bibfnamefont {L.}~\bibnamefont {Childress}},
  \bibinfo {author} {\bibfnamefont {H.}~\bibnamefont {Bernien}}, \bibinfo
  {author} {\bibfnamefont {B.}~\bibnamefont {Hensen}}, \bibinfo {author}
  {\bibfnamefont {P.~F.}\ \bibnamefont {Alkemade}}, \ and\ \bibinfo {author}
  {\bibfnamefont {R.}~\bibnamefont {Hanson}},\ }\href {\doibase
  10.1038/nature10401} {\bibfield  {journal} {\bibinfo  {journal} {Nature}\
  }\textbf {\bibinfo {volume} {477}},\ \bibinfo {pages} {574} (\bibinfo {year}
  {2011})}\BibitemShut {NoStop}%
\bibitem [{\citenamefont {Pla}\ \emph {et~al.}(2013)\citenamefont {Pla},
  \citenamefont {Tan}, \citenamefont {Dehollain}, \citenamefont {Lim},
  \citenamefont {Morton}, \citenamefont {Zwanenburg}, \citenamefont {Jamieson},
  \citenamefont {Dzurak},\ and\ \citenamefont {Morello}}]{pla2013}%
  \BibitemOpen
  \bibfield  {author} {\bibinfo {author} {\bibfnamefont {J.~J.}\ \bibnamefont
  {Pla}}, \bibinfo {author} {\bibfnamefont {K.~Y.}\ \bibnamefont {Tan}},
  \bibinfo {author} {\bibfnamefont {J.~P.}\ \bibnamefont {Dehollain}}, \bibinfo
  {author} {\bibfnamefont {W.~H.}\ \bibnamefont {Lim}}, \bibinfo {author}
  {\bibfnamefont {J.~J.}\ \bibnamefont {Morton}}, \bibinfo {author}
  {\bibfnamefont {F.~A.}\ \bibnamefont {Zwanenburg}}, \bibinfo {author}
  {\bibfnamefont {D.~N.}\ \bibnamefont {Jamieson}}, \bibinfo {author}
  {\bibfnamefont {A.~S.}\ \bibnamefont {Dzurak}}, \ and\ \bibinfo {author}
  {\bibfnamefont {A.}~\bibnamefont {Morello}},\ }\href {\doibase
  10.1038/nature12011} {\bibfield  {journal} {\bibinfo  {journal} {Nature}\
  }\textbf {\bibinfo {volume} {496}},\ \bibinfo {pages} {334} (\bibinfo {year}
  {2013})}\BibitemShut {NoStop}%
\bibitem [{\citenamefont {Nakajima}\ \emph {et~al.}(2017)\citenamefont
  {Nakajima}, \citenamefont {Delbecq}, \citenamefont {Otsuka}, \citenamefont
  {Stano}, \citenamefont {Amaha}, \citenamefont {Yoneda}, \citenamefont
  {Noiri}, \citenamefont {Kawasaki}, \citenamefont {Takeda}, \citenamefont
  {Allison}, \citenamefont {Ludwig}, \citenamefont {Wieck}, \citenamefont
  {Loss},\ and\ \citenamefont {Tarucha}}]{nakajima2017}%
  \BibitemOpen
  \bibfield  {author} {\bibinfo {author} {\bibfnamefont {T.}~\bibnamefont
  {Nakajima}}, \bibinfo {author} {\bibfnamefont {M.~R.}\ \bibnamefont
  {Delbecq}}, \bibinfo {author} {\bibfnamefont {T.}~\bibnamefont {Otsuka}},
  \bibinfo {author} {\bibfnamefont {P.}~\bibnamefont {Stano}}, \bibinfo
  {author} {\bibfnamefont {S.}~\bibnamefont {Amaha}}, \bibinfo {author}
  {\bibfnamefont {J.}~\bibnamefont {Yoneda}}, \bibinfo {author} {\bibfnamefont
  {A.}~\bibnamefont {Noiri}}, \bibinfo {author} {\bibfnamefont
  {K.}~\bibnamefont {Kawasaki}}, \bibinfo {author} {\bibfnamefont
  {K.}~\bibnamefont {Takeda}}, \bibinfo {author} {\bibfnamefont
  {G.}~\bibnamefont {Allison}}, \bibinfo {author} {\bibfnamefont
  {A.}~\bibnamefont {Ludwig}}, \bibinfo {author} {\bibfnamefont {A.~D.}\
  \bibnamefont {Wieck}}, \bibinfo {author} {\bibfnamefont {D.}~\bibnamefont
  {Loss}}, \ and\ \bibinfo {author} {\bibfnamefont {S.}~\bibnamefont
  {Tarucha}},\ }\href {\doibase 10.1103/PhysRevLett.119.017701} {\bibfield
  {journal} {\bibinfo  {journal} {Phys. Rev. Lett.}\ }\textbf {\bibinfo
  {volume} {119}},\ \bibinfo {pages} {017701} (\bibinfo {year}
  {2017})}\BibitemShut {NoStop}%
\bibitem [{\citenamefont {Pakkiam}\ \emph {et~al.}(2018)\citenamefont
  {Pakkiam}, \citenamefont {Timofeev}, \citenamefont {House}, \citenamefont
  {Hogg}, \citenamefont {Kobayashi}, \citenamefont {Koch}, \citenamefont
  {Rogge},\ and\ \citenamefont {Simmons}}]{pakkiam2018}%
  \BibitemOpen
  \bibfield  {author} {\bibinfo {author} {\bibfnamefont {P.}~\bibnamefont
  {Pakkiam}}, \bibinfo {author} {\bibfnamefont {A.~V.}\ \bibnamefont
  {Timofeev}}, \bibinfo {author} {\bibfnamefont {M.~G.}\ \bibnamefont {House}},
  \bibinfo {author} {\bibfnamefont {M.~R.}\ \bibnamefont {Hogg}}, \bibinfo
  {author} {\bibfnamefont {T.}~\bibnamefont {Kobayashi}}, \bibinfo {author}
  {\bibfnamefont {M.}~\bibnamefont {Koch}}, \bibinfo {author} {\bibfnamefont
  {S.}~\bibnamefont {Rogge}}, \ and\ \bibinfo {author} {\bibfnamefont {M.~Y.}\
  \bibnamefont {Simmons}},\ }\href {\doibase 10.1103/PhysRevX.8.041032}
  {\bibfield  {journal} {\bibinfo  {journal} {Phys. Rev. X}\ }\textbf {\bibinfo
  {volume} {8}},\ \bibinfo {pages} {041032} (\bibinfo {year}
  {2018})}\BibitemShut {NoStop}%
\bibitem [{\citenamefont {West}\ \emph {et~al.}(2019)\citenamefont {West},
  \citenamefont {Hensen}, \citenamefont {Jouan}, \citenamefont {Tanttu},
  \citenamefont {Yang}, \citenamefont {Rossi}, \citenamefont {Gonzalez-Zalba},
  \citenamefont {Hudson}, \citenamefont {Morello}, \citenamefont {Reilly},\
  and\ \citenamefont {Dzurak}}]{west2019}%
  \BibitemOpen
  \bibfield  {author} {\bibinfo {author} {\bibfnamefont {A.}~\bibnamefont
  {West}}, \bibinfo {author} {\bibfnamefont {B.}~\bibnamefont {Hensen}},
  \bibinfo {author} {\bibfnamefont {A.}~\bibnamefont {Jouan}}, \bibinfo
  {author} {\bibfnamefont {T.}~\bibnamefont {Tanttu}}, \bibinfo {author}
  {\bibfnamefont {C.-H.}\ \bibnamefont {Yang}}, \bibinfo {author}
  {\bibfnamefont {A.}~\bibnamefont {Rossi}}, \bibinfo {author} {\bibfnamefont
  {M.~F.}\ \bibnamefont {Gonzalez-Zalba}}, \bibinfo {author} {\bibfnamefont
  {F.}~\bibnamefont {Hudson}}, \bibinfo {author} {\bibfnamefont
  {A.}~\bibnamefont {Morello}}, \bibinfo {author} {\bibfnamefont {D.~J.}\
  \bibnamefont {Reilly}}, \ and\ \bibinfo {author} {\bibfnamefont {A.~S.}\
  \bibnamefont {Dzurak}},\ }\href {\doibase 10.1038/s41565-019-0400-7}
  {\bibfield  {journal} {\bibinfo  {journal} {Nat. Nanotech.}\ }\textbf
  {\bibinfo {volume} {14}},\ \bibinfo {pages} {437} (\bibinfo {year}
  {2019})}\BibitemShut {NoStop}%
\bibitem [{\citenamefont {Crippa}\ \emph {et~al.}(2019)\citenamefont {Crippa},
  \citenamefont {Ezzouch}, \citenamefont {Apr{\'a}}, \citenamefont {Amisse},
  \citenamefont {Hutin}, \citenamefont {Bertrand}, \citenamefont {Vinet},
  \citenamefont {Urdampilleta}, \citenamefont {Meunier}, \citenamefont
  {Sanquer}, \citenamefont {Jehl}, \citenamefont {Maurand},\ and\ \citenamefont
  {De~Franceschi}}]{crippa2019}%
  \BibitemOpen
  \bibfield  {author} {\bibinfo {author} {\bibfnamefont {A.}~\bibnamefont
  {Crippa}}, \bibinfo {author} {\bibfnamefont {R.}~\bibnamefont {Ezzouch}},
  \bibinfo {author} {\bibfnamefont {A.}~\bibnamefont {Apr{\'a}}}, \bibinfo
  {author} {\bibfnamefont {A.}~\bibnamefont {Amisse}}, \bibinfo {author}
  {\bibfnamefont {L.}~\bibnamefont {Hutin}}, \bibinfo {author} {\bibfnamefont
  {B.}~\bibnamefont {Bertrand}}, \bibinfo {author} {\bibfnamefont
  {M.}~\bibnamefont {Vinet}}, \bibinfo {author} {\bibfnamefont
  {M.}~\bibnamefont {Urdampilleta}}, \bibinfo {author} {\bibfnamefont
  {T.}~\bibnamefont {Meunier}}, \bibinfo {author} {\bibfnamefont
  {M.}~\bibnamefont {Sanquer}}, \bibinfo {author} {\bibfnamefont
  {X.}~\bibnamefont {Jehl}}, \bibinfo {author} {\bibfnamefont {R.}~\bibnamefont
  {Maurand}}, \ and\ \bibinfo {author} {\bibfnamefont {S.}~\bibnamefont
  {De~Franceschi}},\ }\href {\doibase 10.1038/s41467-019-10848-z} {\bibfield
  {journal} {\bibinfo  {journal} {Nat. Commun.}\ }\textbf {\bibinfo {volume}
  {10}},\ \bibinfo {pages} {2776} (\bibinfo {year} {2019})}\BibitemShut
  {NoStop}%
\bibitem [{\citenamefont {Urdampilleta}\ \emph {et~al.}(2019)\citenamefont
  {Urdampilleta}, \citenamefont {Chanrion}, \citenamefont {Jadot},
  \citenamefont {Spence}, \citenamefont {Mortemousque}, \citenamefont {Hutin},
  \citenamefont {Bertrand}, \citenamefont {Barraud}, \citenamefont {Maurand},
  \citenamefont {Sanquer}, \citenamefont {Jehl}, \citenamefont {De~Franceschi},
  \citenamefont {Vinet},\ and\ \citenamefont {Meunier}}]{urdampilleta2019}%
  \BibitemOpen
  \bibfield  {author} {\bibinfo {author} {\bibfnamefont {D.~J.}\ \bibnamefont
  {Urdampilleta}, \bibfnamefont {Matias und~Niegemann}}, \bibinfo {author}
  {\bibfnamefont {E.}~\bibnamefont {Chanrion}}, \bibinfo {author}
  {\bibfnamefont {B.}~\bibnamefont {Jadot}}, \bibinfo {author} {\bibfnamefont
  {C.}~\bibnamefont {Spence}}, \bibinfo {author} {\bibfnamefont
  {C.}~\bibnamefont {Mortemousque}, \bibfnamefont {Pierre-Andr{\'e}
  und~B{\"a}uerle}}, \bibinfo {author} {\bibfnamefont {L.}~\bibnamefont
  {Hutin}}, \bibinfo {author} {\bibfnamefont {B.}~\bibnamefont {Bertrand}},
  \bibinfo {author} {\bibfnamefont {S.}~\bibnamefont {Barraud}}, \bibinfo
  {author} {\bibfnamefont {R.}~\bibnamefont {Maurand}}, \bibinfo {author}
  {\bibfnamefont {M.}~\bibnamefont {Sanquer}}, \bibinfo {author} {\bibfnamefont
  {X.}~\bibnamefont {Jehl}}, \bibinfo {author} {\bibfnamefont {S.}~\bibnamefont
  {De~Franceschi}}, \bibinfo {author} {\bibfnamefont {M.}~\bibnamefont
  {Vinet}}, \ and\ \bibinfo {author} {\bibfnamefont {T.}~\bibnamefont
  {Meunier}},\ }\href {\doibase 10.1038/s41565-019-0443-9} {\bibfield
  {journal} {\bibinfo  {journal} {Nat. Nanotech.}\ }\textbf {\bibinfo {volume}
  {14}},\ \bibinfo {pages} {737} (\bibinfo {year} {2019})}\BibitemShut
  {NoStop}%
\bibitem [{\citenamefont {Harvey-Collard}\ \emph {et~al.}(2018)\citenamefont
  {Harvey-Collard}, \citenamefont {D'Anjou}, \citenamefont {Rudolph},
  \citenamefont {Jacobson}, \citenamefont {Dominguez}, \citenamefont
  {Ten~Eyck}, \citenamefont {Wendt}, \citenamefont {Pluym}, \citenamefont
  {Lilly}, \citenamefont {Coish}, \citenamefont {Pioro-Ladri{\`e}re},\ and\
  \citenamefont {Carroll}}]{harveycollard2018}%
  \BibitemOpen
  \bibfield  {author} {\bibinfo {author} {\bibfnamefont {P.}~\bibnamefont
  {Harvey-Collard}}, \bibinfo {author} {\bibfnamefont {B.}~\bibnamefont
  {D'Anjou}}, \bibinfo {author} {\bibfnamefont {M.}~\bibnamefont {Rudolph}},
  \bibinfo {author} {\bibfnamefont {N.~T.}\ \bibnamefont {Jacobson}}, \bibinfo
  {author} {\bibfnamefont {J.}~\bibnamefont {Dominguez}}, \bibinfo {author}
  {\bibfnamefont {G.~A.}\ \bibnamefont {Ten~Eyck}}, \bibinfo {author}
  {\bibfnamefont {J.~R.}\ \bibnamefont {Wendt}}, \bibinfo {author}
  {\bibfnamefont {T.}~\bibnamefont {Pluym}}, \bibinfo {author} {\bibfnamefont
  {M.~P.}\ \bibnamefont {Lilly}}, \bibinfo {author} {\bibfnamefont {W.~A.}\
  \bibnamefont {Coish}}, \bibinfo {author} {\bibfnamefont {M.}~\bibnamefont
  {Pioro-Ladri{\`e}re}}, \ and\ \bibinfo {author} {\bibfnamefont {M.~S.}\
  \bibnamefont {Carroll}},\ }\href {\doibase 10.1103/PhysRevX.8.021046}
  {\bibfield  {journal} {\bibinfo  {journal} {Phys. Rev. X}\ }\textbf {\bibinfo
  {volume} {8}},\ \bibinfo {pages} {021046} (\bibinfo {year}
  {2018})}\BibitemShut {NoStop}%
\bibitem [{\citenamefont {Zheng}\ \emph {et~al.}(2019)\citenamefont {Zheng},
  \citenamefont {Samkharadze}, \citenamefont {Noordam}, \citenamefont {Kalhor},
  \citenamefont {Brousse}, \citenamefont {Sammak}, \citenamefont {Scappucci},\
  and\ \citenamefont {Vandersypen}}]{zheng2019}%
  \BibitemOpen
  \bibfield  {author} {\bibinfo {author} {\bibfnamefont {G.}~\bibnamefont
  {Zheng}}, \bibinfo {author} {\bibfnamefont {N.}~\bibnamefont {Samkharadze}},
  \bibinfo {author} {\bibfnamefont {M.~L.}\ \bibnamefont {Noordam}}, \bibinfo
  {author} {\bibfnamefont {N.}~\bibnamefont {Kalhor}}, \bibinfo {author}
  {\bibfnamefont {D.}~\bibnamefont {Brousse}}, \bibinfo {author} {\bibfnamefont
  {A.}~\bibnamefont {Sammak}}, \bibinfo {author} {\bibfnamefont
  {G.}~\bibnamefont {Scappucci}}, \ and\ \bibinfo {author} {\bibfnamefont
  {L.~M.~K.}\ \bibnamefont {Vandersypen}},\ }\href {\doibase
  10.1038/s41565-019-0488-9} {\bibfield  {journal} {\bibinfo  {journal} {Nat.
  Nanotech.}\ }\textbf {\bibinfo {volume} {14}},\ \bibinfo {pages} {742}
  (\bibinfo {year} {2019})}\BibitemShut {NoStop}%
\bibitem [{\citenamefont {Nakajima}\ \emph {et~al.}(2019)\citenamefont
  {Nakajima}, \citenamefont {Noiri}, \citenamefont {Yoneda}, \citenamefont
  {Delbecq}, \citenamefont {Stano}, \citenamefont {Otsuka}, \citenamefont
  {Takeda}, \citenamefont {Amaha}, \citenamefont {Allison}, \citenamefont
  {Kawasaki}, \citenamefont {Ludwig}, \citenamefont {Wieck}, \citenamefont
  {Loss},\ and\ \citenamefont {Tarucha}}]{nakajima2019}%
  \BibitemOpen
  \bibfield  {author} {\bibinfo {author} {\bibfnamefont {T.}~\bibnamefont
  {Nakajima}}, \bibinfo {author} {\bibfnamefont {A.}~\bibnamefont {Noiri}},
  \bibinfo {author} {\bibfnamefont {J.}~\bibnamefont {Yoneda}}, \bibinfo
  {author} {\bibfnamefont {M.~R.}\ \bibnamefont {Delbecq}}, \bibinfo {author}
  {\bibfnamefont {P.}~\bibnamefont {Stano}}, \bibinfo {author} {\bibfnamefont
  {T.}~\bibnamefont {Otsuka}}, \bibinfo {author} {\bibfnamefont
  {K.}~\bibnamefont {Takeda}}, \bibinfo {author} {\bibfnamefont
  {S.}~\bibnamefont {Amaha}}, \bibinfo {author} {\bibfnamefont
  {G.}~\bibnamefont {Allison}}, \bibinfo {author} {\bibfnamefont
  {K.}~\bibnamefont {Kawasaki}}, \bibinfo {author} {\bibfnamefont
  {A.}~\bibnamefont {Ludwig}}, \bibinfo {author} {\bibfnamefont {A.~D.}\
  \bibnamefont {Wieck}}, \bibinfo {author} {\bibfnamefont {D.}~\bibnamefont
  {Loss}}, \ and\ \bibinfo {author} {\bibfnamefont {S.}~\bibnamefont
  {Tarucha}},\ }\href {\doibase 10.1038/s41565-019-0426-x} {\bibfield
  {journal} {\bibinfo  {journal} {Nat. Nanotech.}\ }\textbf {\bibinfo {volume}
  {14}},\ \bibinfo {pages} {555} (\bibinfo {year} {2019})}\BibitemShut
  {NoStop}%
\bibitem [{\citenamefont {Mi}\ \emph {et~al.}(2018)\citenamefont {Mi},
  \citenamefont {Benito}, \citenamefont {Putz}, \citenamefont {Zajac},
  \citenamefont {Taylor}, \citenamefont {Burkard},\ and\ \citenamefont
  {Petta}}]{mi2018}%
  \BibitemOpen
  \bibfield  {author} {\bibinfo {author} {\bibfnamefont {X.}~\bibnamefont
  {Mi}}, \bibinfo {author} {\bibfnamefont {M.}~\bibnamefont {Benito}}, \bibinfo
  {author} {\bibfnamefont {S.}~\bibnamefont {Putz}}, \bibinfo {author}
  {\bibfnamefont {D.~M.}\ \bibnamefont {Zajac}}, \bibinfo {author}
  {\bibfnamefont {J.~M.}\ \bibnamefont {Taylor}}, \bibinfo {author}
  {\bibfnamefont {G.}~\bibnamefont {Burkard}}, \ and\ \bibinfo {author}
  {\bibfnamefont {J.~R.}\ \bibnamefont {Petta}},\ }\href {\doibase
  10.1038/nature25769} {\bibfield  {journal} {\bibinfo  {journal} {Nature}\
  }\textbf {\bibinfo {volume} {555}},\ \bibinfo {pages} {599} (\bibinfo {year}
  {2018})}\BibitemShut {NoStop}%
\bibitem [{\citenamefont {Landig}\ \emph {et~al.}(2018)\citenamefont {Landig},
  \citenamefont {Koski}, \citenamefont {Scarlino}, \citenamefont {Mendes},
  \citenamefont {Blais}, \citenamefont {Reichl}, \citenamefont {Wegscheider},
  \citenamefont {Wallraff}, \citenamefont {Ensslin},\ and\ \citenamefont
  {Ihn}}]{landig2018}%
  \BibitemOpen
  \bibfield  {author} {\bibinfo {author} {\bibfnamefont {A.~J.}\ \bibnamefont
  {Landig}}, \bibinfo {author} {\bibfnamefont {J.~V.}\ \bibnamefont {Koski}},
  \bibinfo {author} {\bibfnamefont {P.}~\bibnamefont {Scarlino}}, \bibinfo
  {author} {\bibfnamefont {U.~C.}\ \bibnamefont {Mendes}}, \bibinfo {author}
  {\bibfnamefont {A.}~\bibnamefont {Blais}}, \bibinfo {author} {\bibfnamefont
  {C.}~\bibnamefont {Reichl}}, \bibinfo {author} {\bibfnamefont
  {W.}~\bibnamefont {Wegscheider}}, \bibinfo {author} {\bibfnamefont
  {A.}~\bibnamefont {Wallraff}}, \bibinfo {author} {\bibfnamefont
  {K.}~\bibnamefont {Ensslin}}, \ and\ \bibinfo {author} {\bibfnamefont
  {T.}~\bibnamefont {Ihn}},\ }\href {\doibase 10.1038/s41586-018-0365-y}
  {\bibfield  {journal} {\bibinfo  {journal} {Nature}\ }\textbf {\bibinfo
  {volume} {560}},\ \bibinfo {pages} {179} (\bibinfo {year}
  {2018})}\BibitemShut {NoStop}%
\bibitem [{\citenamefont {Samkharadze}\ \emph {et~al.}(2018)\citenamefont
  {Samkharadze}, \citenamefont {Zheng}, \citenamefont {Kalhor}, \citenamefont
  {Brousse}, \citenamefont {Sammak}, \citenamefont {Mendes}, \citenamefont
  {Blais}, \citenamefont {Scappucci},\ and\ \citenamefont
  {Vandersypen}}]{samkharadze2018}%
  \BibitemOpen
  \bibfield  {author} {\bibinfo {author} {\bibfnamefont {N.}~\bibnamefont
  {Samkharadze}}, \bibinfo {author} {\bibfnamefont {G.}~\bibnamefont {Zheng}},
  \bibinfo {author} {\bibfnamefont {N.}~\bibnamefont {Kalhor}}, \bibinfo
  {author} {\bibfnamefont {D.}~\bibnamefont {Brousse}}, \bibinfo {author}
  {\bibfnamefont {A.}~\bibnamefont {Sammak}}, \bibinfo {author} {\bibfnamefont
  {U.~C.}\ \bibnamefont {Mendes}}, \bibinfo {author} {\bibfnamefont
  {A.}~\bibnamefont {Blais}}, \bibinfo {author} {\bibfnamefont
  {G.}~\bibnamefont {Scappucci}}, \ and\ \bibinfo {author} {\bibfnamefont
  {L.~M.~K.}\ \bibnamefont {Vandersypen}},\ }\href {\doibase
  10.1126/science.aar4054} {\bibfield  {journal} {\bibinfo  {journal}
  {Science}\ }\textbf {\bibinfo {volume} {359}},\ \bibinfo {pages} {1123}
  (\bibinfo {year} {2018})}\BibitemShut {NoStop}%
\bibitem [{\citenamefont {Cubaynes}\ \emph {et~al.}(2019)\citenamefont
  {Cubaynes}, \citenamefont {Delbecq}, \citenamefont {Dartiailh}, \citenamefont
  {Assouly}, \citenamefont {Desjardins}, \citenamefont {Contamin},
  \citenamefont {Bruhat}, \citenamefont {Leghtas}, \citenamefont {Mallet},
  \citenamefont {Cottet},\ and\ \citenamefont {Kontos}}]{cubaynes2019}%
  \BibitemOpen
  \bibfield  {author} {\bibinfo {author} {\bibfnamefont {T.}~\bibnamefont
  {Cubaynes}}, \bibinfo {author} {\bibfnamefont {M.~R.}\ \bibnamefont
  {Delbecq}}, \bibinfo {author} {\bibfnamefont {M.~C.}\ \bibnamefont
  {Dartiailh}}, \bibinfo {author} {\bibfnamefont {R.}~\bibnamefont {Assouly}},
  \bibinfo {author} {\bibfnamefont {M.~M.}\ \bibnamefont {Desjardins}},
  \bibinfo {author} {\bibfnamefont {L.~C.}\ \bibnamefont {Contamin}}, \bibinfo
  {author} {\bibfnamefont {L.~E.}\ \bibnamefont {Bruhat}}, \bibinfo {author}
  {\bibfnamefont {Z.}~\bibnamefont {Leghtas}}, \bibinfo {author} {\bibfnamefont
  {F.}~\bibnamefont {Mallet}}, \bibinfo {author} {\bibfnamefont
  {A.}~\bibnamefont {Cottet}}, \ and\ \bibinfo {author} {\bibfnamefont
  {T.}~\bibnamefont {Kontos}},\ }\href {\doibase 10.1038/s41534-019-0169-4}
  {\bibfield  {journal} {\bibinfo  {journal} {npj Quantum Inf.}\ }\textbf
  {\bibinfo {volume} {5}},\ \bibinfo {pages} {47} (\bibinfo {year}
  {2019})}\BibitemShut {NoStop}%
\bibitem [{\citenamefont {Kubo}\ \emph {et~al.}(2010)\citenamefont {Kubo},
  \citenamefont {Ong}, \citenamefont {Bertet}, \citenamefont {Vion},
  \citenamefont {Jacques}, \citenamefont {Zheng}, \citenamefont {Dr{\'e}au},
  \citenamefont {Roch}, \citenamefont {Auffeves}, \citenamefont {Jelezko},
  \citenamefont {Wrachtrup}, \citenamefont {Barthe}, \citenamefont {Bergonzo},\
  and\ \citenamefont {Esteve}}]{kubo2010}%
  \BibitemOpen
  \bibfield  {author} {\bibinfo {author} {\bibfnamefont {Y.}~\bibnamefont
  {Kubo}}, \bibinfo {author} {\bibfnamefont {F.~R.}\ \bibnamefont {Ong}},
  \bibinfo {author} {\bibfnamefont {P.}~\bibnamefont {Bertet}}, \bibinfo
  {author} {\bibfnamefont {D.}~\bibnamefont {Vion}}, \bibinfo {author}
  {\bibfnamefont {V.}~\bibnamefont {Jacques}}, \bibinfo {author} {\bibfnamefont
  {D.}~\bibnamefont {Zheng}}, \bibinfo {author} {\bibfnamefont
  {A.}~\bibnamefont {Dr{\'e}au}}, \bibinfo {author} {\bibfnamefont {J.-F.}\
  \bibnamefont {Roch}}, \bibinfo {author} {\bibfnamefont {A.}~\bibnamefont
  {Auffeves}}, \bibinfo {author} {\bibfnamefont {F.}~\bibnamefont {Jelezko}},
  \bibinfo {author} {\bibfnamefont {J.}~\bibnamefont {Wrachtrup}}, \bibinfo
  {author} {\bibfnamefont {M.~F.}\ \bibnamefont {Barthe}}, \bibinfo {author}
  {\bibfnamefont {P.}~\bibnamefont {Bergonzo}}, \ and\ \bibinfo {author}
  {\bibfnamefont {D.}~\bibnamefont {Esteve}},\ }\href {\doibase
  10.1103/PhysRevLett.105.140502} {\bibfield  {journal} {\bibinfo  {journal}
  {Phys. Rev. Lett.}\ }\textbf {\bibinfo {volume} {105}},\ \bibinfo {pages}
  {140502} (\bibinfo {year} {2010})}\BibitemShut {NoStop}%
\bibitem [{\citenamefont {Hou}\ and\ \citenamefont {Liu}(2019)}]{hou2019}%
  \BibitemOpen
  \bibfield  {author} {\bibinfo {author} {\bibfnamefont {J.~T.}\ \bibnamefont
  {Hou}}\ and\ \bibinfo {author} {\bibfnamefont {L.}~\bibnamefont {Liu}},\
  }\href {https://journals.aps.org/prl/abstract/10.1103/PhysRevLett.123.107702}
  {\bibfield  {journal} {\bibinfo  {journal} {Phys. Rev. Lett.}\ }\textbf
  {\bibinfo {volume} {123}},\ \bibinfo {pages} {107702} (\bibinfo {year}
  {2019})}\BibitemShut {NoStop}%
\bibitem [{\citenamefont {Dold}\ \emph {et~al.}(2019)\citenamefont {Dold},
  \citenamefont {Zollitsch}, \citenamefont {O'Sullivan}, \citenamefont
  {Welinski}, \citenamefont {Ferrier}, \citenamefont {Goldner}, \citenamefont
  {de~Graaf}, \citenamefont {Lindstr{\"o}m},\ and\ \citenamefont
  {Morton}}]{dold2019}%
  \BibitemOpen
  \bibfield  {author} {\bibinfo {author} {\bibfnamefont {G.}~\bibnamefont
  {Dold}}, \bibinfo {author} {\bibfnamefont {C.~W.}\ \bibnamefont {Zollitsch}},
  \bibinfo {author} {\bibfnamefont {J.}~\bibnamefont {O'Sullivan}}, \bibinfo
  {author} {\bibfnamefont {S.}~\bibnamefont {Welinski}}, \bibinfo {author}
  {\bibfnamefont {A.}~\bibnamefont {Ferrier}}, \bibinfo {author} {\bibfnamefont
  {P.}~\bibnamefont {Goldner}}, \bibinfo {author} {\bibfnamefont {S.~E.}\
  \bibnamefont {de~Graaf}}, \bibinfo {author} {\bibfnamefont {T.}~\bibnamefont
  {Lindstr{\"o}m}}, \ and\ \bibinfo {author} {\bibfnamefont {J.~J.~L.}\
  \bibnamefont {Morton}},\ }\href {\doibase 0.1103/PhysRevApplied.11.054082}
  {\bibfield  {journal} {\bibinfo  {journal} {Phys. Rev. Appl.}\ }\textbf
  {\bibinfo {volume} {11}},\ \bibinfo {pages} {054082} (\bibinfo {year}
  {2019})}\BibitemShut {NoStop}%
\bibitem [{\citenamefont {Majer}\ \emph {et~al.}(2007)\citenamefont {Majer},
  \citenamefont {Chow}, \citenamefont {Gambetta}, \citenamefont {Koch},
  \citenamefont {Johnson}, \citenamefont {Schreier}, \citenamefont {Frunzio},
  \citenamefont {Schuster}, \citenamefont {Houck}, \citenamefont {Wallraff},
  \citenamefont {Blais}, \citenamefont {Devoret}, \citenamefont {Girvin},\ and\
  \citenamefont {Schoelkopf}}]{majer2007}%
  \BibitemOpen
  \bibfield  {author} {\bibinfo {author} {\bibfnamefont {J.}~\bibnamefont
  {Majer}}, \bibinfo {author} {\bibfnamefont {J.~M.}\ \bibnamefont {Chow}},
  \bibinfo {author} {\bibfnamefont {J.~M.}\ \bibnamefont {Gambetta}}, \bibinfo
  {author} {\bibfnamefont {J.}~\bibnamefont {Koch}}, \bibinfo {author}
  {\bibfnamefont {B.~R.}\ \bibnamefont {Johnson}}, \bibinfo {author}
  {\bibfnamefont {J.~A.}\ \bibnamefont {Schreier}}, \bibinfo {author}
  {\bibfnamefont {L.}~\bibnamefont {Frunzio}}, \bibinfo {author} {\bibfnamefont
  {D.~I.}\ \bibnamefont {Schuster}}, \bibinfo {author} {\bibfnamefont {A.~A.}\
  \bibnamefont {Houck}}, \bibinfo {author} {\bibfnamefont {A.}~\bibnamefont
  {Wallraff}}, \bibinfo {author} {\bibfnamefont {A.}~\bibnamefont {Blais}},
  \bibinfo {author} {\bibfnamefont {M.~H.}\ \bibnamefont {Devoret}}, \bibinfo
  {author} {\bibfnamefont {S.~M.}\ \bibnamefont {Girvin}}, \ and\ \bibinfo
  {author} {\bibfnamefont {R.~J.}\ \bibnamefont {Schoelkopf}},\ }\href
  {\doibase 10.1038/nature06184} {\bibfield  {journal} {\bibinfo  {journal}
  {Nature}\ }\textbf {\bibinfo {volume} {449}},\ \bibinfo {pages} {443}
  (\bibinfo {year} {2007})}\BibitemShut {NoStop}%
\bibitem [{\citenamefont {Sillanp{\"a}{\"a}}\ \emph {et~al.}(2007)\citenamefont
  {Sillanp{\"a}{\"a}}, \citenamefont {Park},\ and\ \citenamefont
  {Simmonds}}]{sillanpaa2007}%
  \BibitemOpen
  \bibfield  {author} {\bibinfo {author} {\bibfnamefont {M.~A.}\ \bibnamefont
  {Sillanp{\"a}{\"a}}}, \bibinfo {author} {\bibfnamefont {J.~I.}\ \bibnamefont
  {Park}}, \ and\ \bibinfo {author} {\bibfnamefont {R.~W.}\ \bibnamefont
  {Simmonds}},\ }\href {\doibase 10.1038/nature06124} {\bibfield  {journal}
  {\bibinfo  {journal} {Nature}\ }\textbf {\bibinfo {volume} {449}},\ \bibinfo
  {pages} {438} (\bibinfo {year} {2007})}\BibitemShut {NoStop}%
\bibitem [{\citenamefont {van Woerkom}\ \emph {et~al.}(2018)\citenamefont {van
  Woerkom}, \citenamefont {Scarlino}, \citenamefont {Ungerer}, \citenamefont
  {M{\" u}ller}, \citenamefont {Koski}, \citenamefont {Landig}, \citenamefont
  {Reichl}, \citenamefont {Wegscheider}, \citenamefont {Ihn}, \citenamefont
  {Ensslin},\ and\ \citenamefont {Wallraff}}]{vanwoerkom2018}%
  \BibitemOpen
  \bibfield  {author} {\bibinfo {author} {\bibfnamefont {D.~J.}\ \bibnamefont
  {van Woerkom}}, \bibinfo {author} {\bibfnamefont {P.}~\bibnamefont
  {Scarlino}}, \bibinfo {author} {\bibfnamefont {J.~H.}\ \bibnamefont
  {Ungerer}}, \bibinfo {author} {\bibfnamefont {C.}~\bibnamefont {M{\"
  u}ller}}, \bibinfo {author} {\bibfnamefont {J.~V.}\ \bibnamefont {Koski}},
  \bibinfo {author} {\bibfnamefont {A.~J.}\ \bibnamefont {Landig}}, \bibinfo
  {author} {\bibfnamefont {C.}~\bibnamefont {Reichl}}, \bibinfo {author}
  {\bibfnamefont {W.}~\bibnamefont {Wegscheider}}, \bibinfo {author}
  {\bibfnamefont {T.}~\bibnamefont {Ihn}}, \bibinfo {author} {\bibfnamefont
  {K.}~\bibnamefont {Ensslin}}, \ and\ \bibinfo {author} {\bibfnamefont
  {A.}~\bibnamefont {Wallraff}},\ }\href {\doibase 10.1103/PhysRevX.8.041018}
  {\bibfield  {journal} {\bibinfo  {journal} {Phys. Rev. X}\ }\textbf {\bibinfo
  {volume} {8}},\ \bibinfo {pages} {041018} (\bibinfo {year}
  {2018})}\BibitemShut {NoStop}%
\bibitem [{\citenamefont {Borjans}\ \emph
  {et~al.}(2019{\natexlab{a}})\citenamefont {Borjans}, \citenamefont {Croot},
  \citenamefont {Mi}, \citenamefont {Gullans},\ and\ \citenamefont
  {Petta}}]{borjans2019-2}%
  \BibitemOpen
  \bibfield  {author} {\bibinfo {author} {\bibfnamefont {F.}~\bibnamefont
  {Borjans}}, \bibinfo {author} {\bibfnamefont {X.~G.}\ \bibnamefont {Croot}},
  \bibinfo {author} {\bibfnamefont {X.}~\bibnamefont {Mi}}, \bibinfo {author}
  {\bibfnamefont {M.~J.}\ \bibnamefont {Gullans}}, \ and\ \bibinfo {author}
  {\bibfnamefont {J.~R.}\ \bibnamefont {Petta}},\ }\href
  {https://arxiv.org/abs/1905.00776} {\bibfield  {journal} {\bibinfo  {journal}
  {arXiv:1905.00776}\ } (\bibinfo {year} {2019}{\natexlab{a}})}\BibitemShut
  {NoStop}%
\bibitem [{\citenamefont {Steffen}\ \emph {et~al.}(2006)\citenamefont
  {Steffen}, \citenamefont {Ansmann}, \citenamefont {Bialczak}, \citenamefont
  {Katz}, \citenamefont {Lucero}, \citenamefont {McDermott}, \citenamefont
  {Neeley}, \citenamefont {Weig}, \citenamefont {Cleland},\ and\ \citenamefont
  {Martinis}}]{steffen2006}%
  \BibitemOpen
  \bibfield  {author} {\bibinfo {author} {\bibfnamefont {M.}~\bibnamefont
  {Steffen}}, \bibinfo {author} {\bibfnamefont {M.}~\bibnamefont {Ansmann}},
  \bibinfo {author} {\bibfnamefont {C.}~\bibnamefont {Bialczak}, \bibfnamefont
  {Radoslaw}}, \bibinfo {author} {\bibfnamefont {N.}~\bibnamefont {Katz}},
  \bibinfo {author} {\bibfnamefont {E.}~\bibnamefont {Lucero}}, \bibinfo
  {author} {\bibfnamefont {R.}~\bibnamefont {McDermott}}, \bibinfo {author}
  {\bibfnamefont {M.}~\bibnamefont {Neeley}}, \bibinfo {author} {\bibfnamefont
  {E.~M.}\ \bibnamefont {Weig}}, \bibinfo {author} {\bibfnamefont {A.~N.}\
  \bibnamefont {Cleland}}, \ and\ \bibinfo {author} {\bibfnamefont {J.~M.}\
  \bibnamefont {Martinis}},\ }\href {\doibase 10.1126/science.1130886}
  {\bibfield  {journal} {\bibinfo  {journal} {Science}\ }\textbf {\bibinfo
  {volume} {313}},\ \bibinfo {pages} {1423} (\bibinfo {year}
  {2006})}\BibitemShut {NoStop}%
\bibitem [{\citenamefont {Lucero}\ \emph {et~al.}(2008)\citenamefont {Lucero},
  \citenamefont {Hofheinz}, \citenamefont {Ansmann}, \citenamefont {Bialczak},
  \citenamefont {Katz}, \citenamefont {Neeley}, \citenamefont {O'Connell},
  \citenamefont {Wang}, \citenamefont {Cleland},\ and\ \citenamefont
  {Martinis}}]{lucero2008}%
  \BibitemOpen
  \bibfield  {author} {\bibinfo {author} {\bibfnamefont {E.}~\bibnamefont
  {Lucero}}, \bibinfo {author} {\bibfnamefont {M.}~\bibnamefont {Hofheinz}},
  \bibinfo {author} {\bibfnamefont {M.}~\bibnamefont {Ansmann}}, \bibinfo
  {author} {\bibfnamefont {R.~C.}\ \bibnamefont {Bialczak}}, \bibinfo {author}
  {\bibfnamefont {N.}~\bibnamefont {Katz}}, \bibinfo {author} {\bibfnamefont
  {M.}~\bibnamefont {Neeley}}, \bibinfo {author} {\bibfnamefont {A.~D.}\
  \bibnamefont {O'Connell}}, \bibinfo {author} {\bibfnamefont {H.}~\bibnamefont
  {Wang}}, \bibinfo {author} {\bibfnamefont {A.~N.}\ \bibnamefont {Cleland}}, \
  and\ \bibinfo {author} {\bibfnamefont {J.~M.}\ \bibnamefont {Martinis}},\
  }\href {\doibase 10.1103/PhysRevLett.100.247001} {\bibfield  {journal}
  {\bibinfo  {journal} {Phys. Rev. Lett.}\ }\textbf {\bibinfo {volume} {100}},\
  \bibinfo {pages} {247001} (\bibinfo {year} {2008})}\BibitemShut {NoStop}%
\bibitem [{\citenamefont {Chow}\ \emph {et~al.}(2009)\citenamefont {Chow},
  \citenamefont {Gambetta}, \citenamefont {Tornberg}, \citenamefont {Koch},
  \citenamefont {Bishop}, \citenamefont {Houck}, \citenamefont {Johnson},
  \citenamefont {Frunzio}, \citenamefont {Girvin},\ and\ \citenamefont
  {Schoelkopf}}]{chow2009}%
  \BibitemOpen
  \bibfield  {author} {\bibinfo {author} {\bibfnamefont {J.~M.}\ \bibnamefont
  {Chow}}, \bibinfo {author} {\bibfnamefont {J.~M.}\ \bibnamefont {Gambetta}},
  \bibinfo {author} {\bibfnamefont {L.}~\bibnamefont {Tornberg}}, \bibinfo
  {author} {\bibfnamefont {J.}~\bibnamefont {Koch}}, \bibinfo {author}
  {\bibfnamefont {L.~S.}\ \bibnamefont {Bishop}}, \bibinfo {author}
  {\bibfnamefont {A.~A.}\ \bibnamefont {Houck}}, \bibinfo {author}
  {\bibfnamefont {B.~R.}\ \bibnamefont {Johnson}}, \bibinfo {author}
  {\bibfnamefont {L.}~\bibnamefont {Frunzio}}, \bibinfo {author} {\bibfnamefont
  {S.~M.}\ \bibnamefont {Girvin}}, \ and\ \bibinfo {author} {\bibfnamefont
  {R.~J.}\ \bibnamefont {Schoelkopf}},\ }\href {\doibase
  10.1103/PhysRevLett.102.090502} {\bibfield  {journal} {\bibinfo  {journal}
  {Phys. Rev. Lett.}\ }\textbf {\bibinfo {volume} {102}},\ \bibinfo {pages}
  {090502} (\bibinfo {year} {2009})}\BibitemShut {NoStop}%
\bibitem [{\citenamefont {Barends}\ \emph {et~al.}(2014)\citenamefont
  {Barends}, \citenamefont {Kelly}, \citenamefont {Megrant}, \citenamefont
  {Veitia}, \citenamefont {Sank}, \citenamefont {Jeffrey}, \citenamefont
  {White}, \citenamefont {Mutus}, \citenamefont {Fowler}, \citenamefont
  {Campbell}, \citenamefont {Chen}, \citenamefont {Chen}, \citenamefont
  {Chiaro}, \citenamefont {Dunsworth}, \citenamefont {Neill}, \citenamefont
  {O'Malley}, \citenamefont {Roushan}, \citenamefont {Vainsencher},
  \citenamefont {Wenner}, \citenamefont {Korotkov}, \citenamefont {Cleland},\
  and\ \citenamefont {Martinis}}]{barends2014}%
  \BibitemOpen
  \bibfield  {author} {\bibinfo {author} {\bibfnamefont {R.}~\bibnamefont
  {Barends}}, \bibinfo {author} {\bibfnamefont {J.}~\bibnamefont {Kelly}},
  \bibinfo {author} {\bibfnamefont {A.}~\bibnamefont {Megrant}}, \bibinfo
  {author} {\bibfnamefont {A.}~\bibnamefont {Veitia}}, \bibinfo {author}
  {\bibfnamefont {D.}~\bibnamefont {Sank}}, \bibinfo {author} {\bibfnamefont
  {E.}~\bibnamefont {Jeffrey}}, \bibinfo {author} {\bibfnamefont {T.~C.}\
  \bibnamefont {White}}, \bibinfo {author} {\bibfnamefont {J.}~\bibnamefont
  {Mutus}}, \bibinfo {author} {\bibfnamefont {A.~G.}\ \bibnamefont {Fowler}},
  \bibinfo {author} {\bibfnamefont {B.}~\bibnamefont {Campbell}}, \bibinfo
  {author} {\bibfnamefont {Y.}~\bibnamefont {Chen}}, \bibinfo {author}
  {\bibfnamefont {Z.}~\bibnamefont {Chen}}, \bibinfo {author} {\bibfnamefont
  {B.}~\bibnamefont {Chiaro}}, \bibinfo {author} {\bibfnamefont
  {A.}~\bibnamefont {Dunsworth}}, \bibinfo {author} {\bibfnamefont
  {C.}~\bibnamefont {Neill}}, \bibinfo {author} {\bibfnamefont
  {P.}~\bibnamefont {O'Malley}}, \bibinfo {author} {\bibfnamefont
  {P.}~\bibnamefont {Roushan}}, \bibinfo {author} {\bibfnamefont
  {A.}~\bibnamefont {Vainsencher}}, \bibinfo {author} {\bibfnamefont
  {J.}~\bibnamefont {Wenner}}, \bibinfo {author} {\bibfnamefont {A.~N.}\
  \bibnamefont {Korotkov}}, \bibinfo {author} {\bibfnamefont {A.~N.}\
  \bibnamefont {Cleland}}, \ and\ \bibinfo {author} {\bibfnamefont {J.~M.}\
  \bibnamefont {Martinis}},\ }\href {\doibase 10.1038/nature13171} {\bibfield
  {journal} {\bibinfo  {journal} {Nature}\ }\textbf {\bibinfo {volume} {508}},\
  \bibinfo {pages} {500} (\bibinfo {year} {2014})}\BibitemShut {NoStop}%
\bibitem [{\citenamefont {Rol}\ \emph {et~al.}(2019)\citenamefont {Rol},
  \citenamefont {Battistel}, \citenamefont {Malinowski}, \citenamefont
  {Bultink}, \citenamefont {Tarasinski}, \citenamefont {Vollmer}, \citenamefont
  {Haider}, \citenamefont {Muthusubramanian}, \citenamefont {Bruno},
  \citenamefont {Terhal},\ and\ \citenamefont {DiCarlo}}]{rol2019}%
  \BibitemOpen
  \bibfield  {author} {\bibinfo {author} {\bibfnamefont {M.~A.}\ \bibnamefont
  {Rol}}, \bibinfo {author} {\bibfnamefont {F.}~\bibnamefont {Battistel}},
  \bibinfo {author} {\bibfnamefont {F.~K.}\ \bibnamefont {Malinowski}},
  \bibinfo {author} {\bibfnamefont {C.~C.}\ \bibnamefont {Bultink}}, \bibinfo
  {author} {\bibfnamefont {B.~M.}\ \bibnamefont {Tarasinski}}, \bibinfo
  {author} {\bibfnamefont {R.}~\bibnamefont {Vollmer}}, \bibinfo {author}
  {\bibfnamefont {N.}~\bibnamefont {Haider}}, \bibinfo {author} {\bibfnamefont
  {N.}~\bibnamefont {Muthusubramanian}}, \bibinfo {author} {\bibfnamefont
  {A.}~\bibnamefont {Bruno}}, \bibinfo {author} {\bibfnamefont {B.~M.}\
  \bibnamefont {Terhal}}, \ and\ \bibinfo {author} {\bibfnamefont
  {L.}~\bibnamefont {DiCarlo}},\ }\href
  {https://journals.aps.org/prl/abstract/10.1103/PhysRevLett.123.120502}
  {\bibfield  {journal} {\bibinfo  {journal} {Phys. Rev. Lett.}\ }\textbf
  {\bibinfo {volume} {123}},\ \bibinfo {pages} {120502} (\bibinfo {year}
  {2019})}\BibitemShut {NoStop}%
\bibitem [{\citenamefont {Lupa{\c{s}}cu}\ \emph {et~al.}(2006)\citenamefont
  {Lupa{\c{s}}cu}, \citenamefont {Driessen}, \citenamefont {Roschier},
  \citenamefont {Harmans},\ and\ \citenamefont {Mooij}}]{lupascu2006}%
  \BibitemOpen
  \bibfield  {author} {\bibinfo {author} {\bibfnamefont {A.}~\bibnamefont
  {Lupa{\c{s}}cu}}, \bibinfo {author} {\bibfnamefont {E.~F.~C.}\ \bibnamefont
  {Driessen}}, \bibinfo {author} {\bibfnamefont {L.}~\bibnamefont {Roschier}},
  \bibinfo {author} {\bibfnamefont {C.~J. P.~M.}\ \bibnamefont {Harmans}}, \
  and\ \bibinfo {author} {\bibfnamefont {J.~E.}\ \bibnamefont {Mooij}},\ }\href
  {\doibase 10.1103/PhysRevLett.96.127003} {\bibfield  {journal} {\bibinfo
  {journal} {Phys. Rev. Lett.}\ }\textbf {\bibinfo {volume} {96}},\ \bibinfo
  {pages} {127003} (\bibinfo {year} {2006})}\BibitemShut {NoStop}%
\bibitem [{\citenamefont {Mallet}\ \emph {et~al.}(2009)\citenamefont {Mallet},
  \citenamefont {Ong}, \citenamefont {Palacios-Laloy}, \citenamefont {Nguyen},
  \citenamefont {Bertet}, \citenamefont {Vion},\ and\ \citenamefont
  {Esteve}}]{mallet2009}%
  \BibitemOpen
  \bibfield  {author} {\bibinfo {author} {\bibfnamefont {F.}~\bibnamefont
  {Mallet}}, \bibinfo {author} {\bibfnamefont {F.~R.}\ \bibnamefont {Ong}},
  \bibinfo {author} {\bibfnamefont {A.}~\bibnamefont {Palacios-Laloy}},
  \bibinfo {author} {\bibfnamefont {F.}~\bibnamefont {Nguyen}}, \bibinfo
  {author} {\bibfnamefont {P.}~\bibnamefont {Bertet}}, \bibinfo {author}
  {\bibfnamefont {D.}~\bibnamefont {Vion}}, \ and\ \bibinfo {author}
  {\bibfnamefont {D.}~\bibnamefont {Esteve}},\ }\href {\doibase
  10.1038/nphys1400} {\bibfield  {journal} {\bibinfo  {journal} {Nat. Phys.}\
  }\textbf {\bibinfo {volume} {5}},\ \bibinfo {pages} {791} (\bibinfo {year}
  {2009})}\BibitemShut {NoStop}%
\bibitem [{\citenamefont {Liu}\ \emph {et~al.}(2014)\citenamefont {Liu},
  \citenamefont {Srinivasan}, \citenamefont {Hover}, \citenamefont {Zhu},
  \citenamefont {McDermott},\ and\ \citenamefont {Houck}}]{liu2014}%
  \BibitemOpen
  \bibfield  {author} {\bibinfo {author} {\bibfnamefont {Y.}~\bibnamefont
  {Liu}}, \bibinfo {author} {\bibfnamefont {S.~J.}\ \bibnamefont {Srinivasan}},
  \bibinfo {author} {\bibfnamefont {D.}~\bibnamefont {Hover}}, \bibinfo
  {author} {\bibfnamefont {S.}~\bibnamefont {Zhu}}, \bibinfo {author}
  {\bibfnamefont {R.}~\bibnamefont {McDermott}}, \ and\ \bibinfo {author}
  {\bibfnamefont {A.~A.}\ \bibnamefont {Houck}},\ }\href {\doibase
  10.1088/1367-2630/16/11/113008} {\bibfield  {journal} {\bibinfo  {journal}
  {New J. Phys.}\ }\textbf {\bibinfo {volume} {16}},\ \bibinfo {pages} {113008}
  (\bibinfo {year} {2014})}\BibitemShut {NoStop}%
\bibitem [{\citenamefont {Hover}\ \emph {et~al.}(2014)\citenamefont {Hover},
  \citenamefont {Zhu}, \citenamefont {Thorbeck}, \citenamefont {Ribeill},
  \citenamefont {Sank}, \citenamefont {Kelly}, \citenamefont {Barends},
  \citenamefont {Martinis},\ and\ \citenamefont {McDermott}}]{hover2014}%
  \BibitemOpen
  \bibfield  {author} {\bibinfo {author} {\bibfnamefont {D.}~\bibnamefont
  {Hover}}, \bibinfo {author} {\bibfnamefont {S.}~\bibnamefont {Zhu}}, \bibinfo
  {author} {\bibfnamefont {T.}~\bibnamefont {Thorbeck}}, \bibinfo {author}
  {\bibfnamefont {G.}~\bibnamefont {Ribeill}}, \bibinfo {author} {\bibfnamefont
  {D.}~\bibnamefont {Sank}}, \bibinfo {author} {\bibfnamefont {J.}~\bibnamefont
  {Kelly}}, \bibinfo {author} {\bibfnamefont {R.}~\bibnamefont {Barends}},
  \bibinfo {author} {\bibfnamefont {J.~M.}\ \bibnamefont {Martinis}}, \ and\
  \bibinfo {author} {\bibfnamefont {R.}~\bibnamefont {McDermott}},\ }\href
  {\doibase 10.1063/1.4871088} {\bibfield  {journal} {\bibinfo  {journal} {App.
  Phys. Lett.}\ }\textbf {\bibinfo {volume} {104}},\ \bibinfo {pages} {152601}
  (\bibinfo {year} {2014})}\BibitemShut {NoStop}%
\bibitem [{\citenamefont {Jeffrey}\ \emph {et~al.}(2014)\citenamefont
  {Jeffrey}, \citenamefont {Sank}, \citenamefont {Mutus}, \citenamefont
  {White}, \citenamefont {Kelly}, \citenamefont {Barends}, \citenamefont
  {Chen}, \citenamefont {Chen}, \citenamefont {Chiaro}, \citenamefont
  {Dunsworth}, \citenamefont {Megrant}, \citenamefont {O'Malley}, \citenamefont
  {Neill}, \citenamefont {Roushan}, \citenamefont {Vainsencher}, \citenamefont
  {Wenner}, \citenamefont {Cleland},\ and\ \citenamefont
  {Martinis}}]{jeffrey2014}%
  \BibitemOpen
  \bibfield  {author} {\bibinfo {author} {\bibfnamefont {E.}~\bibnamefont
  {Jeffrey}}, \bibinfo {author} {\bibfnamefont {D.}~\bibnamefont {Sank}},
  \bibinfo {author} {\bibfnamefont {J.~Y.}\ \bibnamefont {Mutus}}, \bibinfo
  {author} {\bibfnamefont {T.~C.}\ \bibnamefont {White}}, \bibinfo {author}
  {\bibfnamefont {J.}~\bibnamefont {Kelly}}, \bibinfo {author} {\bibfnamefont
  {R.}~\bibnamefont {Barends}}, \bibinfo {author} {\bibfnamefont
  {Y.}~\bibnamefont {Chen}}, \bibinfo {author} {\bibfnamefont {Z.}~\bibnamefont
  {Chen}}, \bibinfo {author} {\bibfnamefont {B.}~\bibnamefont {Chiaro}},
  \bibinfo {author} {\bibfnamefont {A.}~\bibnamefont {Dunsworth}}, \bibinfo
  {author} {\bibfnamefont {A.}~\bibnamefont {Megrant}}, \bibinfo {author}
  {\bibfnamefont {P.~J.~J.}\ \bibnamefont {O'Malley}}, \bibinfo {author}
  {\bibfnamefont {C.}~\bibnamefont {Neill}}, \bibinfo {author} {\bibfnamefont
  {P.}~\bibnamefont {Roushan}}, \bibinfo {author} {\bibfnamefont
  {A.}~\bibnamefont {Vainsencher}}, \bibinfo {author} {\bibfnamefont
  {J.}~\bibnamefont {Wenner}}, \bibinfo {author} {\bibfnamefont {A.~N.}\
  \bibnamefont {Cleland}}, \ and\ \bibinfo {author} {\bibfnamefont {J.~M.}\
  \bibnamefont {Martinis}},\ }\href {\doibase 10.1103/PhysRevLett.112.190504}
  {\bibfield  {journal} {\bibinfo  {journal} {Phys. Rev. Lett.}\ }\textbf
  {\bibinfo {volume} {112}},\ \bibinfo {pages} {190504} (\bibinfo {year}
  {2014})}\BibitemShut {NoStop}%
\bibitem [{\citenamefont {Walter}\ \emph {et~al.}(2017)\citenamefont {Walter},
  \citenamefont {Kurpiers}, \citenamefont {Gasparinetti}, \citenamefont
  {Magnard}, \citenamefont {Poto{\v c}nik}, \citenamefont {Salath{\'e}},
  \citenamefont {Pechal}, \citenamefont {Mondal}, \citenamefont {Oppliger},
  \citenamefont {Eichler},\ and\ \citenamefont {Wallraff}}]{walter2017}%
  \BibitemOpen
  \bibfield  {author} {\bibinfo {author} {\bibfnamefont {T.}~\bibnamefont
  {Walter}}, \bibinfo {author} {\bibfnamefont {P.}~\bibnamefont {Kurpiers}},
  \bibinfo {author} {\bibfnamefont {S.}~\bibnamefont {Gasparinetti}}, \bibinfo
  {author} {\bibfnamefont {P.}~\bibnamefont {Magnard}}, \bibinfo {author}
  {\bibfnamefont {A.}~\bibnamefont {Poto{\v c}nik}}, \bibinfo {author}
  {\bibfnamefont {Y.}~\bibnamefont {Salath{\'e}}}, \bibinfo {author}
  {\bibfnamefont {M.}~\bibnamefont {Pechal}}, \bibinfo {author} {\bibfnamefont
  {M.}~\bibnamefont {Mondal}}, \bibinfo {author} {\bibfnamefont
  {M.}~\bibnamefont {Oppliger}}, \bibinfo {author} {\bibfnamefont
  {C.}~\bibnamefont {Eichler}}, \ and\ \bibinfo {author} {\bibfnamefont
  {A.}~\bibnamefont {Wallraff}},\ }\href {\doibase
  10.1103/PhysRevApplied.7.054020} {\bibfield  {journal} {\bibinfo  {journal}
  {Phys. Rev. Appl.}\ }\textbf {\bibinfo {volume} {7}},\ \bibinfo {pages}
  {054020} (\bibinfo {year} {2017})}\BibitemShut {NoStop}%
\bibitem [{\citenamefont {Blais}\ \emph {et~al.}(2004)\citenamefont {Blais},
  \citenamefont {Huang}, \citenamefont {Wallraff}, \citenamefont {Girvin},\
  and\ \citenamefont {Schoelkopf}}]{blais2004}%
  \BibitemOpen
  \bibfield  {author} {\bibinfo {author} {\bibfnamefont {A.}~\bibnamefont
  {Blais}}, \bibinfo {author} {\bibfnamefont {R.-S.}\ \bibnamefont {Huang}},
  \bibinfo {author} {\bibfnamefont {A.}~\bibnamefont {Wallraff}}, \bibinfo
  {author} {\bibfnamefont {S.~M.}\ \bibnamefont {Girvin}}, \ and\ \bibinfo
  {author} {\bibfnamefont {R.~J.}\ \bibnamefont {Schoelkopf}},\ }\href
  {\doibase 10.1103/PhysRevA.69.062320} {\bibfield  {journal} {\bibinfo
  {journal} {Phys. Rev. A}\ }\textbf {\bibinfo {volume} {69}},\ \bibinfo
  {pages} {062320} (\bibinfo {year} {2004})}\BibitemShut {NoStop}%
\bibitem [{\citenamefont {Beaudoin}\ \emph {et~al.}(2017)\citenamefont
  {Beaudoin}, \citenamefont {Blais},\ and\ \citenamefont
  {Coish}}]{beaudoin2017}%
  \BibitemOpen
  \bibfield  {author} {\bibinfo {author} {\bibfnamefont {F.}~\bibnamefont
  {Beaudoin}}, \bibinfo {author} {\bibfnamefont {A.}~\bibnamefont {Blais}}, \
  and\ \bibinfo {author} {\bibfnamefont {W.~A.}\ \bibnamefont {Coish}},\ }\href
  {\doibase 10.1088/1367-2630/aa5d33} {\bibfield  {journal} {\bibinfo
  {journal} {New J. Phys.}\ }\textbf {\bibinfo {volume} {19}},\ \bibinfo
  {pages} {023041} (\bibinfo {year} {2017})}\BibitemShut {NoStop}%
\bibitem [{\citenamefont {Troiani}(2019)}]{troiani2019}%
  \BibitemOpen
  \bibfield  {author} {\bibinfo {author} {\bibfnamefont {F.}~\bibnamefont
  {Troiani}},\ }\href {\doibase 10.1016/j.physleta.2019.02.016} {\bibfield
  {journal} {\bibinfo  {journal} {Phys. Lett. A}\ }\textbf {\bibinfo {volume}
  {383}},\ \bibinfo {pages} {1536} (\bibinfo {year} {2019})}\BibitemShut
  {NoStop}%
\bibitem [{\citenamefont {Zhang}\ \emph {et~al.}(2019)\citenamefont {Zhang},
  \citenamefont {Wang},\ and\ \citenamefont {You}}]{zhang2019}%
  \BibitemOpen
  \bibfield  {author} {\bibinfo {author} {\bibfnamefont {G.-Q.}\ \bibnamefont
  {Zhang}}, \bibinfo {author} {\bibfnamefont {Y.-P.}\ \bibnamefont {Wang}}, \
  and\ \bibinfo {author} {\bibfnamefont {J.~Q.}\ \bibnamefont {You}},\ }\href
  {\doibase 10.1103/PhysRevA.99.052341} {\bibfield  {journal} {\bibinfo
  {journal} {Phys. Rev. A}\ }\textbf {\bibinfo {volume} {99}},\ \bibinfo
  {pages} {052341} (\bibinfo {year} {2019})}\BibitemShut {NoStop}%
\bibitem [{\citenamefont {Ruskov}\ and\ \citenamefont
  {Tahan}(2019)}]{ruskov2019}%
  \BibitemOpen
  \bibfield  {author} {\bibinfo {author} {\bibfnamefont {R.}~\bibnamefont
  {Ruskov}}\ and\ \bibinfo {author} {\bibfnamefont {C.}~\bibnamefont {Tahan}},\
  }\href {\doibase 10.1103/PhysRevB.99.245306} {\bibfield  {journal} {\bibinfo
  {journal} {Phys. Rev. B}\ }\textbf {\bibinfo {volume} {99}},\ \bibinfo
  {pages} {245306} (\bibinfo {year} {2019})}\BibitemShut {NoStop}%
\bibitem [{\citenamefont {Cottet}\ and\ \citenamefont
  {Kontos}(2010)}]{cottet2010}%
  \BibitemOpen
  \bibfield  {author} {\bibinfo {author} {\bibfnamefont {A.}~\bibnamefont
  {Cottet}}\ and\ \bibinfo {author} {\bibfnamefont {T.}~\bibnamefont
  {Kontos}},\ }\href {\doibase 10.1103/PhysRevLett.105.160502} {\bibfield
  {journal} {\bibinfo  {journal} {Phys. Rev. Lett.}\ }\textbf {\bibinfo
  {volume} {105}},\ \bibinfo {pages} {160502} (\bibinfo {year}
  {2010})}\BibitemShut {NoStop}%
\bibitem [{\citenamefont {Hu}\ \emph {et~al.}(2012)\citenamefont {Hu},
  \citenamefont {Liu},\ and\ \citenamefont {Nori}}]{hu2012}%
  \BibitemOpen
  \bibfield  {author} {\bibinfo {author} {\bibfnamefont {X.}~\bibnamefont
  {Hu}}, \bibinfo {author} {\bibfnamefont {Y.-x.}\ \bibnamefont {Liu}}, \ and\
  \bibinfo {author} {\bibfnamefont {F.}~\bibnamefont {Nori}},\ }\href {\doibase
  10.1103/PhysRevB.86.035314} {\bibfield  {journal} {\bibinfo  {journal} {Phys.
  Rev. B}\ }\textbf {\bibinfo {volume} {86}},\ \bibinfo {pages} {035314}
  (\bibinfo {year} {2012})}\BibitemShut {NoStop}%
\bibitem [{\citenamefont {Beaudoin}\ \emph {et~al.}(2016)\citenamefont
  {Beaudoin}, \citenamefont {Lachance-Quirion}, \citenamefont {A},\ and\
  \citenamefont {Pioro-Ladri{\`e}re}}]{beaudoin2016}%
  \BibitemOpen
  \bibfield  {author} {\bibinfo {author} {\bibfnamefont {F.}~\bibnamefont
  {Beaudoin}}, \bibinfo {author} {\bibfnamefont {D.}~\bibnamefont
  {Lachance-Quirion}}, \bibinfo {author} {\bibfnamefont {C.~W.}\ \bibnamefont
  {A}}, \ and\ \bibinfo {author} {\bibfnamefont {M.}~\bibnamefont
  {Pioro-Ladri{\`e}re}},\ }\href {\doibase 10.1088/0957-4484/27/46/464003}
  {\bibfield  {journal} {\bibinfo  {journal} {Nanotechnology}\ }\textbf
  {\bibinfo {volume} {27}},\ \bibinfo {pages} {464003} (\bibinfo {year}
  {2016})}\BibitemShut {NoStop}%
\bibitem [{\citenamefont {Benito}\ \emph {et~al.}(2017)\citenamefont {Benito},
  \citenamefont {Mi}, \citenamefont {Taylor}, \citenamefont {Petta},\ and\
  \citenamefont {Burkard}}]{benito2017}%
  \BibitemOpen
  \bibfield  {author} {\bibinfo {author} {\bibfnamefont {M.}~\bibnamefont
  {Benito}}, \bibinfo {author} {\bibfnamefont {X.}~\bibnamefont {Mi}}, \bibinfo
  {author} {\bibfnamefont {J.~M.}\ \bibnamefont {Taylor}}, \bibinfo {author}
  {\bibfnamefont {J.~R.}\ \bibnamefont {Petta}}, \ and\ \bibinfo {author}
  {\bibfnamefont {G.}~\bibnamefont {Burkard}},\ }\href {\doibase
  10.1103/PhysRevB.96.235434} {\bibfield  {journal} {\bibinfo  {journal} {Phys.
  Rev. B}\ }\textbf {\bibinfo {volume} {96}},\ \bibinfo {pages} {235434}
  (\bibinfo {year} {2017})}\BibitemShut {NoStop}%
\bibitem [{\citenamefont {Benito}\ \emph
  {et~al.}(2019{\natexlab{a}})\citenamefont {Benito}, \citenamefont {Croot},
  \citenamefont {Adelsberger}, \citenamefont {Putz}, \citenamefont {Mi},
  \citenamefont {Petta},\ and\ \citenamefont {Burkard}}]{benito2019-2}%
  \BibitemOpen
  \bibfield  {author} {\bibinfo {author} {\bibfnamefont {M.}~\bibnamefont
  {Benito}}, \bibinfo {author} {\bibfnamefont {X.}~\bibnamefont {Croot}},
  \bibinfo {author} {\bibfnamefont {C.}~\bibnamefont {Adelsberger}}, \bibinfo
  {author} {\bibfnamefont {S.}~\bibnamefont {Putz}}, \bibinfo {author}
  {\bibfnamefont {X.}~\bibnamefont {Mi}}, \bibinfo {author} {\bibfnamefont
  {J.~R.}\ \bibnamefont {Petta}}, \ and\ \bibinfo {author} {\bibfnamefont
  {G.}~\bibnamefont {Burkard}},\ }\href {\doibase 10.1103/PhysRevB.100.125430}
  {\bibfield  {journal} {\bibinfo  {journal} {Phys. Rev. B}\ }\textbf {\bibinfo
  {volume} {100}},\ \bibinfo {pages} {125430} (\bibinfo {year}
  {2019}{\natexlab{a}})}\BibitemShut {NoStop}%
\bibitem [{\citenamefont {Burkard}\ \emph {et~al.}(2019)\citenamefont
  {Burkard}, \citenamefont {Gullans}, \citenamefont {Mi},\ and\ \citenamefont
  {Petta}}]{burkard2019}%
  \BibitemOpen
  \bibfield  {author} {\bibinfo {author} {\bibfnamefont {G.}~\bibnamefont
  {Burkard}}, \bibinfo {author} {\bibfnamefont {M.~J.}\ \bibnamefont
  {Gullans}}, \bibinfo {author} {\bibfnamefont {X.}~\bibnamefont {Mi}}, \ and\
  \bibinfo {author} {\bibfnamefont {J.~R.}\ \bibnamefont {Petta}},\ }\href
  {https://arxiv.org/abs/1905.01155} {\bibfield  {journal} {\bibinfo  {journal}
  {arXiv:1905.01155}\ } (\bibinfo {year} {2019})}\BibitemShut {NoStop}%
\bibitem [{\citenamefont {Haikka}\ \emph {et~al.}(2017)\citenamefont {Haikka},
  \citenamefont {Kubo}, \citenamefont {Bienfait}, \citenamefont {Bertet},\ and\
  \citenamefont {M{\o}lmer}}]{haikka2017}%
  \BibitemOpen
  \bibfield  {author} {\bibinfo {author} {\bibfnamefont {P.}~\bibnamefont
  {Haikka}}, \bibinfo {author} {\bibfnamefont {Y.}~\bibnamefont {Kubo}},
  \bibinfo {author} {\bibfnamefont {A.}~\bibnamefont {Bienfait}}, \bibinfo
  {author} {\bibfnamefont {P.}~\bibnamefont {Bertet}}, \ and\ \bibinfo {author}
  {\bibfnamefont {K.}~\bibnamefont {M{\o}lmer}},\ }\href {\doibase
  10.1103/PhysRevA.95.022306} {\bibfield  {journal} {\bibinfo  {journal} {Phys.
  Rev. A}\ }\textbf {\bibinfo {volume} {95}},\ \bibinfo {pages} {022306}
  (\bibinfo {year} {2017})}\BibitemShut {NoStop}%
\bibitem [{\citenamefont {Koch}\ \emph {et~al.}(2007)\citenamefont {Koch},
  \citenamefont {Yu}, \citenamefont {Gambetta}, \citenamefont {Houck},
  \citenamefont {Schuster}, \citenamefont {Majer}, \citenamefont {Blais},
  \citenamefont {Devoret}, \citenamefont {Girvin},\ and\ \citenamefont
  {Schoelkopf}}]{koch2007}%
  \BibitemOpen
  \bibfield  {author} {\bibinfo {author} {\bibfnamefont {J.}~\bibnamefont
  {Koch}}, \bibinfo {author} {\bibfnamefont {T.~M.}\ \bibnamefont {Yu}},
  \bibinfo {author} {\bibfnamefont {J.}~\bibnamefont {Gambetta}}, \bibinfo
  {author} {\bibfnamefont {A.~A.}\ \bibnamefont {Houck}}, \bibinfo {author}
  {\bibfnamefont {D.~I.}\ \bibnamefont {Schuster}}, \bibinfo {author}
  {\bibfnamefont {J.}~\bibnamefont {Majer}}, \bibinfo {author} {\bibfnamefont
  {A.}~\bibnamefont {Blais}}, \bibinfo {author} {\bibfnamefont {M.~H.}\
  \bibnamefont {Devoret}}, \bibinfo {author} {\bibfnamefont {S.~M.}\
  \bibnamefont {Girvin}}, \ and\ \bibinfo {author} {\bibfnamefont {R.~J.}\
  \bibnamefont {Schoelkopf}},\ }\href {\doibase PhysRevA.76.042319} {\bibfield
  {journal} {\bibinfo  {journal} {Phys. Rev. A}\ }\textbf {\bibinfo {volume}
  {76}},\ \bibinfo {pages} {042319} (\bibinfo {year} {2007})}\BibitemShut
  {NoStop}%
\bibitem [{\citenamefont {Boissonneault}\ \emph {et~al.}(2012)\citenamefont
  {Boissonneault}, \citenamefont {Gambetta},\ and\ \citenamefont
  {Blais}}]{boissonneault2012}%
  \BibitemOpen
  \bibfield  {author} {\bibinfo {author} {\bibfnamefont {M.}~\bibnamefont
  {Boissonneault}}, \bibinfo {author} {\bibfnamefont {J.~M.}\ \bibnamefont
  {Gambetta}}, \ and\ \bibinfo {author} {\bibfnamefont {A.}~\bibnamefont
  {Blais}},\ }\href {\doibase 10.1103/PhysRevA.86.022326} {\bibfield  {journal}
  {\bibinfo  {journal} {Phys. Rev. A}\ }\textbf {\bibinfo {volume} {86}},\
  \bibinfo {pages} {022326} (\bibinfo {year} {2012})}\BibitemShut {NoStop}%
\bibitem [{\citenamefont {Inomata}\ \emph {et~al.}(2012)\citenamefont
  {Inomata}, \citenamefont {Yamamoto}, \citenamefont {Billangeon},
  \citenamefont {Nakamura},\ and\ \citenamefont {Tsai}}]{inomata2012}%
  \BibitemOpen
  \bibfield  {author} {\bibinfo {author} {\bibfnamefont {K.}~\bibnamefont
  {Inomata}}, \bibinfo {author} {\bibfnamefont {T.}~\bibnamefont {Yamamoto}},
  \bibinfo {author} {\bibfnamefont {P.~M.}\ \bibnamefont {Billangeon}},
  \bibinfo {author} {\bibfnamefont {Y.}~\bibnamefont {Nakamura}}, \ and\
  \bibinfo {author} {\bibfnamefont {J.~S.}\ \bibnamefont {Tsai}},\ }\href
  {\doibase 10.1103/PhysRevB.86.140508} {\bibfield  {journal} {\bibinfo
  {journal} {Phys. Rev. B}\ }\textbf {\bibinfo {volume} {86}},\ \bibinfo
  {pages} {140508(R)} (\bibinfo {year} {2012})}\BibitemShut {NoStop}%
\bibitem [{\citenamefont {Zhu}\ \emph {et~al.}(2013)\citenamefont {Zhu},
  \citenamefont {Ferguson}, \citenamefont {Manucharyan},\ and\ \citenamefont
  {Koch}}]{zhu2013}%
  \BibitemOpen
  \bibfield  {author} {\bibinfo {author} {\bibfnamefont {G.}~\bibnamefont
  {Zhu}}, \bibinfo {author} {\bibfnamefont {D.~G.}\ \bibnamefont {Ferguson}},
  \bibinfo {author} {\bibfnamefont {V.~E.}\ \bibnamefont {Manucharyan}}, \ and\
  \bibinfo {author} {\bibfnamefont {J.}~\bibnamefont {Koch}},\ }\href {\doibase
  10.1103/PhysRevB.87.024510} {\bibfield  {journal} {\bibinfo  {journal} {Phys.
  Rev. B}\ }\textbf {\bibinfo {volume} {87}},\ \bibinfo {pages} {024510}
  (\bibinfo {year} {2013})}\BibitemShut {NoStop}%
\bibitem [{Note1()}]{Note1}%
  \BibitemOpen
  \bibinfo {note} {Any intrinsic resonator losses that result in photons being
  emitted somewhere other than the input and output ports can simply be modeled
  by including an additional port with leakage rate $\kappa _0$. Therefore, the
  formalism presented below remains completely general in the presence of such
  losses. In this manuscript, we assume that the intrinsic losses are small
  compared to the leakage through the input ports.}\BibitemShut {Stop}%
\bibitem [{\citenamefont {Gardiner}\ and\ \citenamefont
  {Collett}(1985)}]{gardiner1985}%
  \BibitemOpen
  \bibfield  {author} {\bibinfo {author} {\bibfnamefont {C.~W.}\ \bibnamefont
  {Gardiner}}\ and\ \bibinfo {author} {\bibfnamefont {M.~J.}\ \bibnamefont
  {Collett}},\ }\href {\doibase 10.1103/PhysRevA.31.3761} {\bibfield  {journal}
  {\bibinfo  {journal} {Phys. Rev. A}\ }\textbf {\bibinfo {volume} {31}},\
  \bibinfo {pages} {3761} (\bibinfo {year} {1985})}\BibitemShut {NoStop}%
\bibitem [{Note2()}]{Note2}%
  \BibitemOpen
  \bibinfo {note} {More precisely, it is assumed that the correlation time of
  the noise is much smaller than the typical timescale for the system evolution
  in the frame rotating at the probe frequency $\omega _\protect \textrm {in}$.
  The correlation time should also be much smaller than the inverse detector
  bandwidth.}\BibitemShut {Stop}%
\bibitem [{Note3()}]{Note3}%
  \BibitemOpen
  \bibinfo {note} {Even though thermal noise is not white, it is typically
  approximately white in the neighborhood of the probe frequency $\omega
  _\protect \textrm {in}$.}\BibitemShut {Stop}%
\bibitem [{Note4()}]{Note4}%
  \BibitemOpen
  \bibinfo {note} {If the value of $\protect \mathaccentV {bar}016{N}$ is not
  the same in both ports, there is a net flow of noise photons from high-noise
  ports to low-noise ports through the resonator. In that case, the output
  noise acquires a finite correlation time $\sim \kappa ^{-1}$ and is a
  function of the transmission and reflection coefficients of the resonator.
  The observed output noise may then be modeled as white provided that the
  detector bandwidth is smaller than $\kappa $.}\BibitemShut {Stop}%
\bibitem [{Note5()}]{Note5}%
  \BibitemOpen
  \bibinfo {note} {The readout can also be performed with the help of a
  heterodyne detector, at the cost of the additional vacuum noise which is
  unavoidably introduced by attempting to simultaneously measure two
  noncommuting quadratures of a quantum field.}\BibitemShut {Stop}%
\bibitem [{Note6()}]{Note6}%
  \BibitemOpen
  \bibinfo {note} {Note that here, $\tau _z$ and $\sigma _z$ are defined such
  that the molecular-like eigenstates and spin-like eigenstates are always
  those which have a mostly molecular character and mostly spin character,
  respectively. An alternative definition is to choose $\tau _z$ and $\sigma
  _z$ such that the two lowest-energy eigenstates always correspond to the same
  value of $\tau _z$, such as in Ref.~\protect \rev@citealpnum {benito2019}. In
  that case, the Hamiltonian takes the form $H_d = \protect \frac {E_\tau
  }{2}\tau _z + \protect \frac {E_\sigma }{2}\sigma _z$, where $E_\tau = E_m$
  and $E_\sigma = E_s$ when $2 t_c > B_z$, and $E_\tau = E_s$ and $E_\sigma =
  E_m$ when $2t_c < B_z$.}\BibitemShut {Stop}%
\bibitem [{\citenamefont {Boissonneault}\ \emph {et~al.}(2009)\citenamefont
  {Boissonneault}, \citenamefont {Gambetta},\ and\ \citenamefont
  {Blais}}]{boissonneault2009}%
  \BibitemOpen
  \bibfield  {author} {\bibinfo {author} {\bibfnamefont {M.}~\bibnamefont
  {Boissonneault}}, \bibinfo {author} {\bibfnamefont {J.~M.}\ \bibnamefont
  {Gambetta}}, \ and\ \bibinfo {author} {\bibfnamefont {A.}~\bibnamefont
  {Blais}},\ }\href {\doibase 10.1103/PhysRevA.79.013819} {\bibfield  {journal}
  {\bibinfo  {journal} {Phys. Rev. A}\ }\textbf {\bibinfo {volume} {79}},\
  \bibinfo {pages} {013819} (\bibinfo {year} {2009})}\BibitemShut {NoStop}%
\bibitem [{Note7()}]{Note7}%
  \BibitemOpen
  \bibinfo {note} {Note that pure dephasing of a transition $j$ in the bare
  double-quantum-dot eigenbasis also enables probe photons to induce
  transitions in the basis dressed by the resonator~\cite
  {boissonneault2008,boissonneault2009,slichter2012}. While the associated
  transition rate is also suppressed in the ratio $\left <n\right > /n_{c,j}$,
  it can in principle be made larger than the relaxation rate of the transition
  if the fluctuations that induce the dephasing have a large spectral weight at
  the frequencies $\pm |\omega _r - E_j|$ (typically several MHz in the present
  work). For low frequency charge and nuclear noise typical of spin-qubit
  environments, these contributions are likely to be small. A detailed analysis
  of these processes is beyond the scope of this work and we thus ignore the
  effect of pure dephasing throughout.}\BibitemShut {Stop}%
\bibitem [{\citenamefont {Borjans}\ \emph
  {et~al.}(2019{\natexlab{b}})\citenamefont {Borjans}, \citenamefont {Zajac},
  \citenamefont {Hazard},\ and\ \citenamefont {Petta}}]{borjans2019}%
  \BibitemOpen
  \bibfield  {author} {\bibinfo {author} {\bibfnamefont {F.}~\bibnamefont
  {Borjans}}, \bibinfo {author} {\bibfnamefont {D.~M.}\ \bibnamefont {Zajac}},
  \bibinfo {author} {\bibfnamefont {T.~M.}\ \bibnamefont {Hazard}}, \ and\
  \bibinfo {author} {\bibfnamefont {J.~R.}\ \bibnamefont {Petta}},\ }\href
  {\doibase 10.1103/PhysRevApplied.11.044063} {\bibfield  {journal} {\bibinfo
  {journal} {Phys. Rev. Appl.}\ }\textbf {\bibinfo {volume} {11}},\ \bibinfo
  {pages} {044063} (\bibinfo {year} {2019}{\natexlab{b}})}\BibitemShut
  {NoStop}%
\bibitem [{\citenamefont {Srinivasa}\ \emph {et~al.}(2013)\citenamefont
  {Srinivasa}, \citenamefont {Nowack}, \citenamefont {Shafiei}, \citenamefont
  {Vandersypen},\ and\ \citenamefont {Taylor}}]{srinivasa2013}%
  \BibitemOpen
  \bibfield  {author} {\bibinfo {author} {\bibfnamefont {V.}~\bibnamefont
  {Srinivasa}}, \bibinfo {author} {\bibfnamefont {K.~C.}\ \bibnamefont
  {Nowack}}, \bibinfo {author} {\bibfnamefont {M.}~\bibnamefont {Shafiei}},
  \bibinfo {author} {\bibfnamefont {L.~M.~K.}\ \bibnamefont {Vandersypen}}, \
  and\ \bibinfo {author} {\bibfnamefont {J.~M.}\ \bibnamefont {Taylor}},\
  }\href {\doibase 10.1103/PhysRevLett.110.196803} {\bibfield  {journal}
  {\bibinfo  {journal} {Phys. Rev. Lett.}\ }\textbf {\bibinfo {volume} {110}},\
  \bibinfo {pages} {196803} (\bibinfo {year} {2013})}\BibitemShut {NoStop}%
\bibitem [{\citenamefont {Gambetta}\ \emph {et~al.}(2007)\citenamefont
  {Gambetta}, \citenamefont {Braff}, \citenamefont {Wallraff}, \citenamefont
  {Girvin},\ and\ \citenamefont {Schoelkopf}}]{gambetta2007}%
  \BibitemOpen
  \bibfield  {author} {\bibinfo {author} {\bibfnamefont {J.}~\bibnamefont
  {Gambetta}}, \bibinfo {author} {\bibfnamefont {W.~A.}\ \bibnamefont {Braff}},
  \bibinfo {author} {\bibfnamefont {A.}~\bibnamefont {Wallraff}}, \bibinfo
  {author} {\bibfnamefont {S.~M.}\ \bibnamefont {Girvin}}, \ and\ \bibinfo
  {author} {\bibfnamefont {R.~J.}\ \bibnamefont {Schoelkopf}},\ }\href
  {\doibase 10.1103/PhysRevA.76.012325} {\bibfield  {journal} {\bibinfo
  {journal} {Phys. Rev. A}\ }\textbf {\bibinfo {volume} {76}},\ \bibinfo
  {pages} {012325} (\bibinfo {year} {2007})}\BibitemShut {NoStop}%
\bibitem [{\citenamefont {Gambetta}\ \emph {et~al.}(2008)\citenamefont
  {Gambetta}, \citenamefont {Blais}, \citenamefont {Boissonneault},
  \citenamefont {Houck}, \citenamefont {Schuster},\ and\ \citenamefont
  {Girvin}}]{gambetta2008}%
  \BibitemOpen
  \bibfield  {author} {\bibinfo {author} {\bibfnamefont {J.}~\bibnamefont
  {Gambetta}}, \bibinfo {author} {\bibfnamefont {A.}~\bibnamefont {Blais}},
  \bibinfo {author} {\bibfnamefont {M.}~\bibnamefont {Boissonneault}}, \bibinfo
  {author} {\bibfnamefont {A.~A.}\ \bibnamefont {Houck}}, \bibinfo {author}
  {\bibfnamefont {D.~I.}\ \bibnamefont {Schuster}}, \ and\ \bibinfo {author}
  {\bibfnamefont {S.~M.}\ \bibnamefont {Girvin}},\ }\href {\doibase
  10.1103/PhysRevA.77.012112} {\bibfield  {journal} {\bibinfo  {journal} {Phys.
  Rev. A}\ }\textbf {\bibinfo {volume} {77}},\ \bibinfo {pages} {012112}
  (\bibinfo {year} {2008})}\BibitemShut {NoStop}%
\bibitem [{\citenamefont {D'Anjou}\ and\ \citenamefont
  {Coish}(2014)}]{danjou2014}%
  \BibitemOpen
  \bibfield  {author} {\bibinfo {author} {\bibfnamefont {B.}~\bibnamefont
  {D'Anjou}}\ and\ \bibinfo {author} {\bibfnamefont {W.~A.}\ \bibnamefont
  {Coish}},\ }\href {\doibase 10.1103/PhysRevA.89.012313} {\bibfield  {journal}
  {\bibinfo  {journal} {Phys. Rev. A}\ }\textbf {\bibinfo {volume} {89}},\
  \bibinfo {pages} {012313} (\bibinfo {year} {2014})}\BibitemShut {NoStop}%
\bibitem [{\citenamefont {D'Anjou}\ and\ \citenamefont
  {Coish}(2017)}]{danjou2017-2}%
  \BibitemOpen
  \bibfield  {author} {\bibinfo {author} {\bibfnamefont {B.}~\bibnamefont
  {D'Anjou}}\ and\ \bibinfo {author} {\bibfnamefont {W.~A.}\ \bibnamefont
  {Coish}},\ }\href {\doibase 10.1103/PhysRevA.96.052321} {\bibfield  {journal}
  {\bibinfo  {journal} {Phys. Rev. A}\ }\textbf {\bibinfo {volume} {96}},\
  \bibinfo {pages} {052321} (\bibinfo {year} {2017})}\BibitemShut {NoStop}%
\bibitem [{\citenamefont {Benito}\ \emph
  {et~al.}(2019{\natexlab{b}})\citenamefont {Benito}, \citenamefont {Petta},\
  and\ \citenamefont {Burkard}}]{benito2019}%
  \BibitemOpen
  \bibfield  {author} {\bibinfo {author} {\bibfnamefont {M.}~\bibnamefont
  {Benito}}, \bibinfo {author} {\bibfnamefont {J.~R.}\ \bibnamefont {Petta}}, \
  and\ \bibinfo {author} {\bibfnamefont {G.}~\bibnamefont {Burkard}},\ }\href
  {\doibase 10.1103/PhysRevB.100.081412} {\bibfield  {journal} {\bibinfo
  {journal} {Phys. Rev. B}\ }\textbf {\bibinfo {volume} {100}},\ \bibinfo
  {pages} {081412(R)} (\bibinfo {year} {2019}{\natexlab{b}})}\BibitemShut
  {NoStop}%
\bibitem [{\citenamefont {Kay}(1998)}]{kay1998}%
  \BibitemOpen
  \bibfield  {author} {\bibinfo {author} {\bibfnamefont {S.~M.}\ \bibnamefont
  {Kay}},\ }\href {http://www.worldcat.org/oclc/223145408} {\emph {\bibinfo
  {title} {Fundamentals of Statistical Signal Processing, Vol. II: {Detection}
  Theory}}},\ Vol.~\bibinfo {volume} {II}\ (\bibinfo  {publisher} {Prentice
  Hall},\ \bibinfo {address} {New Jersey, U.S.A.},\ \bibinfo {year}
  {1998})\BibitemShut {NoStop}%
\bibitem [{\citenamefont {Ryan}\ \emph {et~al.}(2015)\citenamefont {Ryan},
  \citenamefont {Johnson}, \citenamefont {Gambetta}, \citenamefont {Chow},
  \citenamefont {da~Silva}, \citenamefont {Dial},\ and\ \citenamefont
  {Ohki}}]{ryan2015}%
  \BibitemOpen
  \bibfield  {author} {\bibinfo {author} {\bibfnamefont {C.~A.}\ \bibnamefont
  {Ryan}}, \bibinfo {author} {\bibfnamefont {B.~R.}\ \bibnamefont {Johnson}},
  \bibinfo {author} {\bibfnamefont {J.~M.}\ \bibnamefont {Gambetta}}, \bibinfo
  {author} {\bibfnamefont {J.~M.}\ \bibnamefont {Chow}}, \bibinfo {author}
  {\bibfnamefont {M.~P.}\ \bibnamefont {da~Silva}}, \bibinfo {author}
  {\bibfnamefont {O.~E.}\ \bibnamefont {Dial}}, \ and\ \bibinfo {author}
  {\bibfnamefont {T.~A.}\ \bibnamefont {Ohki}},\ }\href {\doibase
  10.1103/PhysRevA.91.022118} {\bibfield  {journal} {\bibinfo  {journal} {Phys.
  Rev. A}\ }\textbf {\bibinfo {volume} {91}},\ \bibinfo {pages} {022118}
  (\bibinfo {year} {2015})}\BibitemShut {NoStop}%
\bibitem [{\citenamefont {Reed}\ \emph {et~al.}(2016)\citenamefont {Reed},
  \citenamefont {Maune}, \citenamefont {Andrews}, \citenamefont {Borselli},
  \citenamefont {Eng}, \citenamefont {Jura}, \citenamefont {Kiselev},
  \citenamefont {Ladd}, \citenamefont {Merkel}, \citenamefont {Milosavljevic},
  \citenamefont {Pritchett}, \citenamefont {Rakher}, \citenamefont {Ross},
  \citenamefont {Schmitz}, \citenamefont {Smith}, \citenamefont {Wright},
  \citenamefont {Gyure},\ and\ \citenamefont {Hunter}}]{reed2016}%
  \BibitemOpen
  \bibfield  {author} {\bibinfo {author} {\bibfnamefont {M.~D.}\ \bibnamefont
  {Reed}}, \bibinfo {author} {\bibfnamefont {B.~M.}\ \bibnamefont {Maune}},
  \bibinfo {author} {\bibfnamefont {R.~W.}\ \bibnamefont {Andrews}}, \bibinfo
  {author} {\bibfnamefont {M.~G.}\ \bibnamefont {Borselli}}, \bibinfo {author}
  {\bibfnamefont {K.}~\bibnamefont {Eng}}, \bibinfo {author} {\bibfnamefont
  {M.~P.}\ \bibnamefont {Jura}}, \bibinfo {author} {\bibfnamefont {A.~A.}\
  \bibnamefont {Kiselev}}, \bibinfo {author} {\bibfnamefont {T.~D.}\
  \bibnamefont {Ladd}}, \bibinfo {author} {\bibfnamefont {S.~T.}\ \bibnamefont
  {Merkel}}, \bibinfo {author} {\bibfnamefont {I.}~\bibnamefont
  {Milosavljevic}}, \bibinfo {author} {\bibfnamefont {E.~J.}\ \bibnamefont
  {Pritchett}}, \bibinfo {author} {\bibfnamefont {M.~T.}\ \bibnamefont
  {Rakher}}, \bibinfo {author} {\bibfnamefont {R.~S.}\ \bibnamefont {Ross}},
  \bibinfo {author} {\bibfnamefont {A.~E.}\ \bibnamefont {Schmitz}}, \bibinfo
  {author} {\bibfnamefont {A.}~\bibnamefont {Smith}}, \bibinfo {author}
  {\bibfnamefont {J.~A.}\ \bibnamefont {Wright}}, \bibinfo {author}
  {\bibfnamefont {M.~F.}\ \bibnamefont {Gyure}}, \ and\ \bibinfo {author}
  {\bibfnamefont {A.~T.}\ \bibnamefont {Hunter}},\ }\href {\doibase
  10.1103/PhysRevLett.116.110402} {\bibfield  {journal} {\bibinfo  {journal}
  {Phys. Rev. Lett.}\ }\textbf {\bibinfo {volume} {116}},\ \bibinfo {pages}
  {110402} (\bibinfo {year} {2016})}\BibitemShut {NoStop}%
\bibitem [{\citenamefont {Martins}\ \emph {et~al.}(2016)\citenamefont
  {Martins}, \citenamefont {Malinowski}, \citenamefont {Nissen}, \citenamefont
  {Barnes}, \citenamefont {Fallahi}, \citenamefont {Gardner}, \citenamefont
  {Manfra}, \citenamefont {Marcus},\ and\ \citenamefont
  {Kuemmeth}}]{martins2016}%
  \BibitemOpen
  \bibfield  {author} {\bibinfo {author} {\bibfnamefont {F.}~\bibnamefont
  {Martins}}, \bibinfo {author} {\bibfnamefont {F.~K.}\ \bibnamefont
  {Malinowski}}, \bibinfo {author} {\bibfnamefont {P.~D.}\ \bibnamefont
  {Nissen}}, \bibinfo {author} {\bibfnamefont {E.}~\bibnamefont {Barnes}},
  \bibinfo {author} {\bibfnamefont {S.}~\bibnamefont {Fallahi}}, \bibinfo
  {author} {\bibfnamefont {G.~C.}\ \bibnamefont {Gardner}}, \bibinfo {author}
  {\bibfnamefont {M.~J.}\ \bibnamefont {Manfra}}, \bibinfo {author}
  {\bibfnamefont {C.~M.}\ \bibnamefont {Marcus}}, \ and\ \bibinfo {author}
  {\bibfnamefont {F.}~\bibnamefont {Kuemmeth}},\ }\href {\doibase
  10.1103/PhysRevLett.116.116801} {\bibfield  {journal} {\bibinfo  {journal}
  {Phys. Rev. Lett.}\ }\textbf {\bibinfo {volume} {116}},\ \bibinfo {pages}
  {116801} (\bibinfo {year} {2016})}\BibitemShut {NoStop}%
\bibitem [{\citenamefont {Clerk}\ \emph {et~al.}(2010)\citenamefont {Clerk},
  \citenamefont {Devoret}, \citenamefont {Girvin}, \citenamefont {Marquardt},\
  and\ \citenamefont {Schoelkopf}}]{clerk2010}%
  \BibitemOpen
  \bibfield  {author} {\bibinfo {author} {\bibfnamefont {A.~A.}\ \bibnamefont
  {Clerk}}, \bibinfo {author} {\bibfnamefont {M.~H.}\ \bibnamefont {Devoret}},
  \bibinfo {author} {\bibfnamefont {S.~M.}\ \bibnamefont {Girvin}}, \bibinfo
  {author} {\bibfnamefont {F.}~\bibnamefont {Marquardt}}, \ and\ \bibinfo
  {author} {\bibfnamefont {R.~J.}\ \bibnamefont {Schoelkopf}},\ }\href
  {\doibase 10.1103/RevModPhys.82.1155} {\bibfield  {journal} {\bibinfo
  {journal} {Rev. Mod. Phys.}\ }\textbf {\bibinfo {volume} {82}},\ \bibinfo
  {pages} {1155} (\bibinfo {year} {2010})}\BibitemShut {NoStop}%
\bibitem [{\citenamefont {Mutus}\ \emph {et~al.}(2014)\citenamefont {Mutus},
  \citenamefont {White}, \citenamefont {Barends}, \citenamefont {Chen},
  \citenamefont {Chen}, \citenamefont {Chiaro}, \citenamefont {Dunsworth},
  \citenamefont {Jeffrey}, \citenamefont {Kelly}, \citenamefont {Megrant},
  \citenamefont {Neill}, \citenamefont {O'Malley}, \citenamefont {Rousha},
  \citenamefont {Sank}, \citenamefont {Vainsencher}, \citenamefont {Wenner},
  \citenamefont {Sundqvist}, \citenamefont {Cleland},\ and\ \citenamefont
  {Martinis}}]{mutus2014}%
  \BibitemOpen
  \bibfield  {author} {\bibinfo {author} {\bibfnamefont {J.~Y.}\ \bibnamefont
  {Mutus}}, \bibinfo {author} {\bibfnamefont {T.~C.}\ \bibnamefont {White}},
  \bibinfo {author} {\bibfnamefont {R.}~\bibnamefont {Barends}}, \bibinfo
  {author} {\bibfnamefont {Y.}~\bibnamefont {Chen}}, \bibinfo {author}
  {\bibfnamefont {Z.}~\bibnamefont {Chen}}, \bibinfo {author} {\bibfnamefont
  {B.}~\bibnamefont {Chiaro}}, \bibinfo {author} {\bibfnamefont
  {A.}~\bibnamefont {Dunsworth}}, \bibinfo {author} {\bibfnamefont
  {E.}~\bibnamefont {Jeffrey}}, \bibinfo {author} {\bibfnamefont
  {J.}~\bibnamefont {Kelly}}, \bibinfo {author} {\bibfnamefont
  {A.}~\bibnamefont {Megrant}}, \bibinfo {author} {\bibfnamefont
  {C.}~\bibnamefont {Neill}}, \bibinfo {author} {\bibfnamefont {P.~J.~J.}\
  \bibnamefont {O'Malley}}, \bibinfo {author} {\bibfnamefont {P.}~\bibnamefont
  {Rousha}}, \bibinfo {author} {\bibfnamefont {D.}~\bibnamefont {Sank}},
  \bibinfo {author} {\bibfnamefont {A.}~\bibnamefont {Vainsencher}}, \bibinfo
  {author} {\bibfnamefont {J.}~\bibnamefont {Wenner}}, \bibinfo {author}
  {\bibfnamefont {K.~M.}\ \bibnamefont {Sundqvist}}, \bibinfo {author}
  {\bibfnamefont {A.~N.}\ \bibnamefont {Cleland}}, \ and\ \bibinfo {author}
  {\bibfnamefont {J.~M.}\ \bibnamefont {Martinis}},\ }\href {\doibase
  10.1063/1.4886408} {\bibfield  {journal} {\bibinfo  {journal} {App. Phys.
  Lett.}\ }\textbf {\bibinfo {volume} {104}},\ \bibinfo {pages} {263513}
  (\bibinfo {year} {2014})}\BibitemShut {NoStop}%
\bibitem [{\citenamefont {Roy}\ \emph {et~al.}(2015)\citenamefont {Roy},
  \citenamefont {Kundu}, \citenamefont {Chand}, \citenamefont {Vadiraj},
  \citenamefont {Ranadive}, \citenamefont {Nehra}, \citenamefont {Patankar},
  \citenamefont {Aumentado}, \citenamefont {Clerk},\ and\ \citenamefont
  {Vijay}}]{roy2015}%
  \BibitemOpen
  \bibfield  {author} {\bibinfo {author} {\bibfnamefont {T.}~\bibnamefont
  {Roy}}, \bibinfo {author} {\bibfnamefont {S.}~\bibnamefont {Kundu}}, \bibinfo
  {author} {\bibfnamefont {M.}~\bibnamefont {Chand}}, \bibinfo {author}
  {\bibfnamefont {A.~M.}\ \bibnamefont {Vadiraj}}, \bibinfo {author}
  {\bibfnamefont {A.}~\bibnamefont {Ranadive}}, \bibinfo {author}
  {\bibfnamefont {N.}~\bibnamefont {Nehra}}, \bibinfo {author} {\bibfnamefont
  {M.~P.}\ \bibnamefont {Patankar}}, \bibinfo {author} {\bibfnamefont
  {J.}~\bibnamefont {Aumentado}}, \bibinfo {author} {\bibfnamefont {A.~A.}\
  \bibnamefont {Clerk}}, \ and\ \bibinfo {author} {\bibfnamefont
  {R.}~\bibnamefont {Vijay}},\ }\href {\doibase 10.1063/1.4939148} {\bibfield
  {journal} {\bibinfo  {journal} {App. Phys. Lett.}\ }\textbf {\bibinfo
  {volume} {107}},\ \bibinfo {pages} {262601} (\bibinfo {year}
  {2015})}\BibitemShut {NoStop}%
\bibitem [{\citenamefont {Sete}\ \emph {et~al.}(2015)\citenamefont {Sete},
  \citenamefont {Martinis},\ and\ \citenamefont {Korotkov}}]{sete2015}%
  \BibitemOpen
  \bibfield  {author} {\bibinfo {author} {\bibfnamefont {E.~A.}\ \bibnamefont
  {Sete}}, \bibinfo {author} {\bibfnamefont {J.~M.}\ \bibnamefont {Martinis}},
  \ and\ \bibinfo {author} {\bibfnamefont {A.~N.}\ \bibnamefont {Korotkov}},\
  }\href {\doibase 10.1103/PhysRevA.92.012325} {\bibfield  {journal} {\bibinfo
  {journal} {Phys. Rev. A}\ }\textbf {\bibinfo {volume} {92}},\ \bibinfo
  {pages} {012325} (\bibinfo {year} {2015})}\BibitemShut {NoStop}%
\bibitem [{\citenamefont {Cleland}\ \emph {et~al.}(2019)\citenamefont
  {Cleland}, \citenamefont {Pechal}, \citenamefont {Stas}, \citenamefont
  {Sarabalis},\ and\ \citenamefont {Savafi-Naeini}}]{cleland2019}%
  \BibitemOpen
  \bibfield  {author} {\bibinfo {author} {\bibfnamefont {A.~Y.}\ \bibnamefont
  {Cleland}}, \bibinfo {author} {\bibfnamefont {M.}~\bibnamefont {Pechal}},
  \bibinfo {author} {\bibfnamefont {P.-J.~C.}\ \bibnamefont {Stas}}, \bibinfo
  {author} {\bibfnamefont {C.~J.}\ \bibnamefont {Sarabalis}}, \ and\ \bibinfo
  {author} {\bibfnamefont {A.~H.}\ \bibnamefont {Savafi-Naeini}},\ }\href
  {https://arxiv.org/abs/1905.08403} {\bibfield  {journal} {\bibinfo  {journal}
  {arXiv:1905.08403}\ } (\bibinfo {year} {2019})}\BibitemShut {NoStop}%
\bibitem [{\citenamefont {Sete}\ \emph {et~al.}(2013)\citenamefont {Sete},
  \citenamefont {Galiautdinov}, \citenamefont {Mlinar}, \citenamefont
  {Martinis},\ and\ \citenamefont {Korotkov}}]{sete2013}%
  \BibitemOpen
  \bibfield  {author} {\bibinfo {author} {\bibfnamefont {E.~A.}\ \bibnamefont
  {Sete}}, \bibinfo {author} {\bibfnamefont {A.}~\bibnamefont {Galiautdinov}},
  \bibinfo {author} {\bibfnamefont {E.}~\bibnamefont {Mlinar}}, \bibinfo
  {author} {\bibfnamefont {J.~M.}\ \bibnamefont {Martinis}}, \ and\ \bibinfo
  {author} {\bibfnamefont {A.~N.}\ \bibnamefont {Korotkov}},\ }\href {\doibase
  10.1103/PhysRevLett.110.210501} {\bibfield  {journal} {\bibinfo  {journal}
  {Phys. Rev. Lett.}\ }\textbf {\bibinfo {volume} {110}},\ \bibinfo {pages}
  {210501} (\bibinfo {year} {2013})}\BibitemShut {NoStop}%
\bibitem [{\citenamefont {Gard}\ \emph {et~al.}(2018)\citenamefont {Gard},
  \citenamefont {Jacobs}, \citenamefont {Aumentado},\ and\ \citenamefont
  {Simmonds}}]{gard2018}%
  \BibitemOpen
  \bibfield  {author} {\bibinfo {author} {\bibfnamefont {B.~T.}\ \bibnamefont
  {Gard}}, \bibinfo {author} {\bibfnamefont {K.}~\bibnamefont {Jacobs}},
  \bibinfo {author} {\bibfnamefont {J.}~\bibnamefont {Aumentado}}, \ and\
  \bibinfo {author} {\bibfnamefont {R.~W.}\ \bibnamefont {Simmonds}},\ }\href
  {https://arxiv.org/abs/1809.02597} {\bibfield  {journal} {\bibinfo  {journal}
  {arXiv:1809.02597}\ } (\bibinfo {year} {2018})}\BibitemShut {NoStop}%
\bibitem [{\citenamefont {Wang}\ \emph {et~al.}(2018)\citenamefont {Wang},
  \citenamefont {Miranowicz},\ and\ \citenamefont {Nori}}]{wang2018}%
  \BibitemOpen
  \bibfield  {author} {\bibinfo {author} {\bibfnamefont {X.}~\bibnamefont
  {Wang}}, \bibinfo {author} {\bibfnamefont {A.}~\bibnamefont {Miranowicz}}, \
  and\ \bibinfo {author} {\bibfnamefont {F.}~\bibnamefont {Nori}},\ }\href
  {https://arxiv.org/abs/1811.09048} {\bibfield  {journal} {\bibinfo  {journal}
  {arXiv:1811.09048}\ } (\bibinfo {year} {2018})}\BibitemShut {NoStop}%
\bibitem [{\citenamefont {Harrington}\ \emph {et~al.}(2019)\citenamefont
  {Harrington}, \citenamefont {Naghiloo}, \citenamefont {Tan},\ and\
  \citenamefont {Murch}}]{harrington2019}%
  \BibitemOpen
  \bibfield  {author} {\bibinfo {author} {\bibfnamefont {P.~M.}\ \bibnamefont
  {Harrington}}, \bibinfo {author} {\bibfnamefont {M.}~\bibnamefont
  {Naghiloo}}, \bibinfo {author} {\bibfnamefont {D.}~\bibnamefont {Tan}}, \
  and\ \bibinfo {author} {\bibfnamefont {K.~W.}\ \bibnamefont {Murch}},\ }\href
  {\doibase 0.1103/PhysRevA.99.052126} {\bibfield  {journal} {\bibinfo
  {journal} {Phys. Rev. A}\ }\textbf {\bibinfo {volume} {99}},\ \bibinfo
  {pages} {052126} (\bibinfo {year} {2019})}\BibitemShut {NoStop}%
\bibitem [{\citenamefont {Peronnin}\ \emph {et~al.}(2019)\citenamefont
  {Peronnin}, \citenamefont {Markovi{\'c}}, \citenamefont {Ficheux},\ and\
  \citenamefont {Huard}}]{peronnin2019}%
  \BibitemOpen
  \bibfield  {author} {\bibinfo {author} {\bibfnamefont {T.}~\bibnamefont
  {Peronnin}}, \bibinfo {author} {\bibfnamefont {D.}~\bibnamefont
  {Markovi{\'c}}}, \bibinfo {author} {\bibfnamefont {Q.}~\bibnamefont
  {Ficheux}}, \ and\ \bibinfo {author} {\bibfnamefont {B.}~\bibnamefont
  {Huard}},\ }\href {https://arxiv.org/abs/1904.04635} {\bibfield  {journal}
  {\bibinfo  {journal} {arXiv:1904.04635}\ } (\bibinfo {year}
  {2019})}\BibitemShut {NoStop}%
\bibitem [{\citenamefont {Dassonneville}\ \emph {et~al.}(2019)\citenamefont
  {Dassonneville}, \citenamefont {Ramos}, \citenamefont {Milchakov},
  \citenamefont {Planat}, \citenamefont {Dumur}, \citenamefont {Foroughi},
  \citenamefont {Puertas}, \citenamefont {Leger}, \citenamefont {Bharadwaj},
  \citenamefont {Delaforce}, \citenamefont {Rafsanjani}, \citenamefont {Naud},
  \citenamefont {Hasch-Guichard}, \citenamefont {Garc{\' i}a-Ripoll},
  \citenamefont {Roch},\ and\ \citenamefont {Buisson}}]{dassonneville2019}%
  \BibitemOpen
  \bibfield  {author} {\bibinfo {author} {\bibfnamefont {R.}~\bibnamefont
  {Dassonneville}}, \bibinfo {author} {\bibfnamefont {T.}~\bibnamefont
  {Ramos}}, \bibinfo {author} {\bibfnamefont {V.}~\bibnamefont {Milchakov}},
  \bibinfo {author} {\bibfnamefont {L.}~\bibnamefont {Planat}}, \bibinfo
  {author} {\bibfnamefont {{\' E}.}~\bibnamefont {Dumur}}, \bibinfo {author}
  {\bibfnamefont {F.}~\bibnamefont {Foroughi}}, \bibinfo {author}
  {\bibfnamefont {J.}~\bibnamefont {Puertas}}, \bibinfo {author} {\bibfnamefont
  {S.}~\bibnamefont {Leger}}, \bibinfo {author} {\bibfnamefont
  {K.}~\bibnamefont {Bharadwaj}}, \bibinfo {author} {\bibfnamefont
  {J.}~\bibnamefont {Delaforce}}, \bibinfo {author} {\bibfnamefont
  {K.}~\bibnamefont {Rafsanjani}}, \bibinfo {author} {\bibfnamefont
  {C.}~\bibnamefont {Naud}}, \bibinfo {author} {\bibfnamefont {W.}~\bibnamefont
  {Hasch-Guichard}}, \bibinfo {author} {\bibfnamefont {J.~J.}\ \bibnamefont
  {Garc{\' i}a-Ripoll}}, \bibinfo {author} {\bibfnamefont {N.}~\bibnamefont
  {Roch}}, \ and\ \bibinfo {author} {\bibfnamefont {O.}~\bibnamefont
  {Buisson}},\ }\href {https://arxiv.org/abs/1905.00271} {\bibfield  {journal}
  {\bibinfo  {journal} {arXiv:1905.00271}\ } (\bibinfo {year}
  {2019})}\BibitemShut {NoStop}%
\bibitem [{\citenamefont {Agarwal}\ \emph {et~al.}(2013)\citenamefont
  {Agarwal}, \citenamefont {Martin}, \citenamefont {Lukin},\ and\ \citenamefont
  {Demler}}]{agarwal2013}%
  \BibitemOpen
  \bibfield  {author} {\bibinfo {author} {\bibfnamefont {K.}~\bibnamefont
  {Agarwal}}, \bibinfo {author} {\bibfnamefont {I.}~\bibnamefont {Martin}},
  \bibinfo {author} {\bibfnamefont {M.~D.}\ \bibnamefont {Lukin}}, \ and\
  \bibinfo {author} {\bibfnamefont {E.}~\bibnamefont {Demler}},\ }\href
  {\doibase 10.1103/PhysRevB.87.144201} {\bibfield  {journal} {\bibinfo
  {journal} {Phys. Rev. B}\ }\textbf {\bibinfo {volume} {87}},\ \bibinfo
  {pages} {144201} (\bibinfo {year} {2013})}\BibitemShut {NoStop}%
\bibitem [{\citenamefont {Rosen}\ \emph {et~al.}(2019)\citenamefont {Rosen},
  \citenamefont {Horsley}, \citenamefont {Harrison}, \citenamefont {Holland},
  \citenamefont {Chang}, \citenamefont {Bond},\ and\ \citenamefont
  {DuBois}}]{rosen2019}%
  \BibitemOpen
  \bibfield  {author} {\bibinfo {author} {\bibfnamefont {Y.~J.}\ \bibnamefont
  {Rosen}}, \bibinfo {author} {\bibfnamefont {M.~A.}\ \bibnamefont {Horsley}},
  \bibinfo {author} {\bibfnamefont {S.~E.}\ \bibnamefont {Harrison}}, \bibinfo
  {author} {\bibfnamefont {E.~T.}\ \bibnamefont {Holland}}, \bibinfo {author}
  {\bibfnamefont {A.~S.}\ \bibnamefont {Chang}}, \bibinfo {author}
  {\bibfnamefont {T.}~\bibnamefont {Bond}}, \ and\ \bibinfo {author}
  {\bibfnamefont {J.~L.}\ \bibnamefont {DuBois}},\ }\href {\doibase
  10.1063/1.5096182} {\bibfield  {journal} {\bibinfo  {journal} {App. Phys.
  Lett.}\ }\textbf {\bibinfo {volume} {114}},\ \bibinfo {pages} {202601}
  (\bibinfo {year} {2019})}\BibitemShut {NoStop}%
\bibitem [{\citenamefont {Castellanos-Beltran}\ \emph
  {et~al.}(2008)\citenamefont {Castellanos-Beltran}, \citenamefont {Irwin},
  \citenamefont {Hilton}, \citenamefont {Vale},\ and\ \citenamefont
  {Lehnert}}]{castellanosbeltran2008}%
  \BibitemOpen
  \bibfield  {author} {\bibinfo {author} {\bibfnamefont {M.~A.}\ \bibnamefont
  {Castellanos-Beltran}}, \bibinfo {author} {\bibfnamefont {K.~D.}\
  \bibnamefont {Irwin}}, \bibinfo {author} {\bibfnamefont {G.~C.}\ \bibnamefont
  {Hilton}}, \bibinfo {author} {\bibfnamefont {L.~R.}\ \bibnamefont {Vale}}, \
  and\ \bibinfo {author} {\bibfnamefont {K.~W.}\ \bibnamefont {Lehnert}},\
  }\href {\doibase 10.1038/nphys1090} {\bibfield  {journal} {\bibinfo
  {journal} {Nat. Phys.}\ }\textbf {\bibinfo {volume} {4}},\ \bibinfo {pages}
  {929} (\bibinfo {year} {2008})}\BibitemShut {NoStop}%
\bibitem [{\citenamefont {Eddins}\ \emph {et~al.}(2019)\citenamefont {Eddins},
  \citenamefont {Kreikebaum}, \citenamefont {Toyli}, \citenamefont
  {Levenson-Falk}, \citenamefont {Dove}, \citenamefont {Livingston},
  \citenamefont {Levitan}, \citenamefont {Govia}, \citenamefont {Clerk},\ and\
  \citenamefont {Siddiqi}}]{eddins2019}%
  \BibitemOpen
  \bibfield  {author} {\bibinfo {author} {\bibfnamefont {A.}~\bibnamefont
  {Eddins}}, \bibinfo {author} {\bibfnamefont {J.~M.}\ \bibnamefont
  {Kreikebaum}}, \bibinfo {author} {\bibfnamefont {D.~M.}\ \bibnamefont
  {Toyli}}, \bibinfo {author} {\bibfnamefont {E.~M.}\ \bibnamefont
  {Levenson-Falk}}, \bibinfo {author} {\bibfnamefont {A.}~\bibnamefont {Dove}},
  \bibinfo {author} {\bibfnamefont {W.~P.}\ \bibnamefont {Livingston}},
  \bibinfo {author} {\bibfnamefont {B.~A.}\ \bibnamefont {Levitan}}, \bibinfo
  {author} {\bibfnamefont {L.~C.~G.}\ \bibnamefont {Govia}}, \bibinfo {author}
  {\bibfnamefont {A.~A.}\ \bibnamefont {Clerk}}, \ and\ \bibinfo {author}
  {\bibfnamefont {I.}~\bibnamefont {Siddiqi}},\ }\href {\doibase
  10.1103/PhysRevX.9.011004} {\bibfield  {journal} {\bibinfo  {journal} {Phys.
  Rev. X}\ }\textbf {\bibinfo {volume} {9}},\ \bibinfo {pages} {011004}
  (\bibinfo {year} {2019})}\BibitemShut {NoStop}%
\bibitem [{\citenamefont {Motzoi}\ \emph {et~al.}(2018)\citenamefont {Motzoi},
  \citenamefont {Buchmann},\ and\ \citenamefont {Dickel}}]{motzoi2018}%
  \BibitemOpen
  \bibfield  {author} {\bibinfo {author} {\bibfnamefont {F.}~\bibnamefont
  {Motzoi}}, \bibinfo {author} {\bibfnamefont {L.}~\bibnamefont {Buchmann}}, \
  and\ \bibinfo {author} {\bibfnamefont {C.}~\bibnamefont {Dickel}},\ }\href
  {https://arxiv.org/abs/1809.04116} {\bibfield  {journal} {\bibinfo  {journal}
  {arXiv:1809.04116}\ } (\bibinfo {year} {2018})}\BibitemShut {NoStop}%
\bibitem [{\citenamefont {Hagmann}(2005)}]{hagmann2005}%
  \BibitemOpen
  \bibfield  {author} {\bibinfo {author} {\bibfnamefont {M.~J.}\ \bibnamefont
  {Hagmann}},\ }\href {\doibase 10.1109/TNANO.2004.842040} {\bibfield
  {journal} {\bibinfo  {journal} {IEEE Trans. Nanotechnol.}\ }\textbf {\bibinfo
  {volume} {4}},\ \bibinfo {pages} {289} (\bibinfo {year} {2005})}\BibitemShut
  {NoStop}%
\bibitem [{\citenamefont {Altimiras}\ \emph {et~al.}(2013)\citenamefont
  {Altimiras}, \citenamefont {Parlavecchio}, \citenamefont {Joyez},
  \citenamefont {Vion}, \citenamefont {Roche}, \citenamefont {Esteve},\ and\
  \citenamefont {Portier}}]{altimiras2013}%
  \BibitemOpen
  \bibfield  {author} {\bibinfo {author} {\bibfnamefont {C.}~\bibnamefont
  {Altimiras}}, \bibinfo {author} {\bibfnamefont {O.}~\bibnamefont
  {Parlavecchio}}, \bibinfo {author} {\bibfnamefont {P.}~\bibnamefont {Joyez}},
  \bibinfo {author} {\bibfnamefont {D.}~\bibnamefont {Vion}}, \bibinfo {author}
  {\bibfnamefont {P.}~\bibnamefont {Roche}}, \bibinfo {author} {\bibfnamefont
  {D.}~\bibnamefont {Esteve}}, \ and\ \bibinfo {author} {\bibfnamefont
  {F.}~\bibnamefont {Portier}},\ }\href {\doibase 10.1063/1.4832074} {\bibfield
   {journal} {\bibinfo  {journal} {App. Phys. Lett.}\ }\textbf {\bibinfo
  {volume} {103}},\ \bibinfo {pages} {212601} (\bibinfo {year}
  {2013})}\BibitemShut {NoStop}%
\bibitem [{\citenamefont {Stockklauser}\ \emph {et~al.}(2017)\citenamefont
  {Stockklauser}, \citenamefont {Scarlino}, \citenamefont {Koski},
  \citenamefont {Gasparinetti}, \citenamefont {Andersen}, \citenamefont
  {Reichl}, \citenamefont {Wegscheider}, \citenamefont {Ihn}, \citenamefont
  {Ensslin},\ and\ \citenamefont {Wallraff}}]{stockklauser2017}%
  \BibitemOpen
  \bibfield  {author} {\bibinfo {author} {\bibfnamefont {A.}~\bibnamefont
  {Stockklauser}}, \bibinfo {author} {\bibfnamefont {P.}~\bibnamefont
  {Scarlino}}, \bibinfo {author} {\bibfnamefont {J.~V.}\ \bibnamefont {Koski}},
  \bibinfo {author} {\bibfnamefont {S.}~\bibnamefont {Gasparinetti}}, \bibinfo
  {author} {\bibfnamefont {C.~K.}\ \bibnamefont {Andersen}}, \bibinfo {author}
  {\bibfnamefont {C.}~\bibnamefont {Reichl}}, \bibinfo {author} {\bibfnamefont
  {W.}~\bibnamefont {Wegscheider}}, \bibinfo {author} {\bibfnamefont
  {T.}~\bibnamefont {Ihn}}, \bibinfo {author} {\bibfnamefont {K.}~\bibnamefont
  {Ensslin}}, \ and\ \bibinfo {author} {\bibfnamefont {A.}~\bibnamefont
  {Wallraff}},\ }\href {\doibase 10.1103/PhysRevX.7.011030} {\bibfield
  {journal} {\bibinfo  {journal} {Phys. Rev. X}\ }\textbf {\bibinfo {volume}
  {7}},\ \bibinfo {pages} {011030} (\bibinfo {year} {2017})}\BibitemShut
  {NoStop}%
\bibitem [{\citenamefont {Bosco}\ \emph {et~al.}(2019)\citenamefont {Bosco},
  \citenamefont {DiVincenzo},\ and\ \citenamefont {Reilly}}]{bosco2019}%
  \BibitemOpen
  \bibfield  {author} {\bibinfo {author} {\bibfnamefont {S.}~\bibnamefont
  {Bosco}}, \bibinfo {author} {\bibfnamefont {D.~P.}\ \bibnamefont
  {DiVincenzo}}, \ and\ \bibinfo {author} {\bibfnamefont {D.~J.}\ \bibnamefont
  {Reilly}},\ }\href {\doibase 10.1103/PhysRevApplied.12.014030} {\bibfield
  {journal} {\bibinfo  {journal} {Phys. Rev. Appl.}\ }\textbf {\bibinfo
  {volume} {12}},\ \bibinfo {pages} {014030} (\bibinfo {year}
  {2019})}\BibitemShut {NoStop}%
\bibitem [{\citenamefont {Burkard}\ and\ \citenamefont
  {Petta}(2016)}]{burkard2016}%
  \BibitemOpen
  \bibfield  {author} {\bibinfo {author} {\bibfnamefont {G.}~\bibnamefont
  {Burkard}}\ and\ \bibinfo {author} {\bibfnamefont {J.~R.}\ \bibnamefont
  {Petta}},\ }\href {\doibase 10.1103/PhysRevB.94.195305} {\bibfield  {journal}
  {\bibinfo  {journal} {Phys. Rev. B}\ }\textbf {\bibinfo {volume} {94}},\
  \bibinfo {pages} {195305} (\bibinfo {year} {2016})}\BibitemShut {NoStop}%
\bibitem [{\citenamefont {Boissonneault}\ \emph {et~al.}(2008)\citenamefont
  {Boissonneault}, \citenamefont {Gambetta},\ and\ \citenamefont
  {Blais}}]{boissonneault2008}%
  \BibitemOpen
  \bibfield  {author} {\bibinfo {author} {\bibfnamefont {M.}~\bibnamefont
  {Boissonneault}}, \bibinfo {author} {\bibfnamefont {J.~M.}\ \bibnamefont
  {Gambetta}}, \ and\ \bibinfo {author} {\bibfnamefont {A.}~\bibnamefont
  {Blais}},\ }\href {\doibase 10.1103/PhysRevA.77.060305} {\bibfield  {journal}
  {\bibinfo  {journal} {Phys. Rev. A}\ }\textbf {\bibinfo {volume} {77}},\
  \bibinfo {pages} {060305(R)} (\bibinfo {year} {2008})}\BibitemShut {NoStop}%
\bibitem [{\citenamefont {Slichter}\ \emph {et~al.}(2012)\citenamefont
  {Slichter}, \citenamefont {Vijay}, \citenamefont {Weber}, \citenamefont
  {Boutin}, \citenamefont {Boissonneault}, \citenamefont {Gambetta},
  \citenamefont {Blais},\ and\ \citenamefont {Siddiqi}}]{slichter2012}%
  \BibitemOpen
  \bibfield  {author} {\bibinfo {author} {\bibfnamefont {D.~H.}\ \bibnamefont
  {Slichter}}, \bibinfo {author} {\bibfnamefont {R.}~\bibnamefont {Vijay}},
  \bibinfo {author} {\bibfnamefont {S.~J.}\ \bibnamefont {Weber}}, \bibinfo
  {author} {\bibfnamefont {S.}~\bibnamefont {Boutin}}, \bibinfo {author}
  {\bibfnamefont {M.}~\bibnamefont {Boissonneault}}, \bibinfo {author}
  {\bibfnamefont {J.~M.}\ \bibnamefont {Gambetta}}, \bibinfo {author}
  {\bibfnamefont {A.}~\bibnamefont {Blais}}, \ and\ \bibinfo {author}
  {\bibfnamefont {I.}~\bibnamefont {Siddiqi}},\ }\href {\doibase
  10.1103/PhysRevLett.109.153601} {\bibfield  {journal} {\bibinfo  {journal}
  {Phys. Rev. Lett.}\ }\textbf {\bibinfo {volume} {109}},\ \bibinfo {pages}
  {153601} (\bibinfo {year} {2012})}\BibitemShut {NoStop}%
\end{thebibliography}%

\end{document}